\documentclass[useAMS,usenatbib]{mnras}
\usepackage{graphicx,amsmath,url,natbib}

\usepackage{multirow}
\usepackage[flushleft]{threeparttable}

\newcommand{\gtrsim}{\,\raisebox{-0.4ex}{$\stackrel{>}{\scriptstyle\sim}$}\,}
\newcommand{\lesssim}{\,\raisebox{-0.4ex}{$\stackrel{<}{\scriptstyle\sim}$}\,}
\newcommand{\mach}{\mathcal{M}}
\newcommand{\ergs}{\mbox{ erg s}^{-1}}
\newcommand{\cc}{\mbox{ cm}^{-3}}
\newcommand{\kms}{\mbox{ km s}^{-1}}

\begin{document}

\title[Relativistic jet feedback]{Relativistic jet feedback in high-redshift galaxies I: Dynamics}
\author[Mukherjee et al.]{Dipanjan Mukherjee$^1$\thanks{dipanjan.mukherjee@anu.edu.au},
Geoffrey V. Bicknell$^1$, Ralph Sutherland$^1$ \& Alex Wagner$^2$ \\
$^1$Research School of Astronomy and Astrophysics, Australian National University, Canberra, ACT 2611, Australia \\
$^2$Center for Computational Sciences, University of Tsukuba, 1-1-1 Tennodai, Tsukuba, Ibaraki, 305-8577}
\date{\today}
\pagerange{\pageref{firstpage}--\pageref{lastpage}} 
\pubyear{2015}
\maketitle

\begin{abstract}
We present the results of three dimensional relativistic hydrodynamic simulations of interaction of AGN jets with a dense turbulent two-phase interstellar medium, which would be typical of high redshift galaxies. We describe the effect of the jet on the evolution of the density  of the turbulent ISM. The jet driven energy bubble affects the gas to distances up to several kiloparsecs from the injection region. The shocks resulting from such interactions create a multi-phase ISM and radial outflows. One of the striking result of this work is that low power jets ($P_{\rm jet} \lesssim 10^{43} \ergs$) although less efficient in accelerating clouds, are trapped in the ISM for a longer time and hence affect the ISM over a larger volume. Jets of higher power drill through with relative ease. Although the relativistic jets launch strong outflows, there is little net mass ejection to very large distances, supporting a galactic fountain scenario for local feedback. 
\end{abstract}


\section{Introduction}\label{sec:intro}
Feedback from Active Galactic Nuclei (AGN) has long been identified as playing an important role in the evolution of galaxies \citep[e.g.][]{silk98a,dimatteo05a,bower06a,croton06a,schawinski07a}. It has been proposed that momentum-driven or energy-driven jets and winds powered by the central blackhole significantly affect the gas content and star formation of galaxies \citep[e.g.][]{silk98a,dimatteo05a,murray05,ciotti10a,dubois13a}. However, only a few papers have addressed the complex nature of the interaction of such winds with a dense multiphase ISM \citep[see for example][]{hopkins10a,gabor14a,hopkins16a}. \citet{oppenheimer10a} and \citet{dave12a} have considered a galactic fountain scenario in which there is a recurrent cycle of blow out of gas and its subsequent infall. However for galaxies of masses $\gtrsim 10^{11} M_\odot$, \citet{oppenheimer10a} find the need for an additional quenching mechanism, possibly due to AGN, to suppress excess accretion of gas and star formation in order to match the galaxy stellar mass function.

Our work concentrates on the role of radio galaxies in AGN feedback, specifically the role played by their relativistic jets. This is motivated by investigations of the radio/optical luminosity function, which have shown that the probability of a galaxy being a radio source increases with optical luminosity \citep{auriemma77a,sadler89a,ledlow96a,best05a,mauch07a}. It is the most luminous part of the optical luminosity function where discrepancies between hierarchical models of galaxy formation and observation are most apparent \citep[e.g.][]{croton06a} and considerations of the radio-optical luminosity function indicate that it is in the optically luminous galaxies in which radio sources are most likely to play an important role. This reinforces the potential role of relativistic jets in AGN feedback. Nevertheless, not every optically luminous galaxy is a radio galaxy, indicating that radio galaxies are an intermittent phenomenon and that jet feedback is also necessarily intermittent. 

Effects of feedback from relativistic jets have been investigated more in the context of heating the intra-cluster medium to prevent catastrophic cooling and accretion of gas to the cluster centre \citep[e.g.][]{binney95a,soker01a,gaspari12a}. However, relativistic jets are also expected to be one of the major drivers of feedback on galactic scale, as supported by several observational evidences of jet-ISM interaction \citep[a few recent works being ][]{mahony16,collet16a,dasyra15a,morganti15a,ogle14a,dasyra14a,morganti13a,lanz10a,tadhunter14a,nesvadba11c,nesvadba08b,nesvadba07a}. However, only a few theoretical papers \citep{sutherland07a,gaibler11a,wagner11b,gaibler12a,wagner12a} have addressed the question of how a relativistic jet interacts with a multi-phase ISM of the host galaxy, and over what scales such interactions are relevant.

In this work we extend the results presented in \citet[hereafter WB11]{wagner11a} and \citet[hereafter WBU12]{wagner12a}. The simulations presented in those papers consist of gas distributed on a scale $\sim 1 \> \rm kpc$ in the form of a two phase turbulent ISM, modeled as a fractal with a log normal density distribution and a Kolmogorov spectrum. A number of useful parameters were derived: the average radial velocity of the clouds, the fraction of jet energy transferred to the kinetic energy of the clouds and the mechanical advantage of the interaction implied by the ratio of cloud momentum to jet momentum. The dependences of these quantities on cloud density and filling factor were examined. The simulations confirmed the results derived in \citet{saxton05a}, namely that the originally well-directed jets form an energy-driven almost spherical bubble, which processes $4 \pi$ steradians of the galaxy atmosphere. Moreover, the outflow velocities derived from the simulations compare well with numerous observations of radio galaxies \citep{wagner12a}. The conditions under which the thermal gas would be dispersed were also established.

In WB11 and WBU12 gas was considered to be ``dispersed" when its radial velocity exceeds the velocity dispersion of the host galaxy. However, while useful, this approach does not fully address the ultimate fate of potentially star-forming gas interacting with relativistic jets. Is it completely ejected from the atmosphere of the host -- or does it simply become turbulent -- impeding for a time, but not indefinitely, the formation of new stars? What happens when the jet breaks free of the dense gas surrounding the nucleus? Also, in WB11 and WBU12, the dense clouds were static without any associated velocity dispersion. 

Hence, in this paper we present the next step in this program of simulations, adding the following significant features: (1) A gravitational field typical of that of an elliptical galaxy consisting of luminous and dark matter \citep[see][]{sutherland07a} (2) An internal velocity dispersion for the dense thermal gas; this is used to establish an initial turbulent interstellar medium (ISM) consistent with observations of high redshift elliptical galaxies \citep{forster09,wisnioski15}. (3) Both phases of the ISM, consisting of hot gas at around the virial temperature and the warm gas at a temperature of about $10^4 \> \rm K$ are distributed consistently with the gravitational field. This restricts the dense, warm gas to a region of order the core radius of the stellar distribution, thereby defining the region for jet break and the timescale over which the jet significantly affects the distribution and kinematics of the dense gas. (4) A scale of 5~kpc for the simulations, significantly larger than the 1~kpc scale in our previous work.

In the following section (Sec.~\ref{sec.setup}) we describe the simulation setup in detail and in Sec.~\ref{sec.settling} we document how the ISM is settled from its initial configuration. In Sec.~\ref{sec.jetsims} we examine the impact of a relativistic jet with power $10^{45} \ergs$ on the turbulent ISM. We describe the evolution of the ISM density and the multiphase nature of the ISM resulting from shocks driven by the jet. In Sec.~\ref{sec.diffpower}, we examine the dependence of morphology on jet power by comparing the results of four jet simulations with jet kinetic powers ranging from $10^{43}$ to $10^{45} \> \rm ergs \> s^{-1}$. We emphasise the efficiency of low power jets in coupling with the ISM. In Sec.~\ref{sec.energetics} we consider the energetics of the disturbed ISM, including a discussion of the galactic fountains revealed by our simulations and summarise the results in Sec.~\ref{sec.summary}.

\section{Simulation setup}\label{sec.setup}
\begin{table}
\centering
\caption{ Parameters of the ambient gas and gravitational potential common to all simulations. }
\label{tab.params}
\begin{tabular}{| l | l | l |}
\hline
\multicolumn{2}{c|}{Parameters} 	& Value   \\
\hline
Baryonic core radius  & $r_B$		& 1 kpc			\\
Baryonic velocity dispersion  & $\sigma _B$	& 250 km s$^{-1}$	\\
Ratio of DM to Baryonic & $\lambda$		& 2	\\
core radius 	&& 		\\
Ratio of DM to Baryonic & $\kappa$		& 10			\\
velocity dispersion && \\
Halo Temperature  & $T_h$		& $10^7$ K		\\
Halo density at r=0  & $n_0$		& $0.5$ cm$^{-3}$ 	\\
Turbulent velocity dispersion$^a$  & $\sigma _t$	& 250 km s$^{-1}$	\\
of warm clouds && \\
\hline
\end{tabular} \\
\begin{tablenotes}
{\small
\item (a) Defines the extent of the cloud distribution in Eq.~\ref{eq.nw}  }
\end{tablenotes}
\end{table}

\subsection{Gravitational potential}
We model the gravity of the host galaxy by prescribing a spherically symmetric isothermal potential as a function of radius $r$ for both the dark matter and baryonic components. Let the velocity dispersions of the dark and baryonic matter be 
$\sigma_{\rm D}$ and $\sigma_{\rm B}$ respectively, the dark matter core radius $r_{\rm D}$, the baryonic core radius $r_{\rm B}$ and the normalised radius $r^\prime = r/r_{\rm D}$. Let $\phi$ be the gravitational potential and $\psi = \phi/\sigma_{\rm D}^2$ the normalised potential.
The net gravitational potential for both components is obtained by solving Poisson's equation  \citep[see][for a detailed derivation]{sutherland07a}:
\begin{equation}
\label{eq.poisson}
\frac{d^2 \psi}{dr'^2} + \frac{2}{r}\frac{d \psi}{d r'} = 9 \left[\exp\left(-\psi\right) + 
\frac{\lambda ^2}{\kappa ^2} \exp \left(-\kappa ^2 \psi \right) \right]
\end{equation}
The normalised potential in eq.~\ref{eq.poisson} is characterised by two parameters: $\lambda=r_D/r_B$ and $\kappa = \sigma _D/\sigma _B$. In our simulations we use $\kappa =2$ and $\lambda = 10$. 

\subsection{Initialisation of the simulation}\label{sec.initialisation}
We initialize the simulation domain with a two phase, spherically distributed medium following the approach of \citet{sutherland07a,wagner11a,wagner12a}. The density consists of an isothermal hot at $T \sim 10^7$ K \citep[typical of galaxy clusters and elliptical galaxies][]{allen06a,croston08a,diehl08a,maughan12a,goulding16a} halo in hydrostatic equilibrium and a dense warm ($T \lesssim 3.4\times 10^4$ K) turbulent and inhomogeneous gas. 
The density of the hot halo is described by: 
\begin{equation}\label{eq.nhot}
n_{\rm h}=n_0\exp\left(-\frac{\mu m_a \phi}{k_BT}\right)
\end{equation}
where $n_0$ is the central number density, $\mu=0.6165$ is the mean molecular weight and $m_a$ is the atomic mass unit. For this work we choose $n_0=0.5 \cc$, similar to values of central gas densities inferred from X-ray observations of diffuse halos around elliptical galaxies and galaxy clusters \citep{allen06a,croston08a,goulding16a}. The pressure, $p = n_{\rm h} k_{\rm B} T$, is evaluated from the specified temperature and eq.~(\ref{eq.nhot}). 

The density in the warm phase is distributed as a fractal with a single point lognormal density distribution and a Kolmogorov power spectrum 
\begin{equation}
D(k) = \int 4\pi k^2 F(\mathbf{k})F^*(\mathbf{k})dk \propto k^{-5/3},
\end{equation}
$F(\mathbf{k})$ being the Fourier transform. The fractal density distribution is created using the publicly available pyFC\footnote{\url{https://pypi.python.org/pypi/pyFC}} routine (written by AYW) with mean $\mu_{\rm PDF}=1$ and $\sigma _{\rm PDF} ^2 =5$. The resultant distribution ($n_{\rm fractal}$) is then apodized to represent a spherical isotropic, turbulent distribution in the gravitational potential, as follows. Let $\sigma_{\rm t}$ be the turbulent velocity dispersion of the warm gas and $T_{\rm w}$ its temperature, with $\sigma_{\rm t}^2 \gg 3 k_{\rm B}/\mu m$. Then the density distribution of warm gas is given by:
\begin{equation}\label{eq.nw}
n_w(r,z) = n_{\rm fractal} \times n_{w0}\exp\left[-\frac{\phi(r,z)-\phi(0,0)}{\sigma_t^2}\right]
\end{equation} 
$n_{w0}$ is the number density at $(0,0)$ \citep[see][for details]{sutherland07a}. For our simulations we assume the mean central density of the warm clouds to be $\sim 100-300 \cc$, which is consistent with typical densities of ISM inferred in high Z galaxies \citep[see e.g.][]{shirazi14a,sanders16a}.  Table~\ref{tab.params} and Table~\ref{tab.jetparams} present the values of the parameters used in our simulations. The warm phase is initialised to be at the same pressure as the hot halo at a given location, so that the entire domain is in pressure equilibrium. A lower bound is placed on the density of the warm phase corresponding to a temperature of $T_{\rm crit}=3.4 \times 10^4$ K, beyond which the clouds are considered to be thermally unstable; gas in those cells is replaced by hot gas.

\subsection{Jet Parameters}\label{sec.jetparams}
Following \citet{bicknell95a}, we express the kinetic jet power as follows. The jet parameters are the pressure $p_{\rm jet}$, velocity with respect to the speed of light $\beta=v/c$, Lorentz factor $\Gamma =1/\sqrt{1-\beta ^2}$,
cross-sectional area $A_{\rm jet}$, adiabatic index $\gamma _{\rm ad}$ and the density parameter $\chi$, which is the ratio of the rest mass energy to the enthalpy. The jet power is
\begin{equation}\label{eq.jetpower}
P_{\rm jet} =\frac{\gamma _{\rm ad}}{\gamma _{\rm ad} -1} c p_{\rm jet} \Gamma ^2 \beta A_{\rm jet} \left(1 + \frac{\Gamma -1}{\Gamma} \chi \right).
\end{equation}
The parameter $\chi$ is given by
\begin{equation}
\chi =\left(\frac{\gamma _{\rm ad} -1}{\gamma _{\rm ad}}\right)\frac{\rho c^2}{ p_{\rm jet}}.
\end{equation}
In our simulations we inject the jet at the lower $z$ boundary of the computation domain, $z=z_0$, assuming conical expansion between $z=0$, the location of the black hole and $z=z_0$. The radius of the inlet region is constrained by the grid resolution such that at least 10 cells cover the jet inlet. For our simulations we set the inlet radius to be 30 pc. The jet is injected with a half opening angle of $10^\circ$ so that $z_0 \doteq 170 \> \rm pc$.  We initialise the jet with a given kinetic power ($P_{\rm jet}$) and Lorentz factor ($\Gamma$). We assume the jet to be in pressure equilibrium with the gas at the inlet, thus constraining the parameter $\chi$ from eq.~\ref{eq.jetpower}; this in turn defines the jet density. For the simulations presented here, $\chi \gtrsim 5$. We assume an ideal equation of state with $\gamma _{\rm ad}=5/3$, which is a reasonable approximation for a relativistic gas in with $\chi >> 1$ \citep{synge57,mignone07a}.  Also, a non-relativistic ideal equation of state is a better descriptor of the thermal gas inside the simulation domain and it is mainly the effect on the thermal gas in which we are interested. 

Adoption of such a value of $\chi$ raises questions about jet composition, which is not very well constrained for extragalactic jets \citep[see, the discussion in][]{worrall09a}. An electron-positron jet would have $\chi \sim 1$ and this value would be obtained if the jet were over-pressured with respect to the ISM by a factor of 5. On the other hand, if the jet is in pressure equilibrium with the ISM, then it may entrain some thermal material as it travels to $\sim 170$ pc from the nucleus, which is the starting point of our simulation. 

As \citet{worrall09a} has noted, the dominant contribution in radio power, comes
from sources around the FRI/FRII break, corresponding to the peak of the curve $P
\Phi(P)$, where $P$ is the radio power and $\Phi(P)$ is the number density of
sources per unit $\log P$. The relationship between radio power and jet power is
not straighforward  \citep[see, e.g.][]{godfrey16a}. Nevertheless, \citet{rawlings91a} identified
$10^{43} \> \rm ergs \> s^{-1}$ as the low end of the FRII population;
\citet{bicknell95a} found the  FRI/FRII break jet power to be $\sim 2 \times 10^{42} \> \rm ergs \> s^{-1}$. Hence, in this paper, we concentrate on jet powers ranging from $10^{43}$ to $10^{45} \> \rm ergs \> s^{-1}$, whilst noting that investigations of jets of both lower and higher powers are certainly of interest.

\subsection{PLUTO setup}
We perform 3D relativistic hydrodynamic simulations using the PLUTO code's Relativistic Hydrodynamic (RHD) module \citep{mignone07}. We use a Cartesian geometry with a uniform grid of resolution 6 pc for the central 3 kpc, followed by a geometrically stretched grid with stretching ratio of $\sim 1.0128$, extending up to $\pm 2.4$ kpc along the x--y directions and $\sim 5.2$ kpc in the z direction. The total number of grid points along the x-y-z directions are $668\times668\times640$. We use the piecewise parabolic method \citep{colella84,marti96} for the reconstruction step of the Godunov scheme, which is well suited for non-uniform grids. The time evolution is carried out using third order dimensionally unsplit Runge-Kutta method. The hllc Riemann solver \citep{toro08} is used for solving the hydrodynamic equations.

The non-equilibrium cooling function was evaluated from the 
Mappings 5.1 code (Sutherland et al. 2016, in prep.)  This code is the latest version of the
Mappings 4.0 code described in \citet{Nicholls2013,Dopita2013}, and
includes numerous upgrades to both the input atomic physics \citet[CHIANTI v8][]{delZanna15}
and new methods of solution.  MAPPINGS~V includes up to 30
elements from H to Zn, of which about 10-15 provide most of the cooling. For most temperatures,
Oxygen and Iron dominate the cooling except in some temperature regimes
(very hot and ~$10^4$~K) where collisional cooling of Hydrogen and Helium are
key.  In these models we have adopted solar abundances from \citep{asplund09a}
as representative of metallicities of larger host galaxies.

The cooling function is constructed by having the plasma initially fully
ionised at an extremely high temperature, $10^9$~K, where the thermal cooling
is primarily free--free emission, and the ions are fully stripped.   This high
temperature is outside the range expected in the simulations, and in a regime
where cooling is unimportant.  Without more detailed microphysics, such
as a fully relativistic treatment of free-free emission for example, the
MAPPINGS cooling functions above $\sim 10^{8.5}$~K or so are not intended for
detailed model fitting, but serve as a smooth upper boundary to the cooling
which improves the numerical properties of the cooling treatment. For gas below $T<10^4$ K 
cooling was deactivated.

The plasma in the MAPPINGS model is allowed to cool in a time dependent isobaric way,
similar to a post shock flow \citep{Sutherland1993,Allen2008}.  
Cooling down to ~$10^6$~K proceeds with equilibrium ionisation and
cooling, until the cooling becomes rapid compared to the recombination
timescales and the ionisation lags behind, being more ionised at a given
temperature than in equilibrium.  Below $10^6$~K, the cooling rates increase to
a maximum around $\sim 10^5$~K, before falling rapidly below $10^4$~K.  At
each point, the full ionisation state including electron densities and atomic level
populations, are solved, allowing the cooling and a simple equation of state
to be inferred from the self-consistently changing mean molecular weight
$\mu(T)$.  The gas is assumed to be atomic, and to have an ideal adiabatic
index of $5/3$. The temperature dependent cooling function 
and mean molecular weight thus obtained were tabulated as a function of the ratio of pressure
and density ($p/\rho$), which are PLUTO primitive variables.  The cooling losses $\left(\left[ \rho/\mu(T)\right]^2 \Lambda(T)\right)$ for each cell 
in the PLUTO domain were applied by interpolating the cooling function and mean molecular weight 
from the tabulated list.

\section{Settling of ISM}\label{sec.settling}
\begin{figure*}
	\centering
	\includegraphics[width = 8.5cm, keepaspectratio] {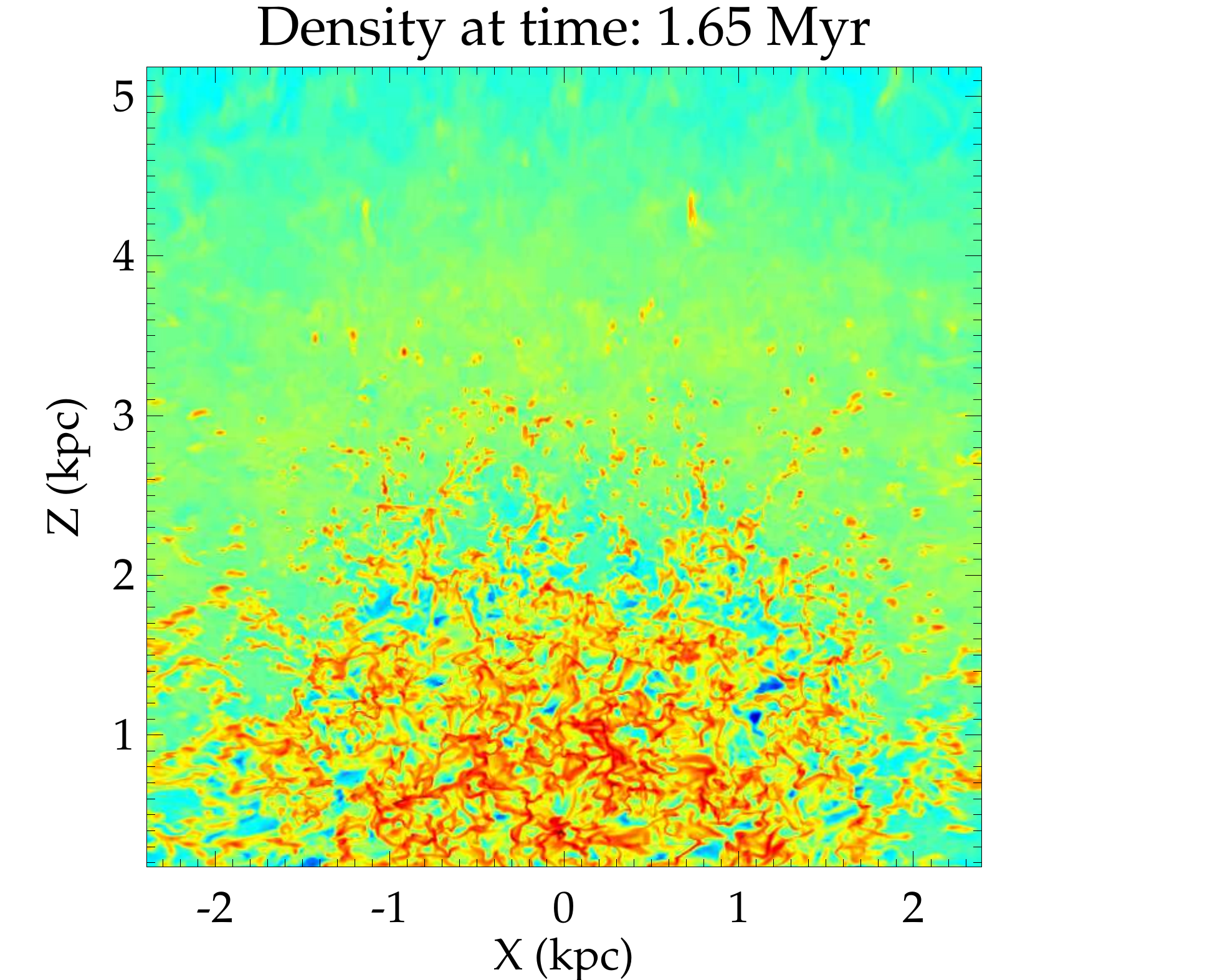}\hspace{-2.7cm}
	\includegraphics[width = 8.5cm, keepaspectratio] {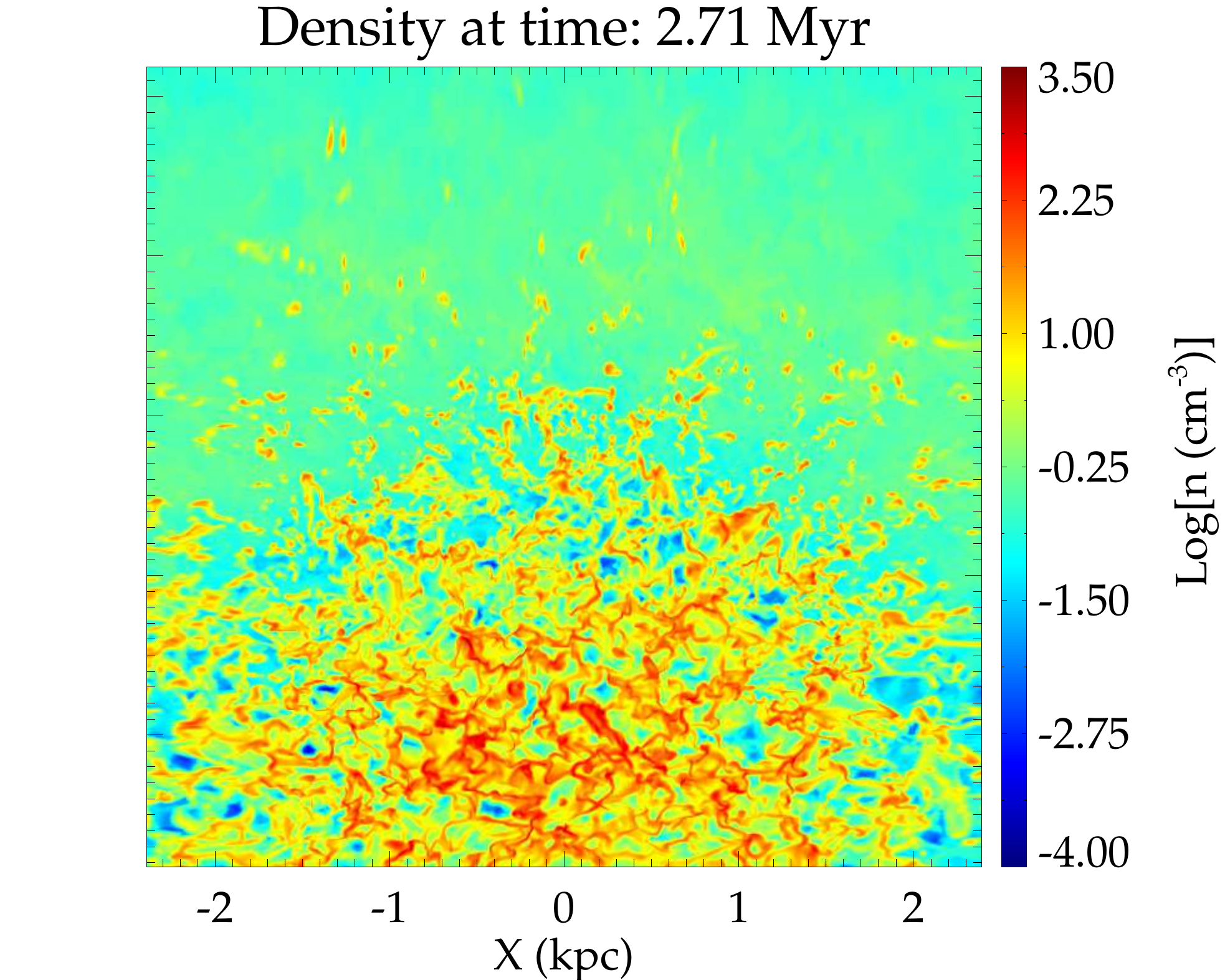}
	\caption{\small Density ($\log[n (\cc)]$) in the $x-z$ plane for settling turbulent ISM at two different times. The ISM develops a filamentary structure, typical of a turbulent medium. }
	\label{fig.cldfilaments}
\end{figure*}
\begin{figure}
	\centering
	\includegraphics[width = 7.cm, height = 7.cm,keepaspectratio] {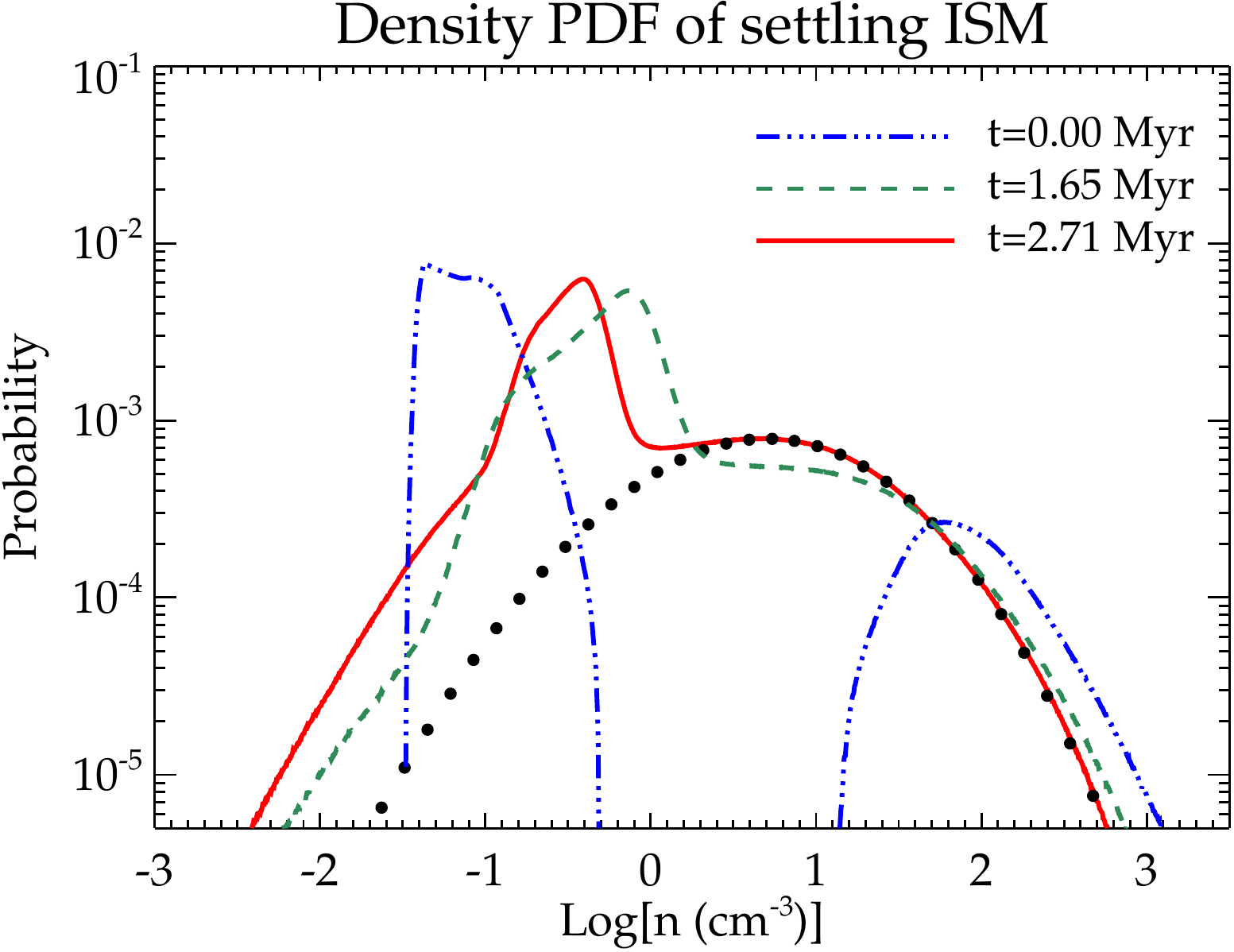}
	\caption{\small The  probability distribution function (PDF) of the density ($\log[n (\cc)]$) for a settling ISM without jet, at $t=0$ (dashed-dotted in blue), $t=1.65$ Myr (dashed in green) and $t=2.66$ Myr (solid red). The black  dotted lines represent a fit to the high density end  of the PDF following eq.~\ref{eq.hopkinsfit}.}
	\label{fig.fitpdf}
\end{figure}
\begin{figure}
	\centering
	\includegraphics[width = 7cm, height = 7cm,keepaspectratio] {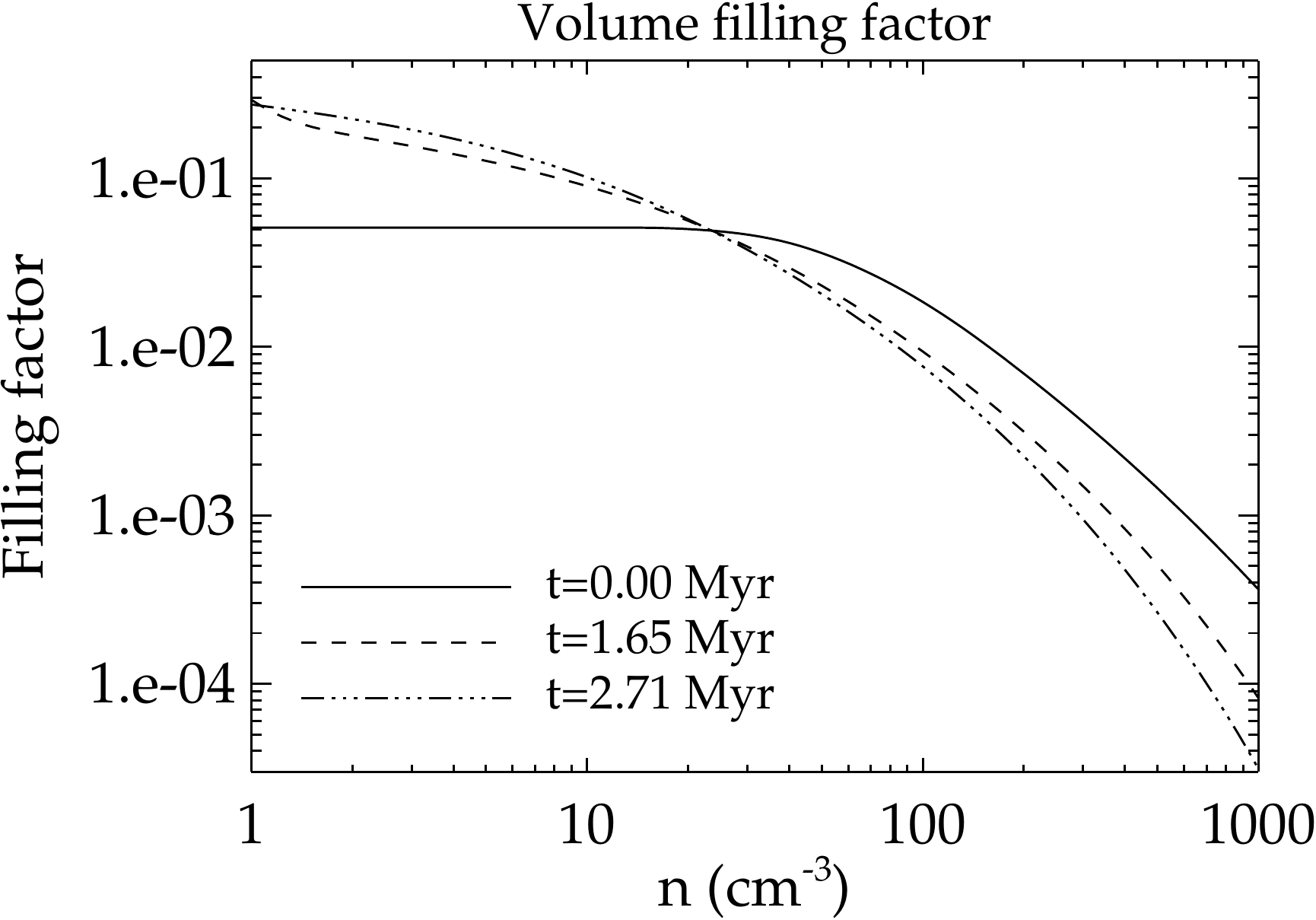}
	\caption{\small The volume filling factor as a function of density (in $\cc$) for a settling ISM.}
	\label{fig.fillfactor}
\end{figure}
\begin{figure}
	\centering
	\includegraphics[width = 7cm, height = 7cm,keepaspectratio] {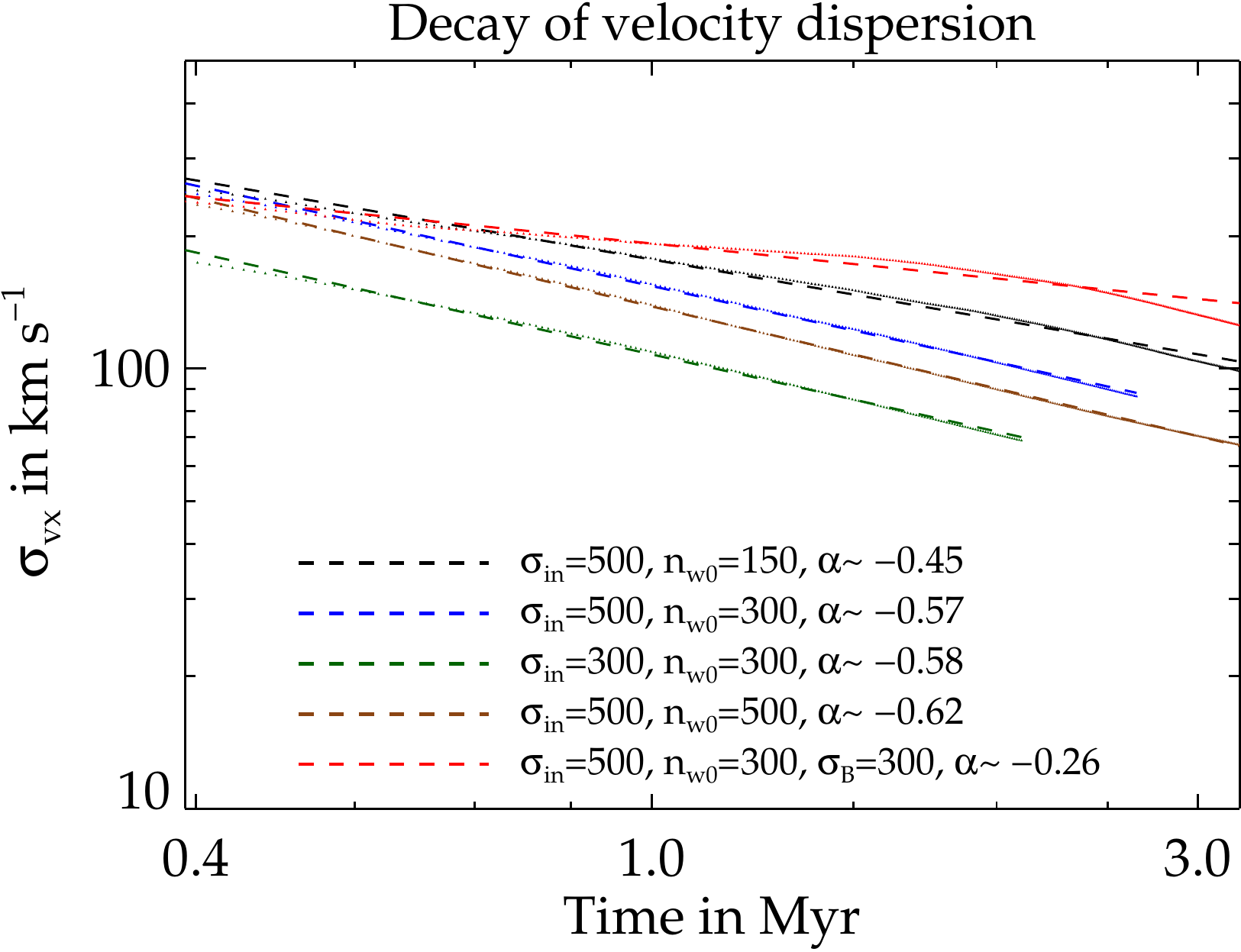}
	\caption{\small  Comparison of the dispersion of $V_x$ (in $\kms$) for different simulations. $\sigma _{\rm in}$ (in $\kms$) is the initial velocity dispersion , $n_{w0}$ (in $\cc$) the mean central density of the warm phase , $\sigma _B$ (in $\kms$) is the dispersion of the baryonic component of the gravitational potential and $\alpha$ is the exponent for a power-law fit to the profiles ($\sigma_{vx} \propto t^\alpha$).}
	\label{fig.sigcompare}
\end{figure}
\begin{itemize}
\item\emph{Filamentary ISM:}
We initialise the velocity in the warm cloudy medium with a turbulent velocity distribution, modelled as a random Gaussian variate for the three velocity components with a Kolmogorov spectrum in Fourier space. We set the velocity dispersion of the clouds higher than that of the baryonic dispersion of the galaxy and let the clouds settle in the potential. The fractal clouds disperse and shear due to the turbulent motions and cloud-cloud collisions. The clouds eventually condense into filaments, typical of a turbulent medium \citep[e.g.][]{federrath16a,federrath15} as shown in Fig.~\ref{fig.cldfilaments}.  After $\sim 1$ Myr, the turbulent distribution settles into a two phase medium (see Fig.~\ref{fig.cldfilaments}) characterised by a distribution of warm filaments extending up to $\sim 2$ kpc and a hot halo extending to larger radii. 

\item\emph{Density PDF and two phase ISM:}
In Fig.~\ref{fig.fitpdf} we show the volume weighted probability distribution function (PDF)\footnote{The volume weighted PDF of variable is constructed by evaluating the histogram and counting the fractional volume of a simulation cell as the weights for a histogram bin.} of the density at different times of the simulation. Initially ($t=0$), the density PDF shows two distinct distributions: a) the warm fractal clouds with high density, b) a low density hot halo. As the ISM evolves under the influence of the turbulent velocity field and gravity, the density PDF changes due to shearing of the clouds. The density PDF converges well after 1 Myr into a two component PDF corresponding to a two phase medium. The turbulent motions result in stripping of the dense clouds, lowering the mean of the high density component of the PDF. The dispersed cloud mass forms a denser halo of gas near the central region.  

Density structures formed as a result of hierarchical turbulent processes are expected to exhibit a log-normal probability distribution \citep[e.g.][]{vazquez94a,padoan97a,klessen2000a}. However recent simulations have shown significant deviation from a log-normal behaviour in the tail of the distribution \citep{federrath10,konstandin12a,federrath13a,federrath15}. \citet{hopkins13a} has shown that  a lognormal-like distribution modified by the influence of intermittency from turbulent shocks is a better descriptor of the PDF in such cases (see Appendix~\ref{sec.hopkinsfit} for a brief summary of the analytical expressions). For our simulations, we find the Hopkins function to provide a good fit to the high density component of the PDFs. For example, the black-dotted line in Fig.~\ref{fig.fitpdf} shows the Hopkins function with parameters $\bar{\rho}=15 \cc$, $\sigma _\rho=39.78 \cc$, $\eta=0.08$, which provides a good fit to the high density portion of the PDF at $t\sim 2.7$ Myr. For the rest of the paper, we have used the Hopkins function to fit the high density end of the density PDF and compare the statistics of the ISM under different conditions.

\item\emph{Volume filling factor:} 
In Fig.~\ref{fig.fillfactor} we show the change in the volume filling factor, defined as the total volume occupied by the gas beyond a threshold density, plotted as a function of density. At $t=0$, the volume filling factor for $n > 10$ is $\lesssim 0.045$. As the clouds shear and settle into filaments, the filling factor is lowered for the high density cores as the some of the warm gas is dispersed. 

\item\emph{Decay of turbulence:}
As the clouds settle, the velocity dispersion decreases as a power law with time. A power-law decay of velocity dispersion (with exponent $\sim 1.2-2$) is typical of hydrodynamic supersonic turbulence \citep{stone98a,maclow98}. The rate of decay does not depend on the initial velocity dispersion. For example, as shown in Fig.~\ref{fig.sigcompare}, the rate of decay for $\sigma _{\rm in}=500$ and $\sigma _{\rm in}=300$ for the same $n_w=300$ is similar (with the power-law exponent $\alpha \sim -0.58$.  However, as a result of the presence of atomic cooling and external gravity, the rate of settling depends on the mean cloud density and the potential of the galaxy. The settling rate is slower for lower mean cloud density and a gravitational potential with higher stellar dispersion.

\end{itemize}


\section{Jet simulations}\label{sec.jetsims} 
\subsection{Evolution of the density of the ISM}
\begin{table}
\centering
\caption{Jet simulations}\label{tab.jetparams}
\begin{tabular}{| l | l | l |}
\hline
Sim. label & Power 		& $n_{w0}$    \\
	   & (in ergs$^{-1}$)	& (in cm$^{-3}$) \\
\hline
A  & $10^{45}$			& 300 \\
B  & $10^{45}$			& 150 \\
C  & $10^{44}$			& 150 \\
D  & $10^{43}$			& 300 \\
\hline
\end{tabular} 
\end{table}
\begin{figure*}
	\centering
	\includegraphics[width = 7.5cm, keepaspectratio] 
	{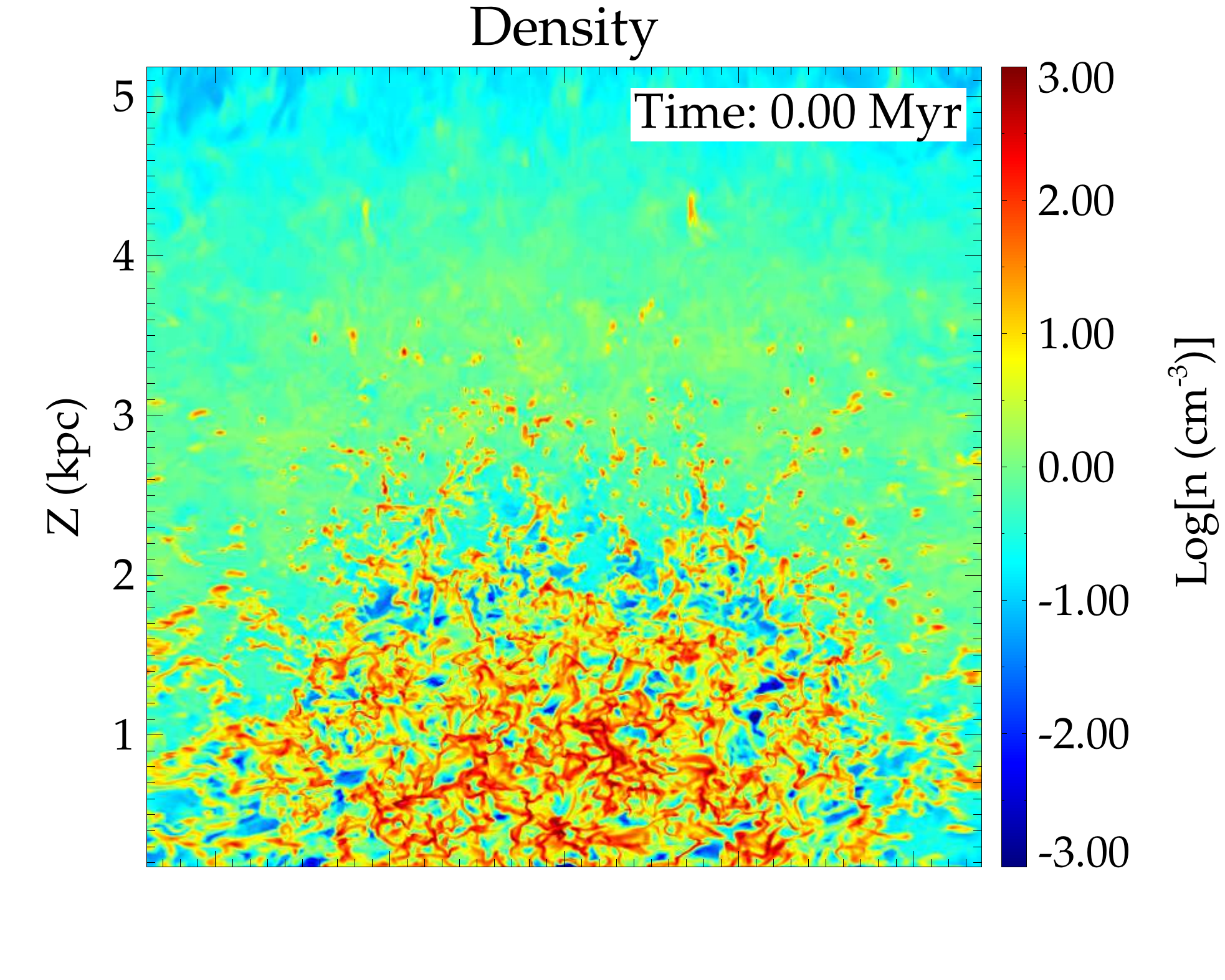}\vspace{-0.52cm}
	\includegraphics[width = 7.5cm, keepaspectratio] 
	{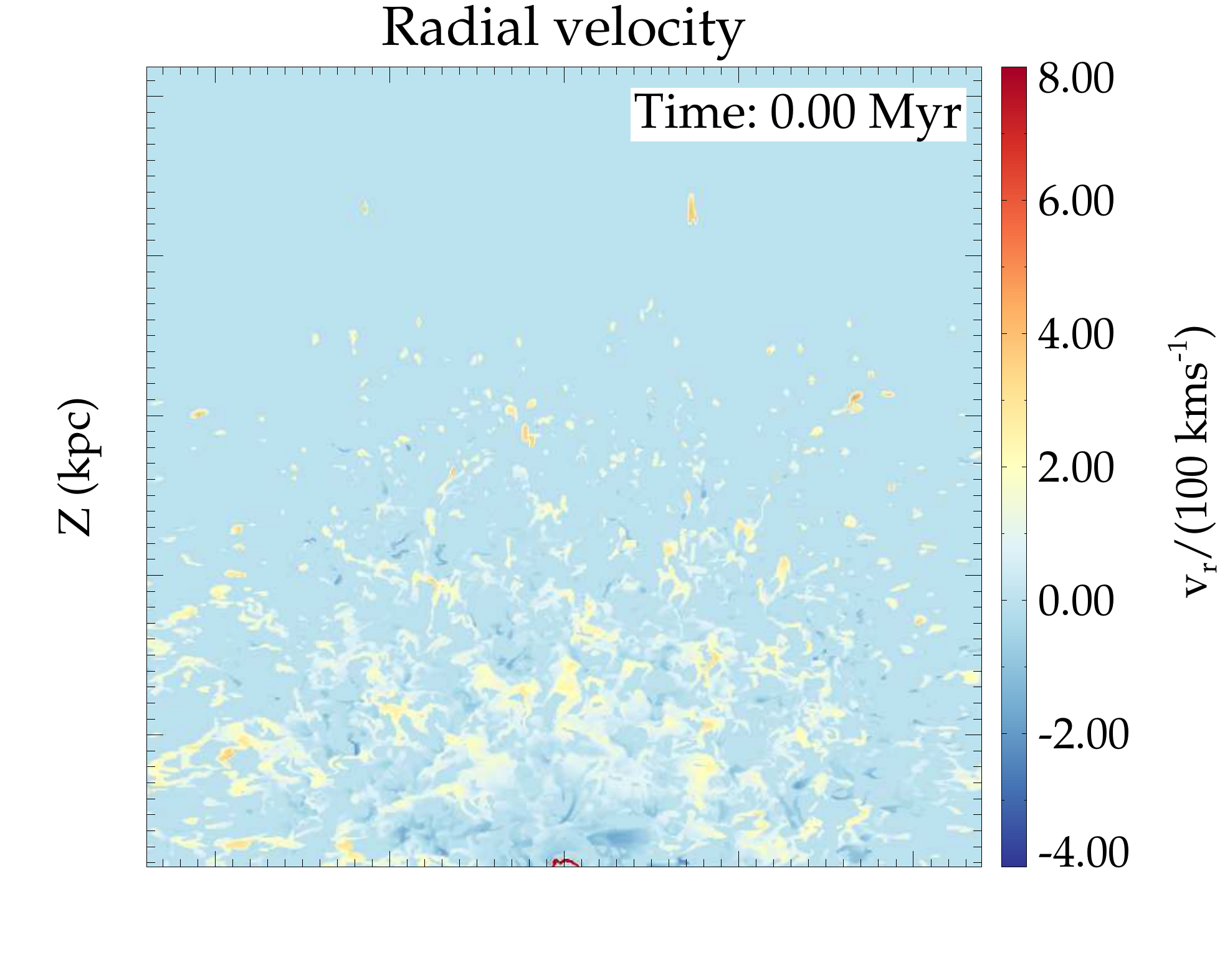}\vspace{-0.52cm}\linebreak
	\includegraphics[width = 7.5cm, keepaspectratio] 
	{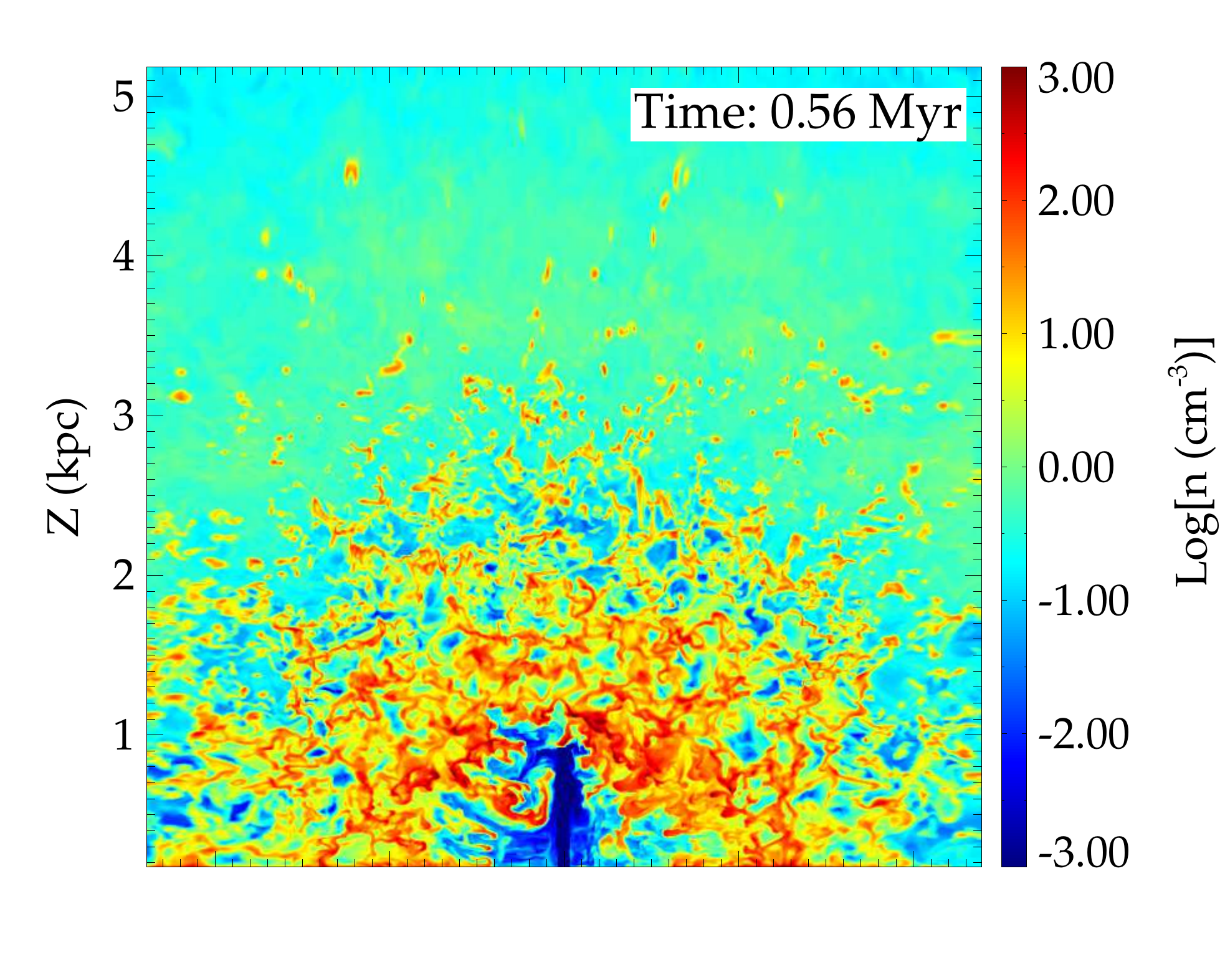}\vspace{-0.52cm}
	\includegraphics[width = 7.5cm, keepaspectratio] 
	{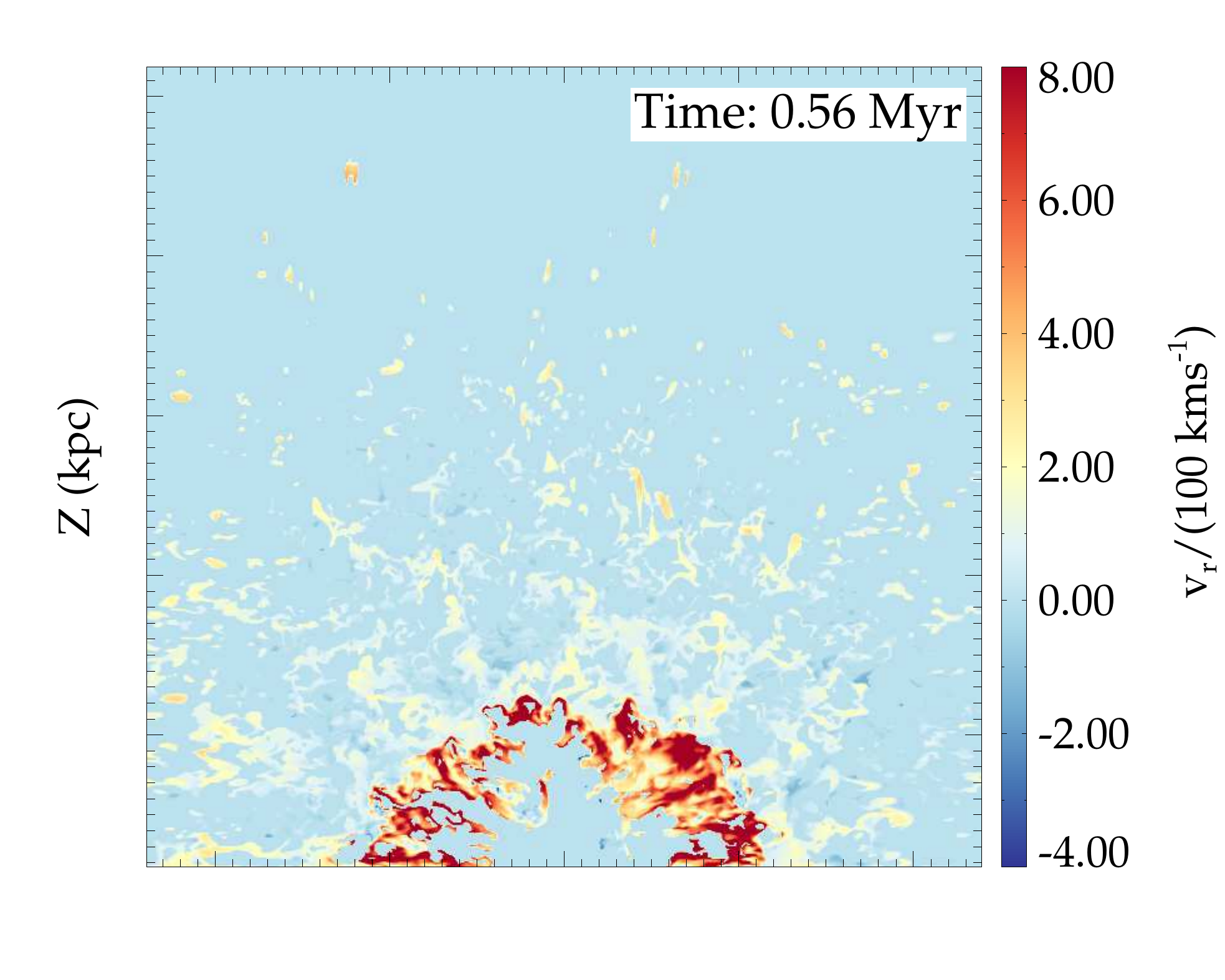}\vspace{-0.52cm}\linebreak
	\includegraphics[width = 7.5cm, keepaspectratio] 
	{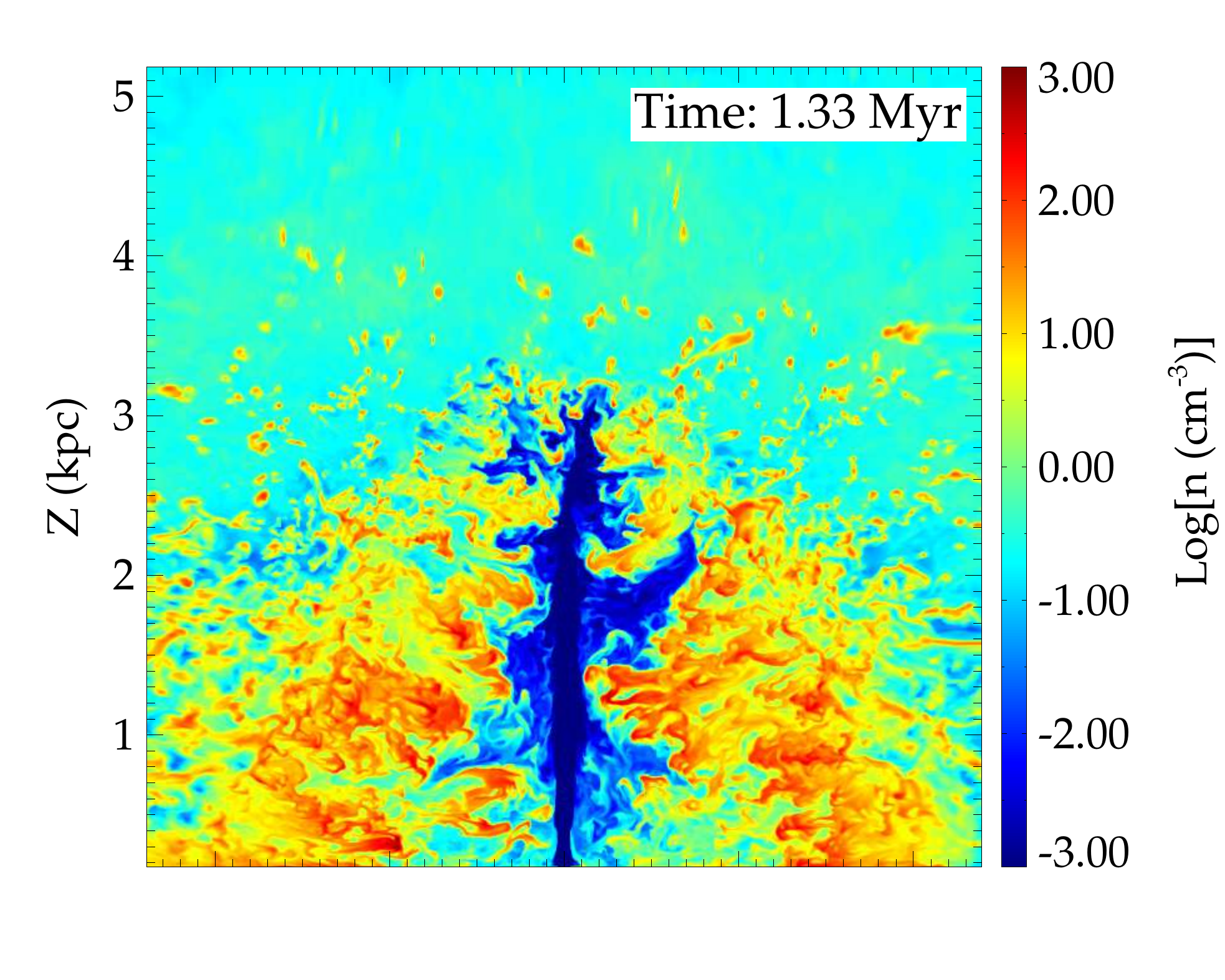}\vspace{-0.52cm}
	\includegraphics[width = 7.5cm, keepaspectratio] 
	{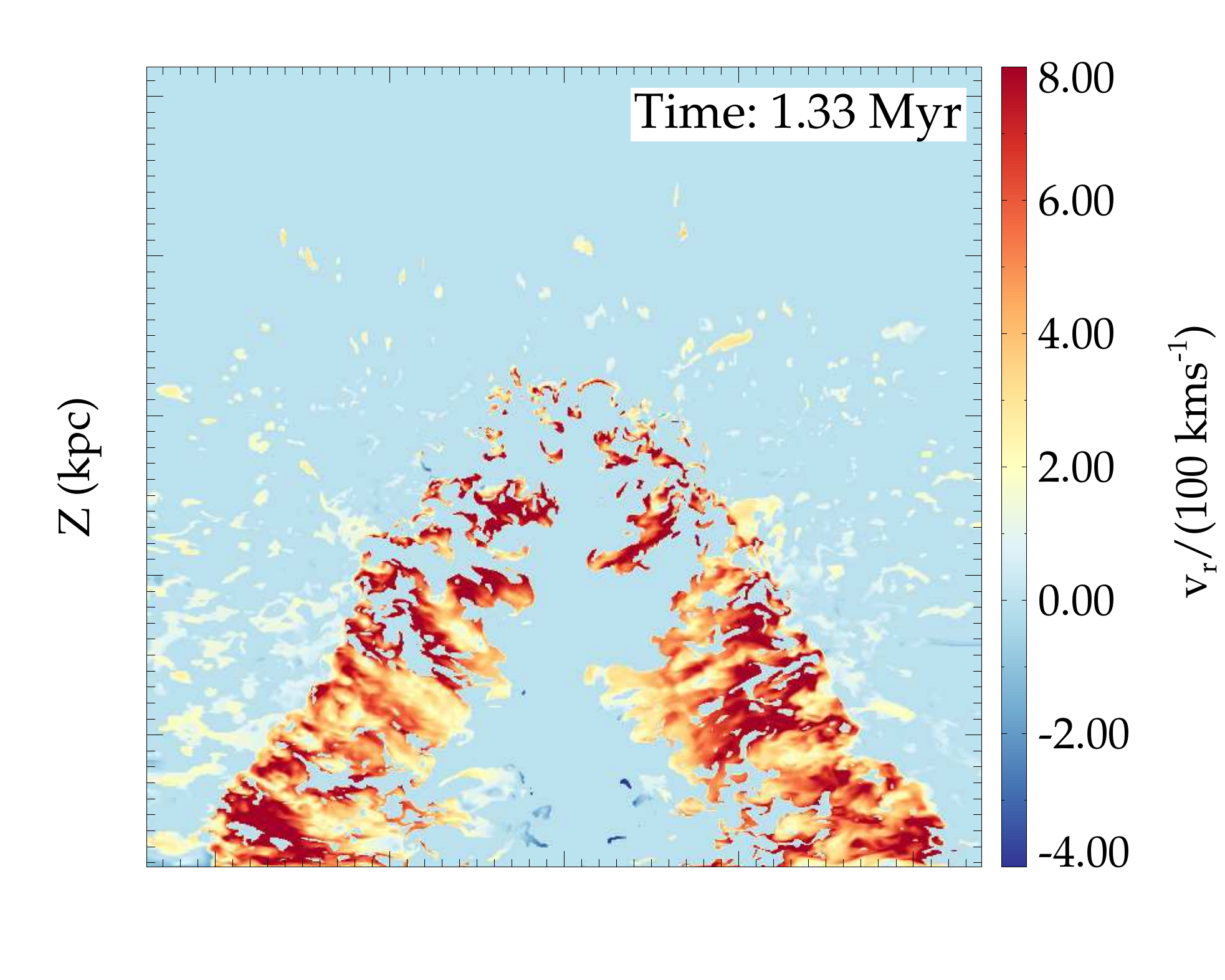}\vspace{-0.52cm}\linebreak
	\includegraphics[width = 7.5cm, keepaspectratio] 
	{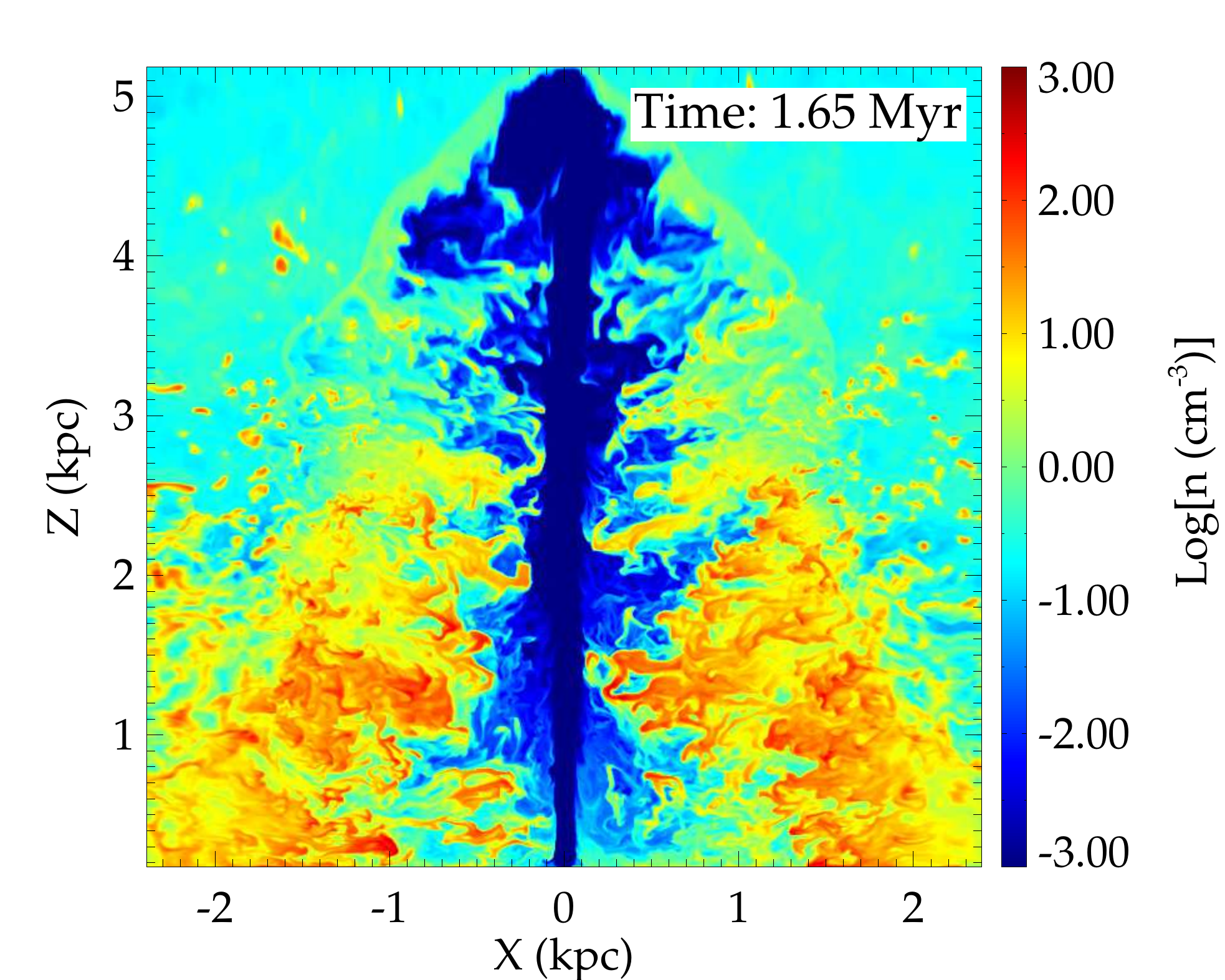}\vspace{-0.2cm}
	\includegraphics[width = 7.5cm, keepaspectratio] 
	{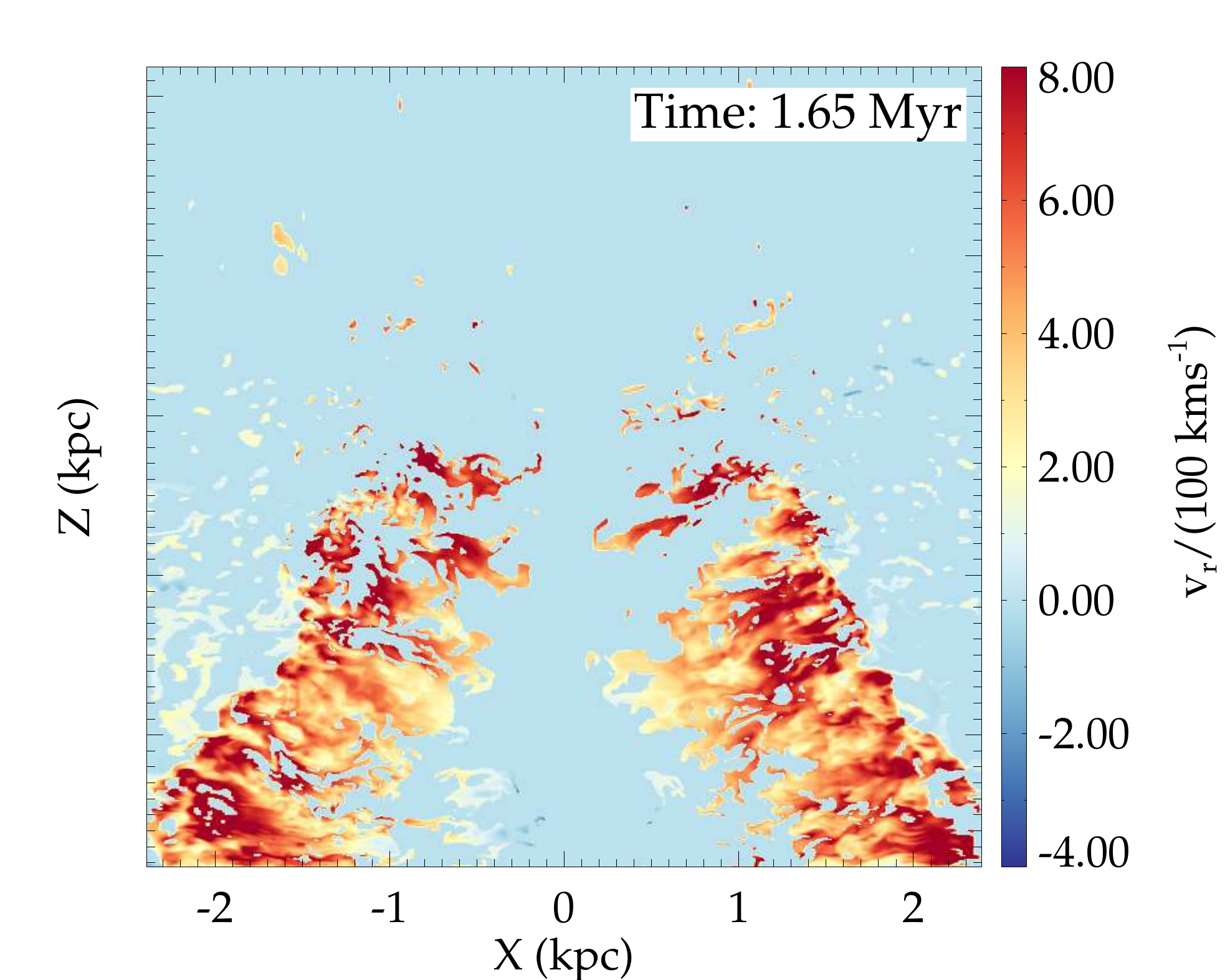}\vspace{-0.2cm}\linebreak
	\caption{\small Left: Density (in $\cc$) in the $x-z$ plane at different times for simulation A (Table~\ref{tab.jetparams}). Right: Radial velocity (in units of 100 $\kms$) at the same times as left. Dense clouds are pushed radially outwards to several hundred $\kms$. Low density ablated cloud mass is accelerated to speeds exceeding $\gtrsim 1000 \kms$.}
	\label{fig.jetsim.rho}
\end{figure*}
\begin{figure*}
	\centering
	\includegraphics[width = 7.5cm, keepaspectratio] 
	{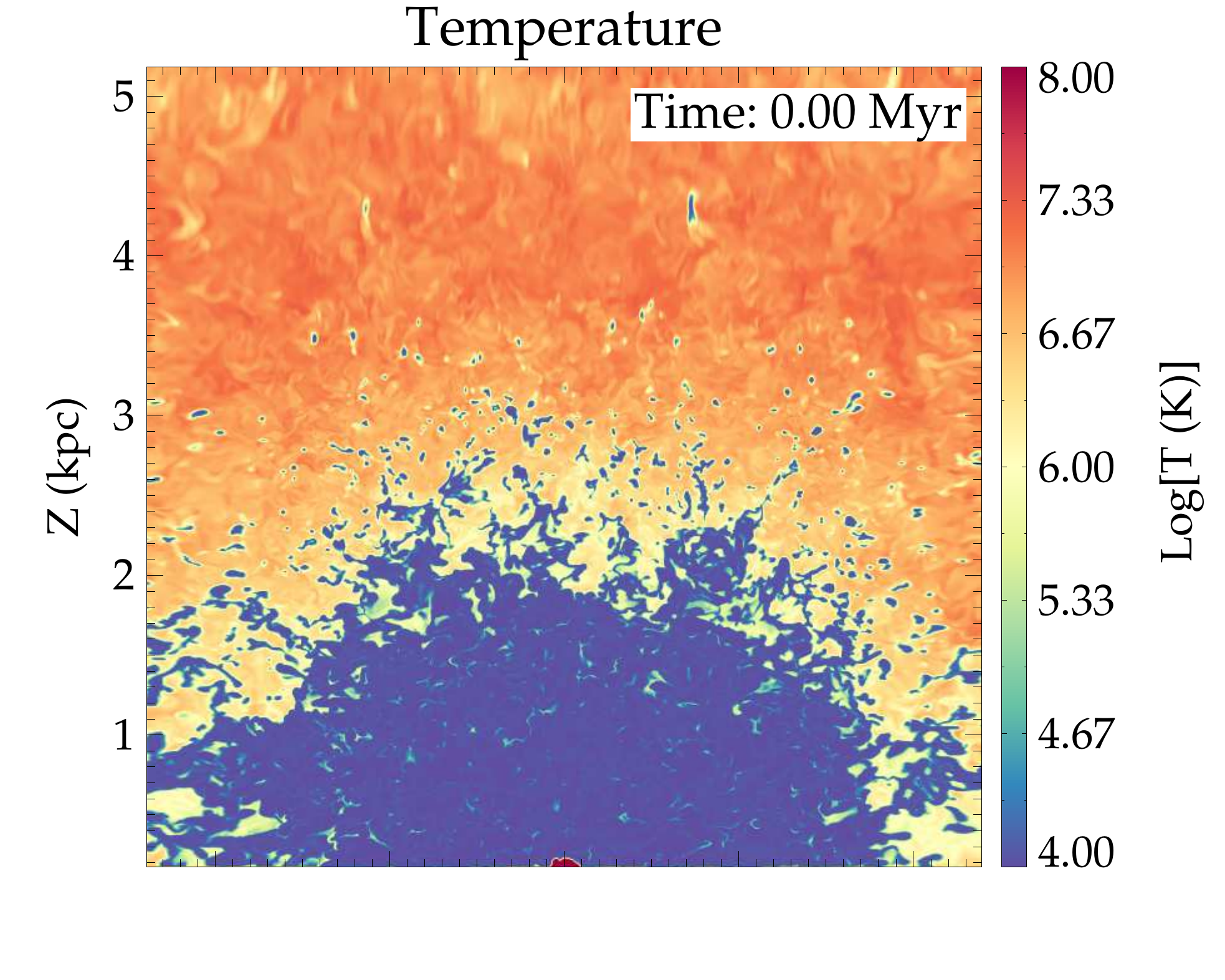}\vspace{-0.52cm}
	\includegraphics[width = 7.5cm, keepaspectratio] 
	{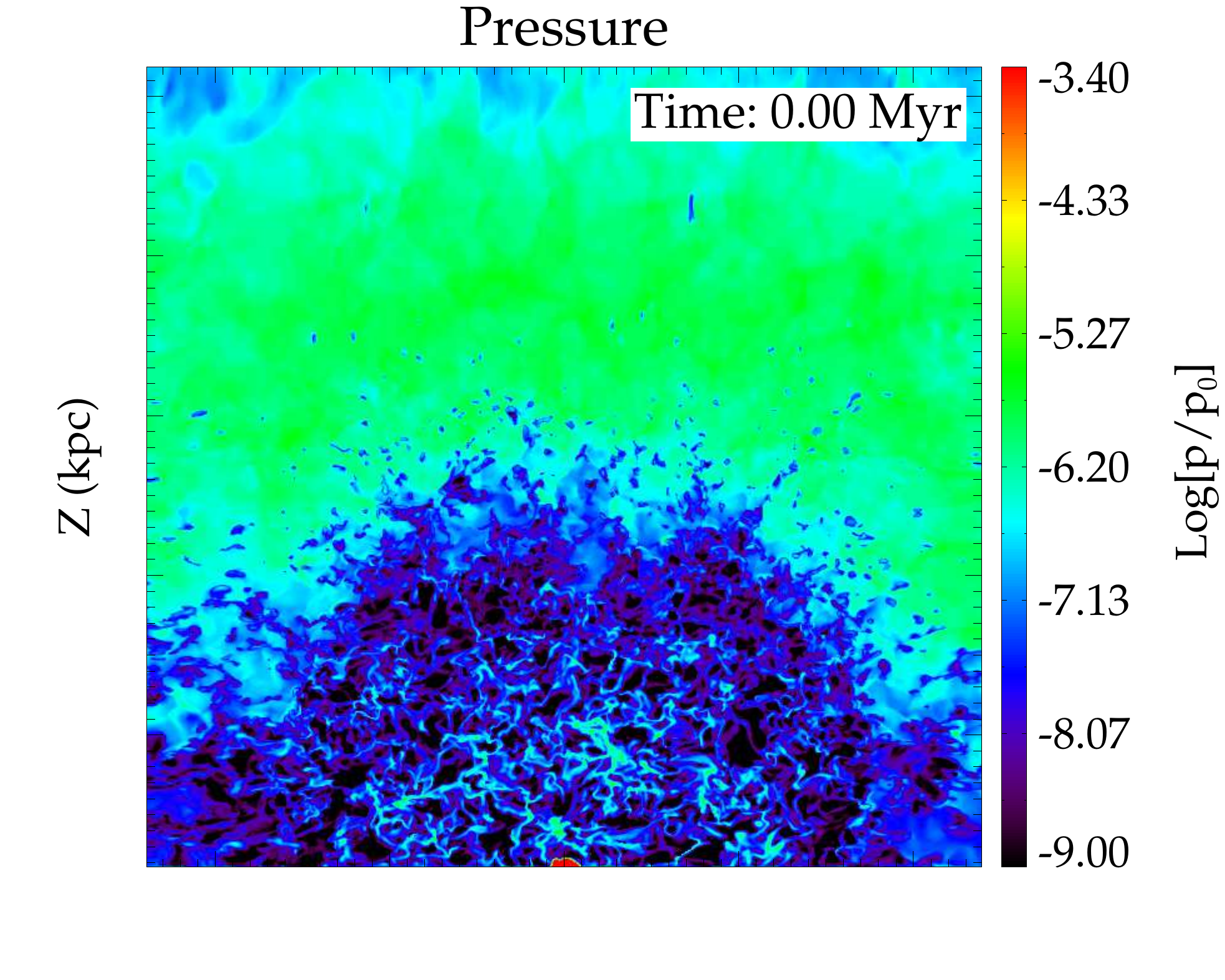}\vspace{-0.52cm}\linebreak
	\includegraphics[width = 7.5cm, keepaspectratio] 
	{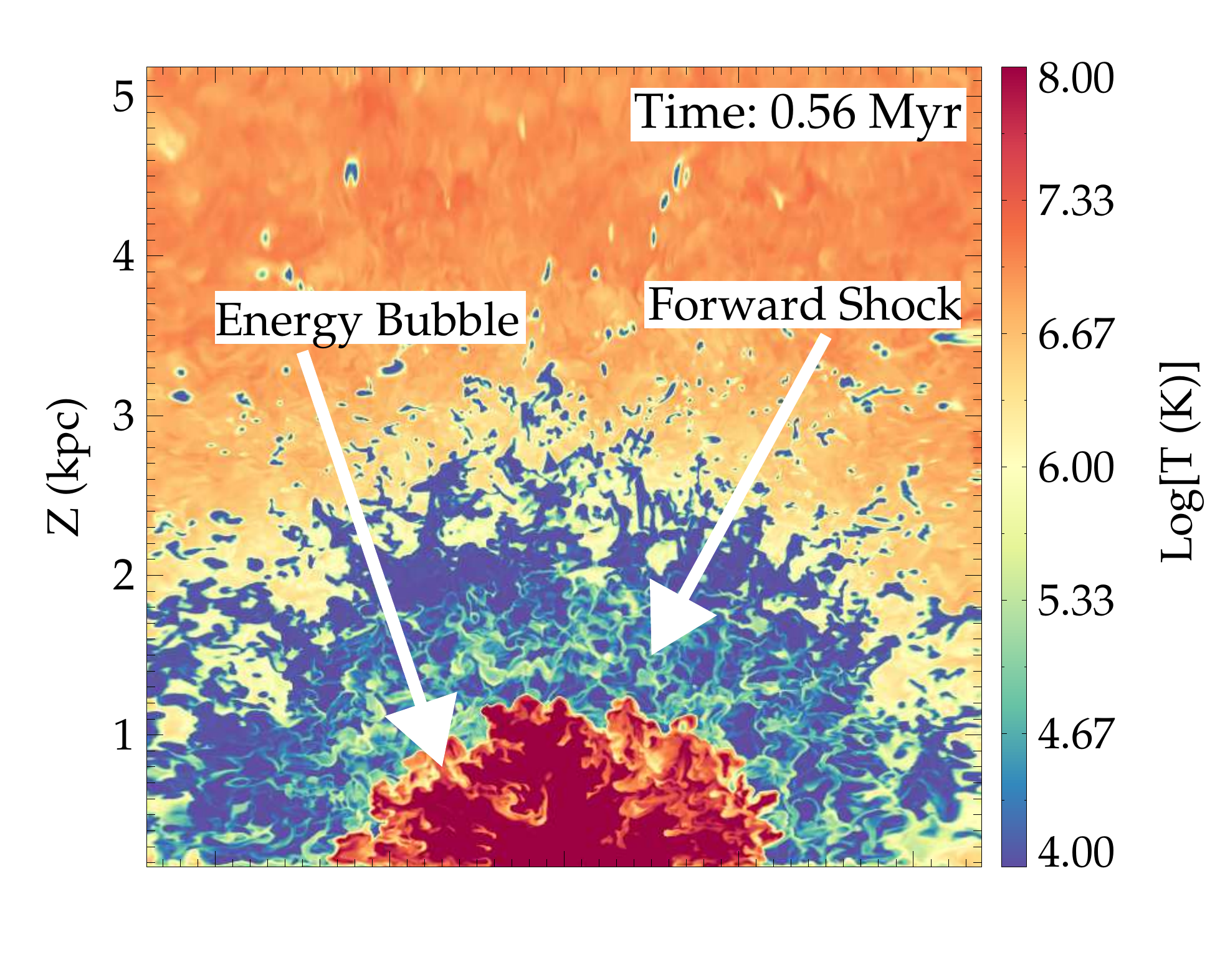}\vspace{-0.52cm}
	\includegraphics[width = 7.5cm, keepaspectratio] 
	{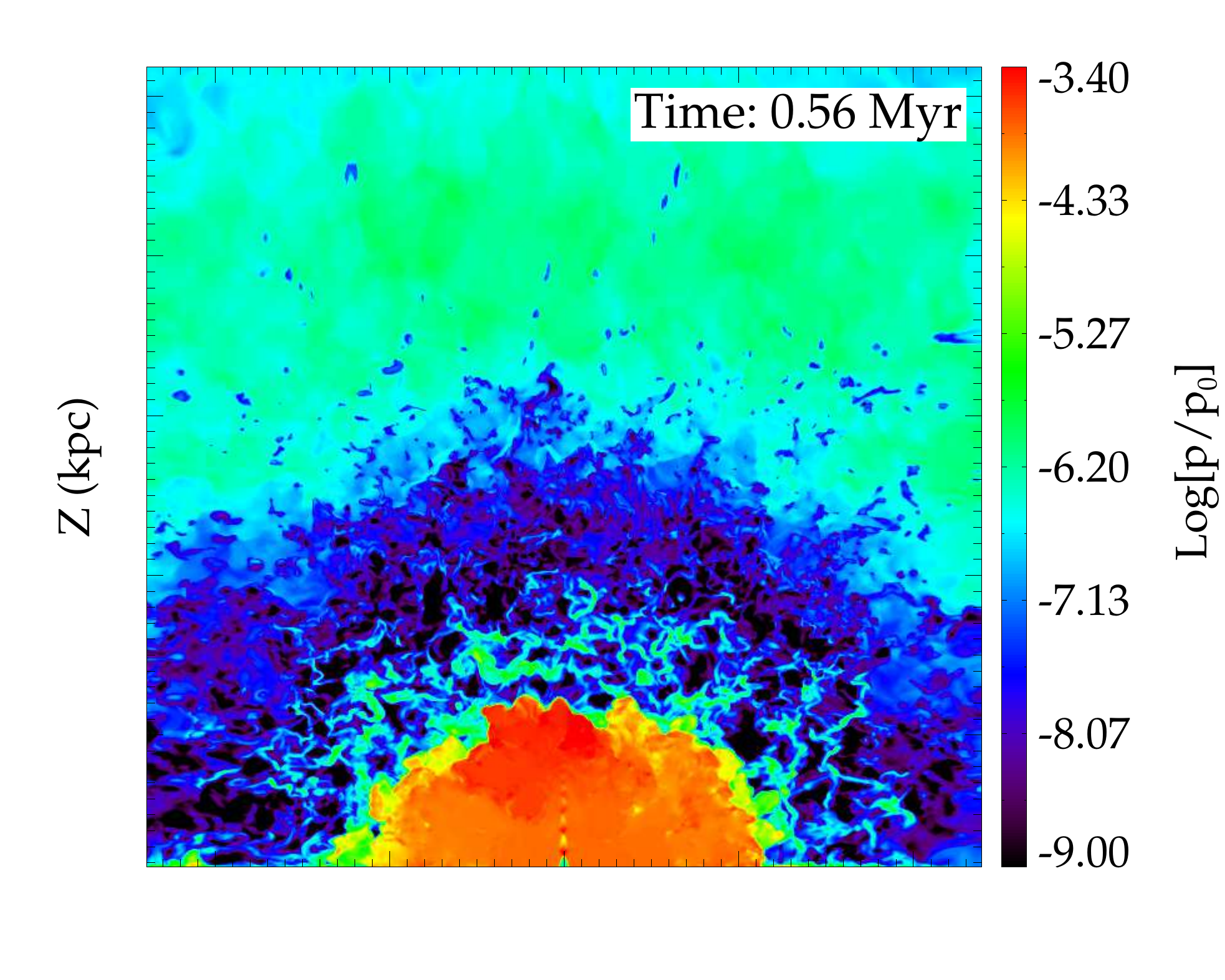}\vspace{-0.52cm}\linebreak
	\includegraphics[width = 7.5cm, keepaspectratio] 
	{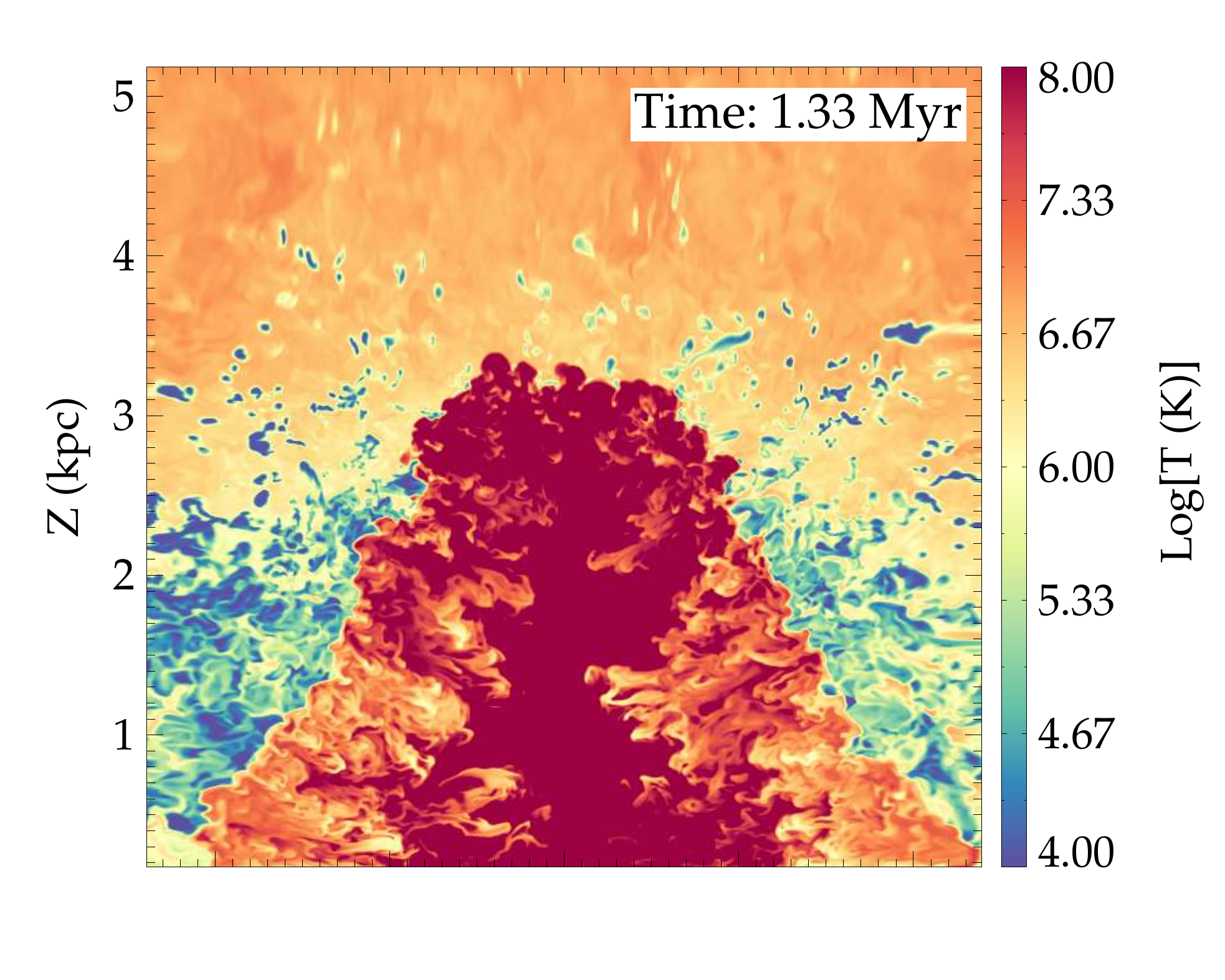}\vspace{-0.52cm}
	\includegraphics[width = 7.5cm, keepaspectratio] 
	{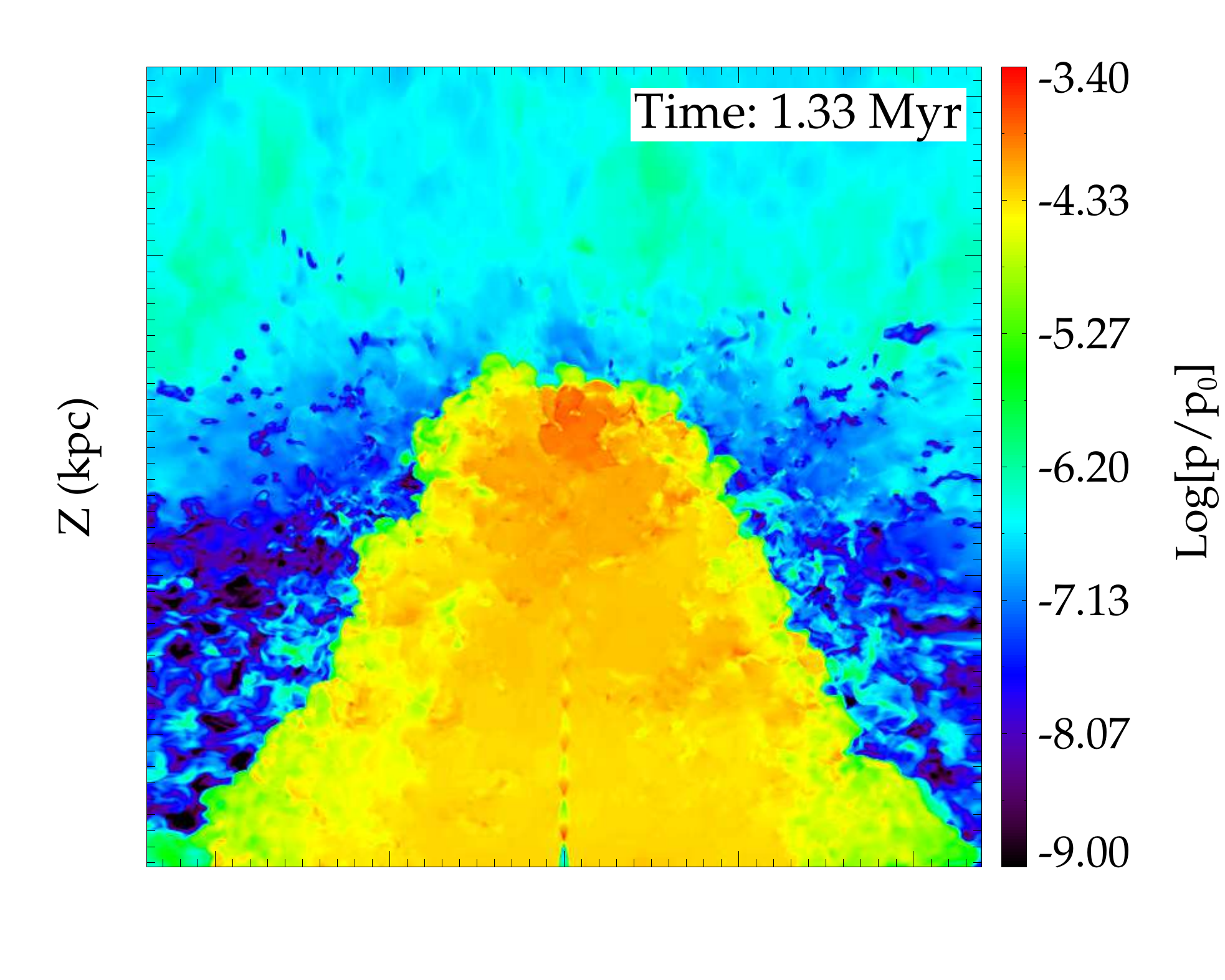}\vspace{-0.52cm}\linebreak
	\includegraphics[width = 7.5cm, keepaspectratio] 
	{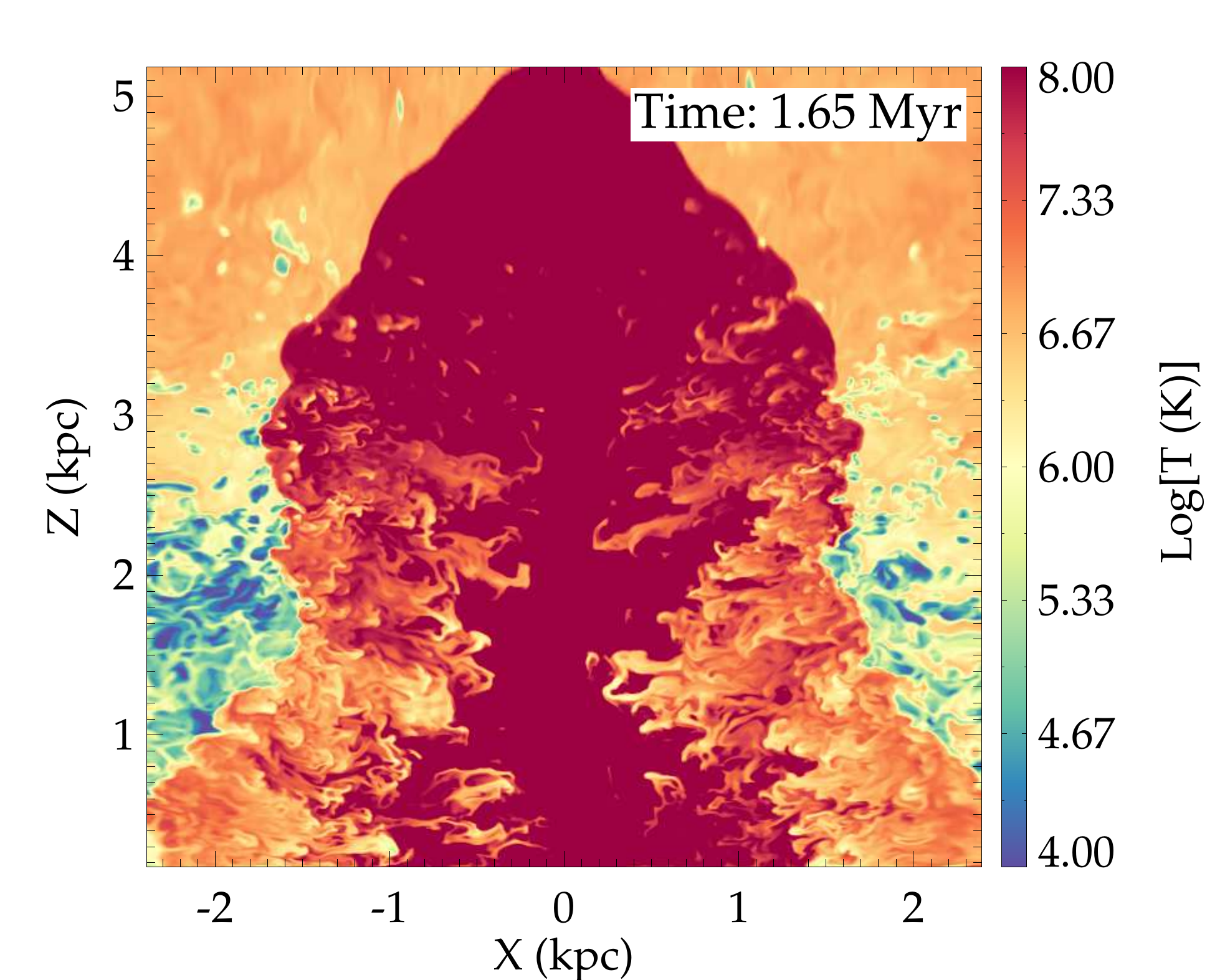}\vspace{-0.2cm}
	\includegraphics[width = 7.5cm, keepaspectratio] 
	{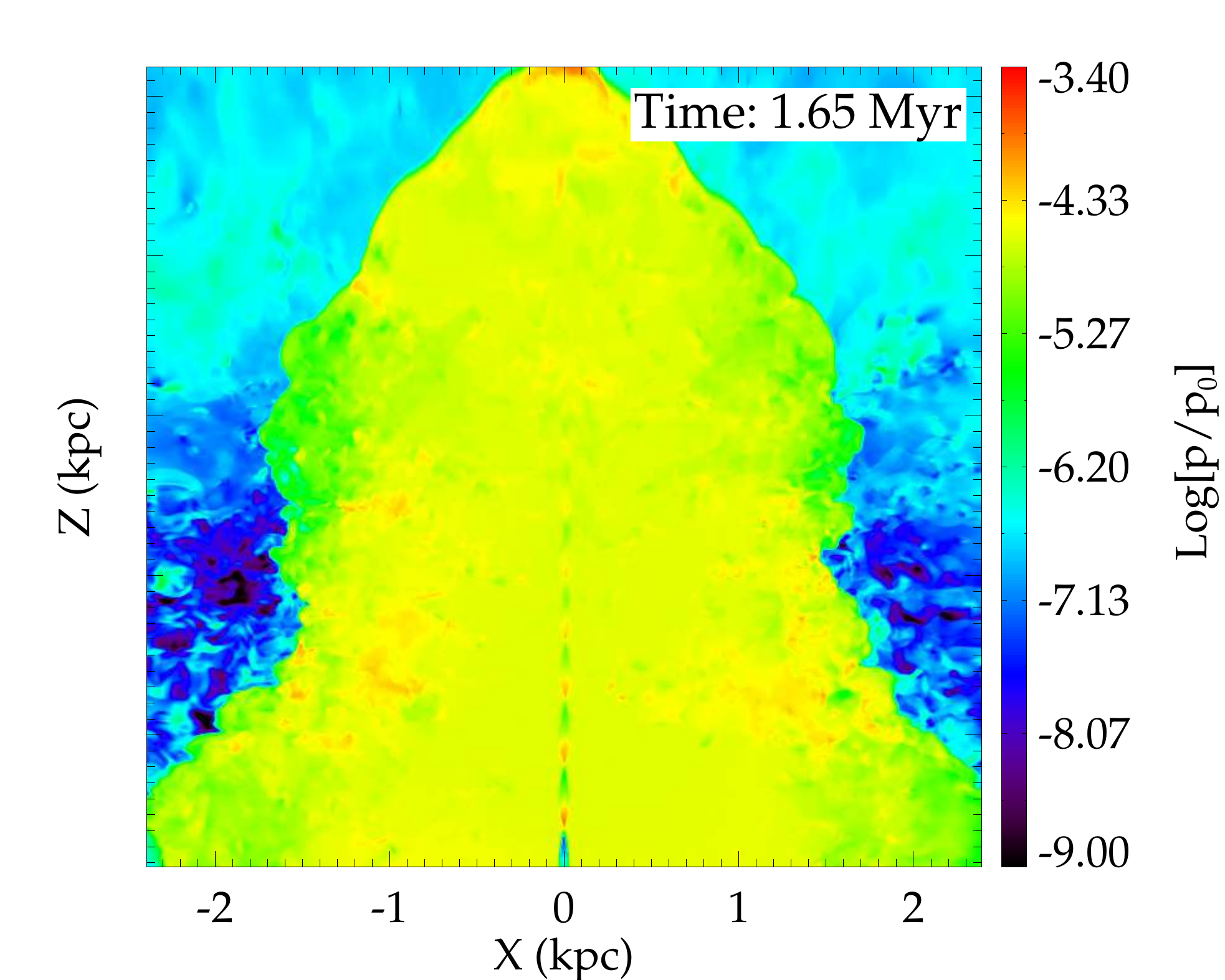}\vspace{-0.2cm}\linebreak
	\caption{\small Left: Temperature ($\log(T)$, with T in Kelvin) in the $x-z$ plane at times same as in Fig.~\ref{fig.jetsim.rho}. Right: Pressure ($\log(p/p_0)$, $p_0=9.2\times 10^{-4} \mbox{dynes cm}^{-2}$). The second panel on the left show the location of the forward shock ($T\sim 10^5$K) preceded by the energy bubble ($T > 10^6$). Corresponding features can be identified in the plot of the pressure on the right. High pressure knots from recollimation shocks can be clearly identified in the third and fourth panels on the right.}
	\label{fig.jetsim.temp}
\end{figure*}
\begin{figure}
	\centering
	\includegraphics[width = 6.8cm, height = 6.8cm,keepaspectratio] {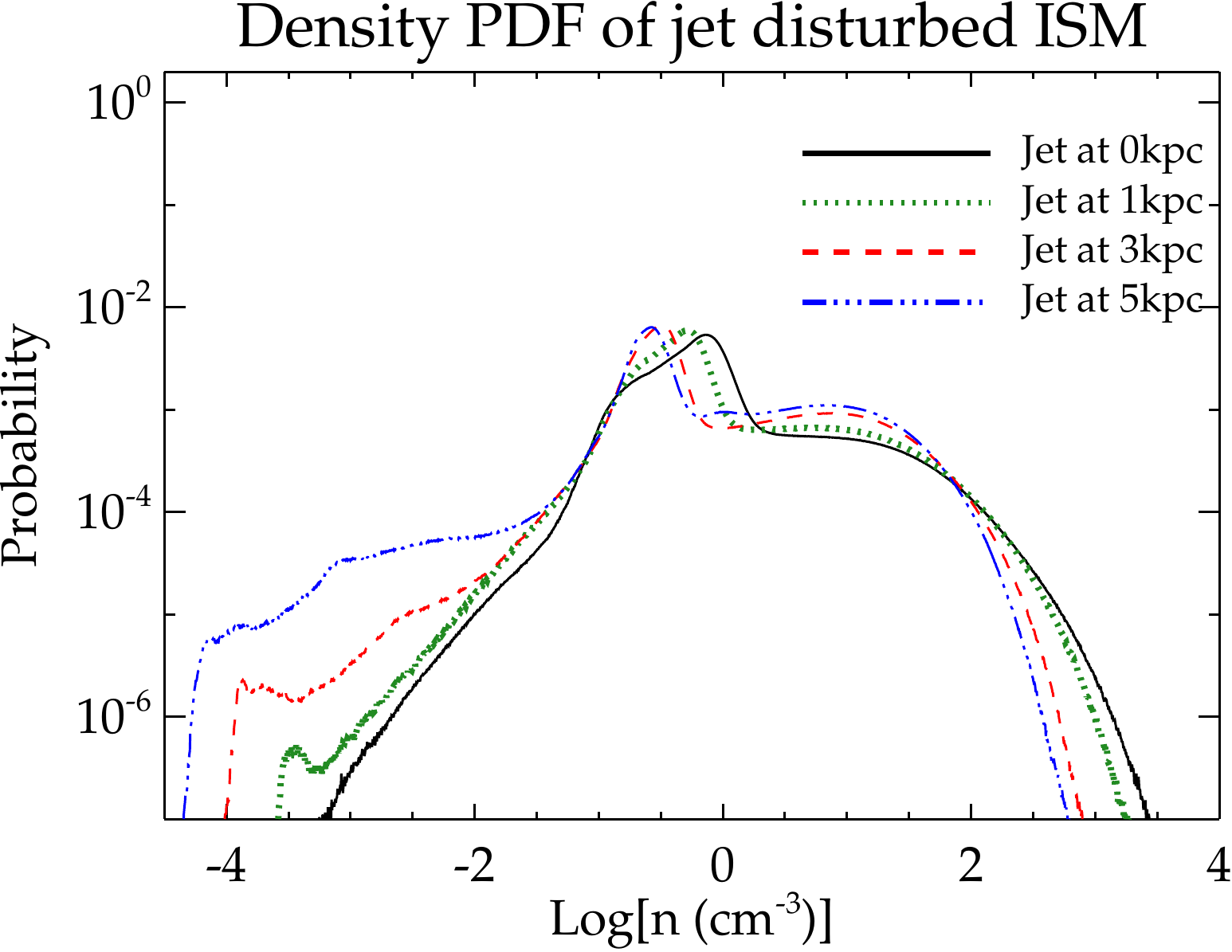}
	\caption{\small The evolution of the density PDF as the jet evolves with time and breaks out of the dense central region after a height of $z\sim 3$ kpc. The density PDF converges after jet break out as the jet decouples from the ISM.}
	\label{fig.densepdf_jet}
\end{figure}
\begin{table}
\centering
\caption{ 
Coefficients of the fit to the density PDF in Fig.~\ref{fig.densepdf_jet} following eq.~\ref{eq.hopkinsfit}}
\label{tab.paramsF45}
\begin{tabular}{| l | l | l | l | l |}
\hline	
  $Z_{\rm jet head}$	&  $\bar{\rho}$  	& $\sigma _\rho$  	& $\eta$   & $\tilde{\rho} _{\small >10}$    \\
  (in kpc)		& (in $\cc$)$^a$	& (in $\cc$) $^b$	&	   & (in $\cc$)$^c$		\\
\hline
 0			& 21.38     		&  52.68		&  0.14	   & 50			\\
 1			& 18.42     		&  44.17		&  0.14    & 43.96		\\
 3 			& 15.73	    		&  27.9                 &  0.12	   & 33.56		\\
 5			& 13.21	    		&  20.52		&  0.17    & 29.5		\\
\hline
\end{tabular} 
\flushleft
\begin{tablenotes}
{\small $^a$ Mean density for the PDF \\
 	$^b$ Standard deviation of variable $\rho$, which is related to the standard deviation in $s=\ln \rho$ as in eq.~\ref{eq.var.LN2}\\
	$^c$ Volume weighted mean of density for $n > 10 \cc$. This gives a measure of the mean density of the high density filaments.}
\end{tablenotes}
\end{table}
We first let the ISM settle into a turbulent filamentary structure with a velocity dispersion $\sim 100-150$ km s$^{-1}$, which is typical of high red-shift galaxies \citep{forster09,wisnioski15}. We then inject the relativistic jet whose parameters are described in Sec.~\ref{sec.jetparams}. We list the simulations performed with jets interacting with the turbulent ISM in  Table~\ref{tab.jetparams}. In Fig.~\ref{fig.jetsim.rho} and Fig.~\ref{fig.jetsim.temp} we present the evolution of the density, radial velocity, pressure and temperature of the ISM for simulation A. The jet is initially impeded by the dense filaments (Fig.~\ref{fig.jetsim.rho}) and passes through a flood a channel phase as described in previous works \citep{sutherland07a,wagner11a,wagner12a}, seeking the paths of least resistance as it drives an energy bubble through the ISM.

The jet shears the dense filaments as it clears its path. The effect of shearing of the dense cores is well demonstrated through the evolution of the density PDF shown in Fig.~\ref{fig.densepdf_jet}. The high density tail of the PDF is significantly reduced as the jet driven bubble shears the clouds. There is a subsequent enhancement of the PDF at $n \sim 10-100 \cc$, which occurs both as a result of compression of low density gas from the forward shock and also of fragmentation of the dense filaments. Coefficients of fits (following eq.~\ref{eq.hopkinsfit}) to the high density end of the PDF are listed in Table~\ref{tab.paramsF45}. The density PDF converges after the jet breaks out (jet head $\gtrsim$ 3 kpc) and decouples from the ISM, proceeding along the cleared path.

\begin{figure*}
	\centering
	\includegraphics[width = 7.5cm,keepaspectratio] {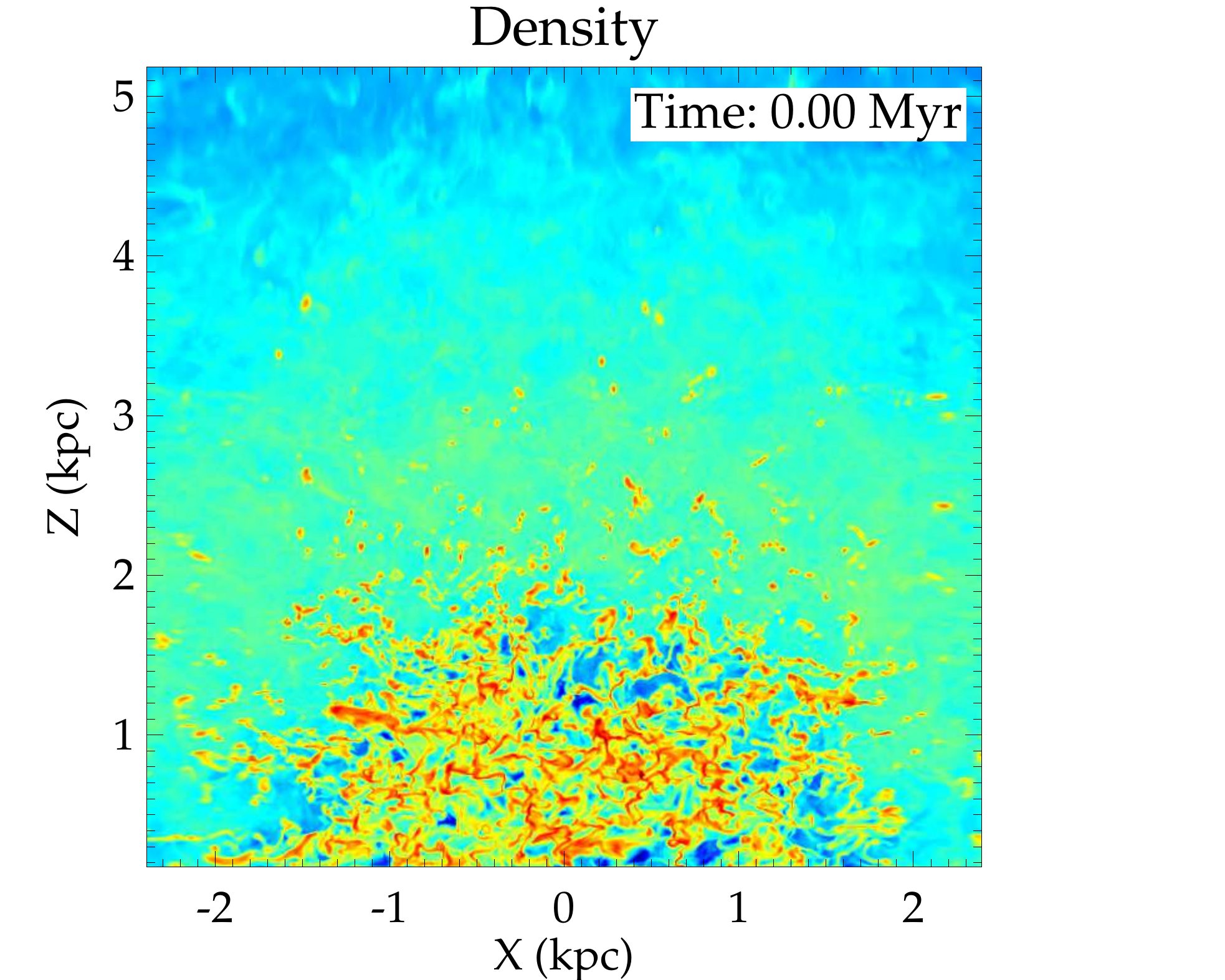}\hspace{-2.4cm}
	\includegraphics[width = 7.5cm,keepaspectratio] {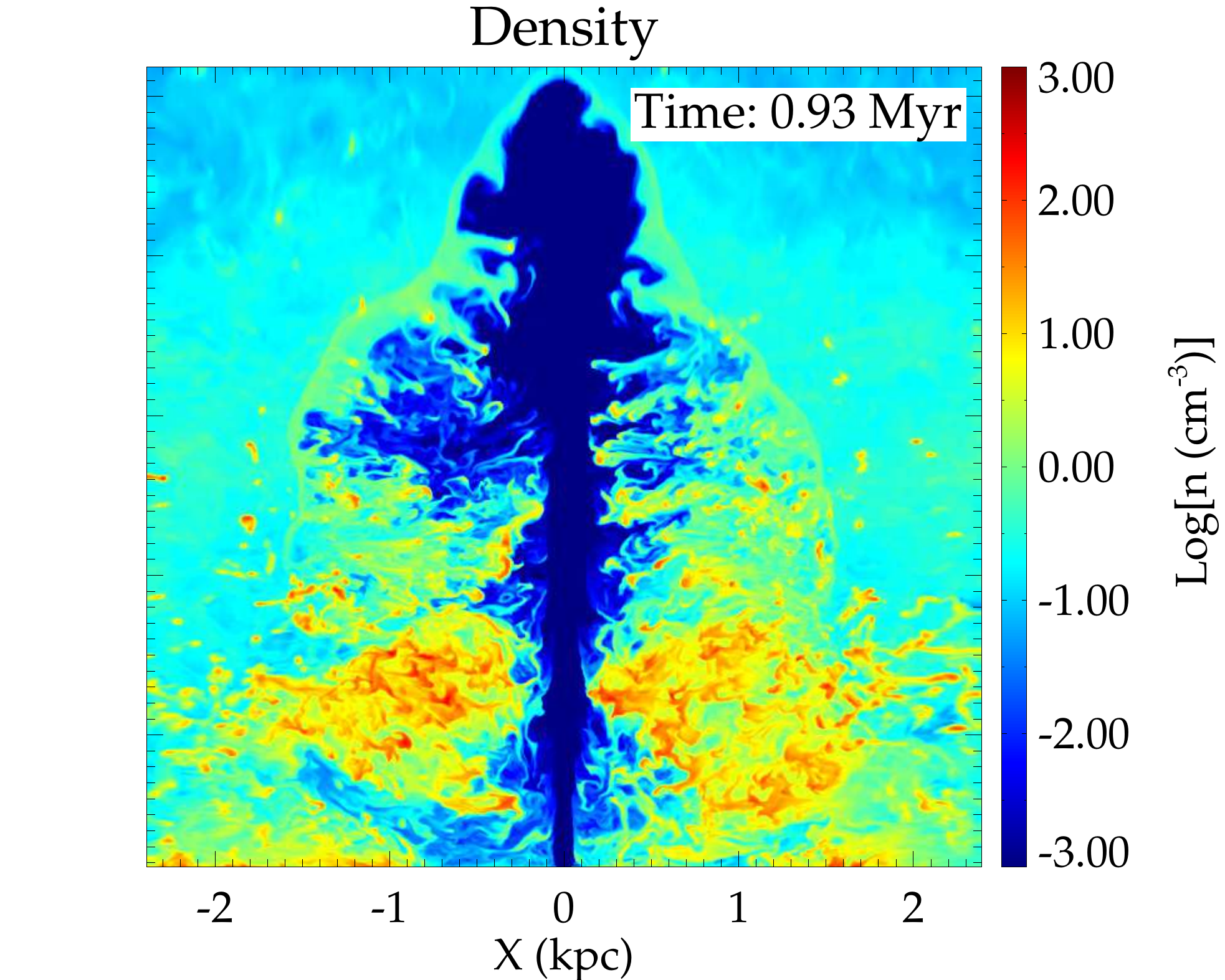}
	\caption{\small Density ($\log[n (\cc)]$) in the $x-z$ plane for simulation B.}
	\label{fig.jetsim.rhonw150}
\end{figure*}
The duration of the flood-channel phase of the jet evolution depends upon the mean density and filling factor of the ISM, as previously discussed in \citet{wagner12a}. Clouds with denser cores are less ablated and  more efficiently  impede the progress of the jet. In Fig.~\ref{fig.jetsim.rhonw150} we present the density for simulation~B (Table~\ref{tab.jetparams}) where a jet of power $10^{45} \ergs$ passes through a medium with lower mean density ($n_{w0}=150 \cc$) compared to that of simulation~A. We note that the jet breaks out of the warm dense gas much faster, in comparison to simulation A. 

\subsection{Evolution of the energy bubble}
\begin{figure}
	\centering
	\includegraphics[width = 6.8cm, height = 6.8cm,keepaspectratio] {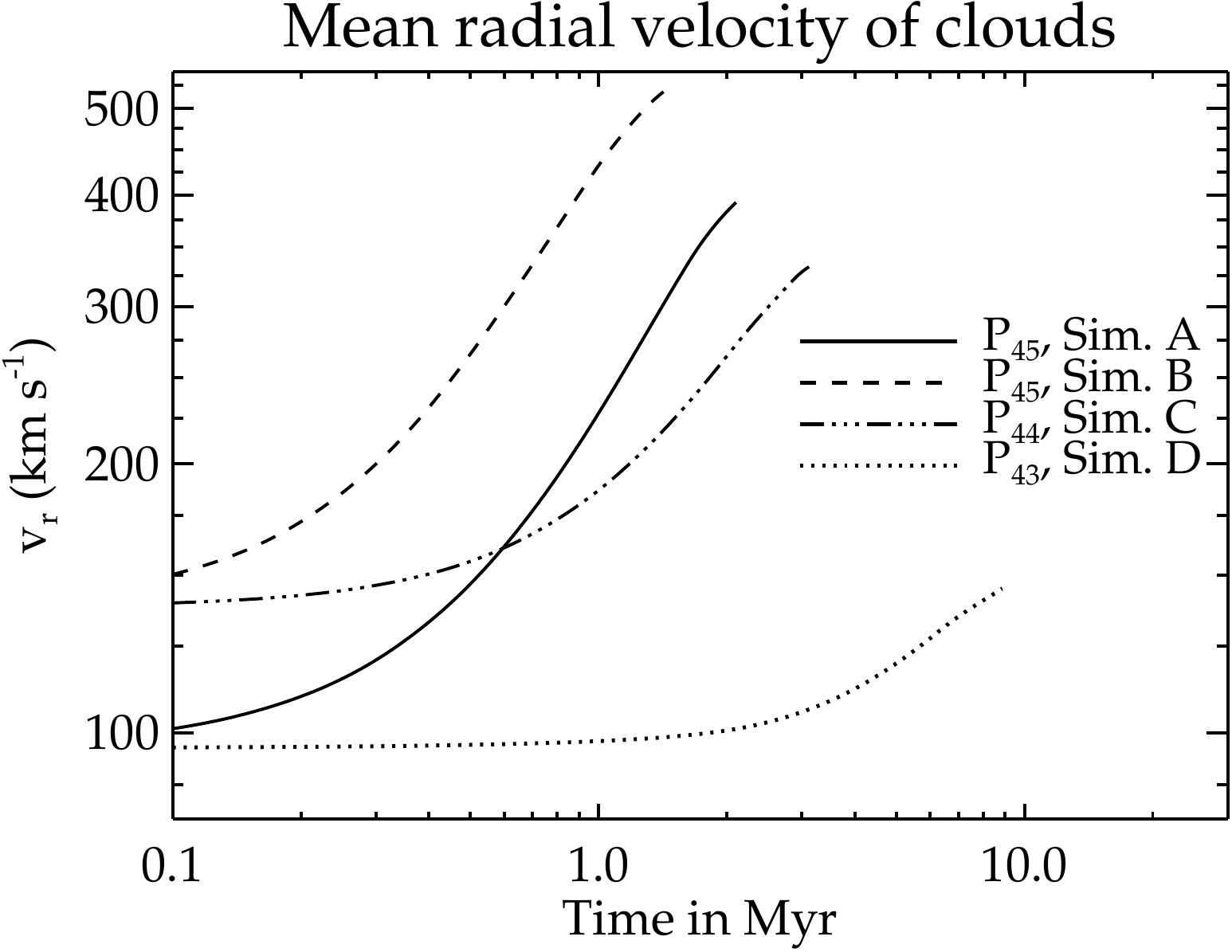}
	\caption{\small The mass weighted mean radial velocity ($\int \rho v_r d^3x/\int \rho d^3x$)  of the warm clouds driven out by the expanding jet-driven bubble.}
	\label{fig.meanVr}
\end{figure}
\begin{figure}
	\centering
	\includegraphics[width = 7.cm, height = 7.cm,keepaspectratio] {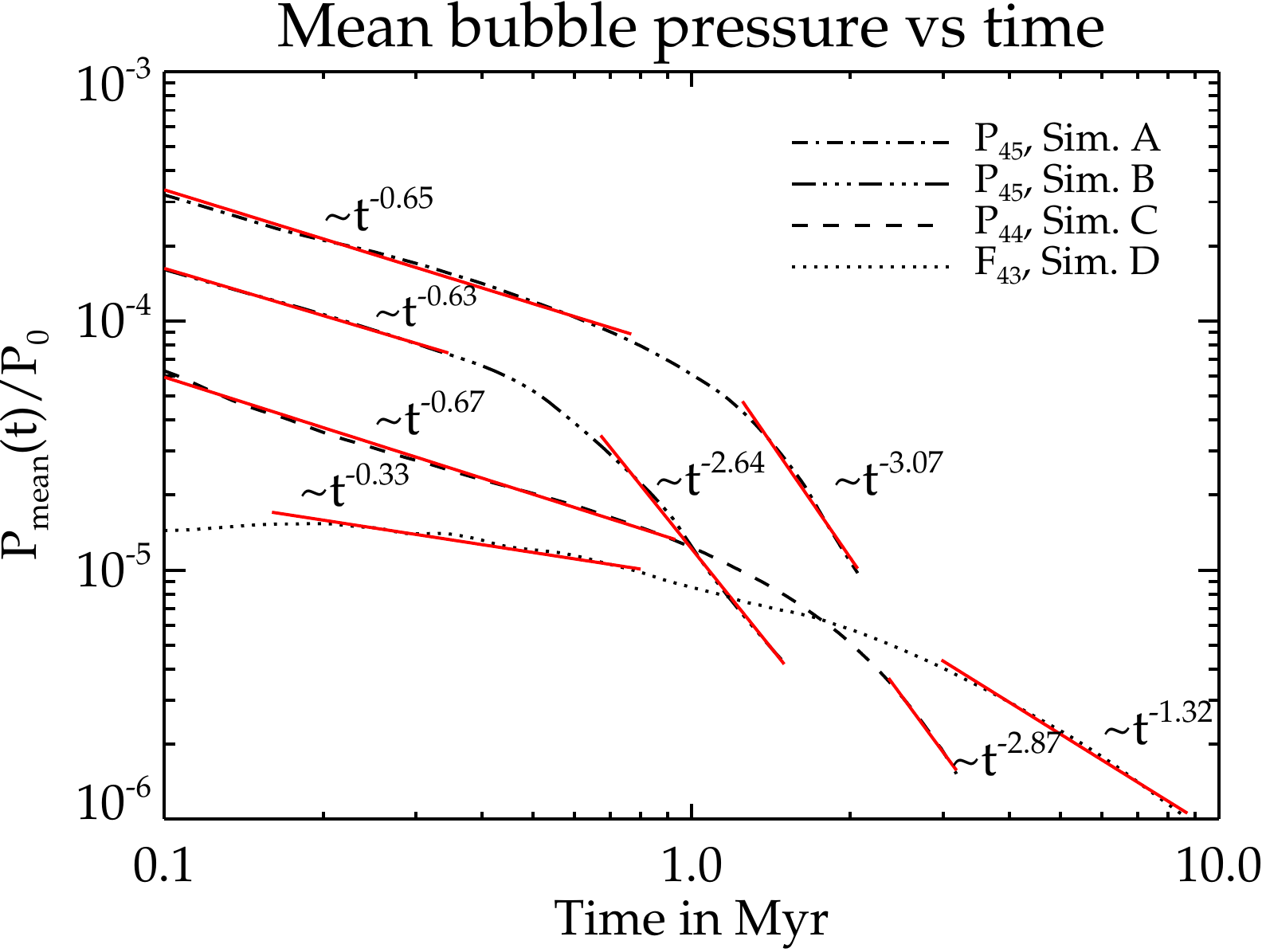}
	\caption{\small Evolution of the mean pressure inside the energy bubble as a function of time. The red solid lines represent power-law fits to the curves to sections representing before and after the jet break out. The slopes obtained from the fits are noted above the curves.}
	\label{fig.meanpres}
\end{figure}
\begin{figure}
	\centering
	\includegraphics[width = 7.cm, height = 7.cm,keepaspectratio] {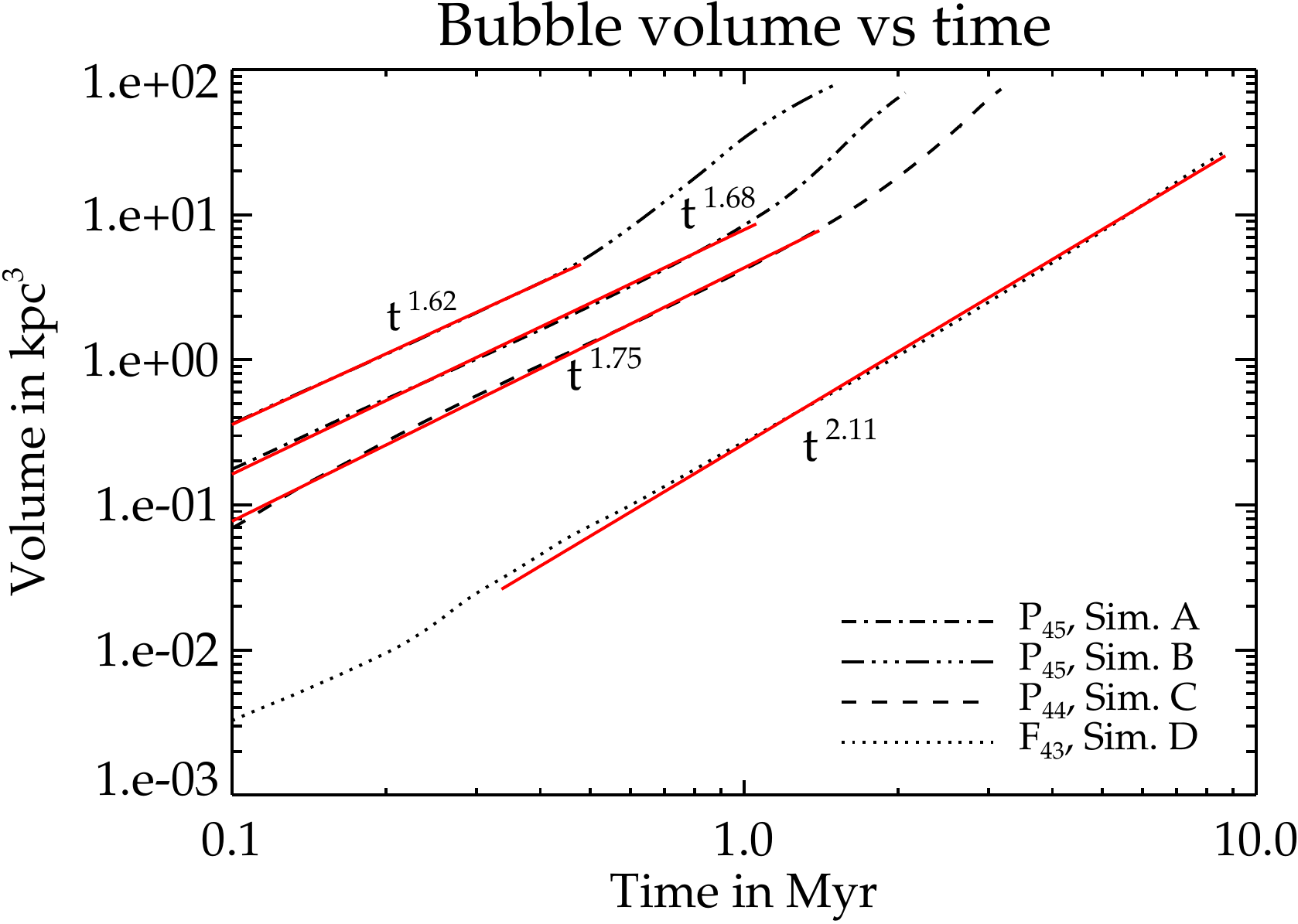}
	\caption{\small Evolution of the volume of the bubble with time for different simulations. The red solid lines represent power-law fits.}
	\label{fig.volbubble}
\end{figure}
The  confined jet creates an expanding energy bubble. A fast moving forward shock first sweeps through the filaments heating the gas to $\sim 10^5$ K (see Fig.~\ref{fig.jetsim.temp}). The forward shock is followed by a slower moving region of nearly homogeneous high pressure, defining the energy bubble (see Fig.~\ref{fig.jetsim.temp}). The  energy bubble shears the dense filaments as it expands into the ISM, accelerating the dense clouds to radial velocities in excess of $\sim 300 \kms$ (as shown in Fig.~\ref{fig.meanVr}), as also reported earlier in \citet{wagner12a}. Some of the clouds are dragged radially inwards by the backflow near the jet axis. The low density ablated cloud mass is swept up by the expanding bubble to velocities higher than $\gtrsim 1000 \kms$, flowing out through the channels between the denser clouds (see Fig.~\ref{fig.jetsim.rho}).

In Fig.~\ref{fig.meanpres} and Fig.~\ref{fig.volbubble} we show the evolution of the mean pressure inside the energy bubble and the volume of the bubble. For high power jets ($P_{\rm jet} \gtrsim 10^{44} \ergs$), initially the high pressure bubble expands nearly adiabatically. The power law evolution of the pressure in Fig.~\ref{fig.meanpres} can be understood by solving the energy equation for an adiabatically expanding bubble, whose volume evolves as in Fig.~\ref{fig.volbubble} (see Appendix~\ref{append.bubble} for detailed derivation). The evolution of the volume is slower than that of a freely expanding self-similar flow \citep{castor75a,weaver77a} due to the complex interaction of the bubble with the multiphase ISM. After jet break out, the mean pressure falls more rapidly due to free expansion in the halo.  The evolution of the bubble for sim.~D with $P_{\rm jet} =10^{43}\ergs$ is much slower compared to the cases with higher $P_{\rm jet}$. The bubble is only weakly over pressured and remains trapped in the ISM for a longer time. This facilitates energy losses via cooling, resulting in slower growth.

\subsection{Evolution of the phase space}
\begin{table*}
\centering
\caption{Summary of different ISM phases}
\label{tab.ism.phases}
\begin{tabular}{| l | l | l |}
\hline
Phase 	& Description 	& Characteristics    \\
acronym	&		&  		\\
\hline
HH  & Hot Halo		& The ambient low density hot halo. Temperature: $T\sim 10^7$ K \\
WC  & Warm Clouds	& Turbulent warm ($T \sim 10^4$K) dense filaments  \\
			&& describing the initial ISM. \\
FS  & Forward Shock	& Shocked ISM behind the initial forward shock. Temperature: $T\sim10^5$ K. \\
			&& Very little acceleration from initial turbulent velocity \\
EB  & Energy Bubble	& High pressure energy bubble expanding nearly adiabatically \\
			&& for $P_{\rm jet} \gtrsim 10^{44} \ergs$. Temperature: $T\sim10^6-10^7$ K. \\
			&& Accelerates clouds to outward radial velocities $\gtrsim 500 \kms$ \\
RSG & Remnant Shocked gas & ISM shocked by the forward shock which remains after jet \\
			&& has decoupled. These represent cooling dense cores of the clouds. \\
\hline
\end{tabular} \\
\end{table*}
\begin{itemize}
\item \emph{Multi-phase ISM:}
\begin{figure*}
	\centering
	\includegraphics[width = 7.5cm, height = 5.6cm,keepaspectratio]
	{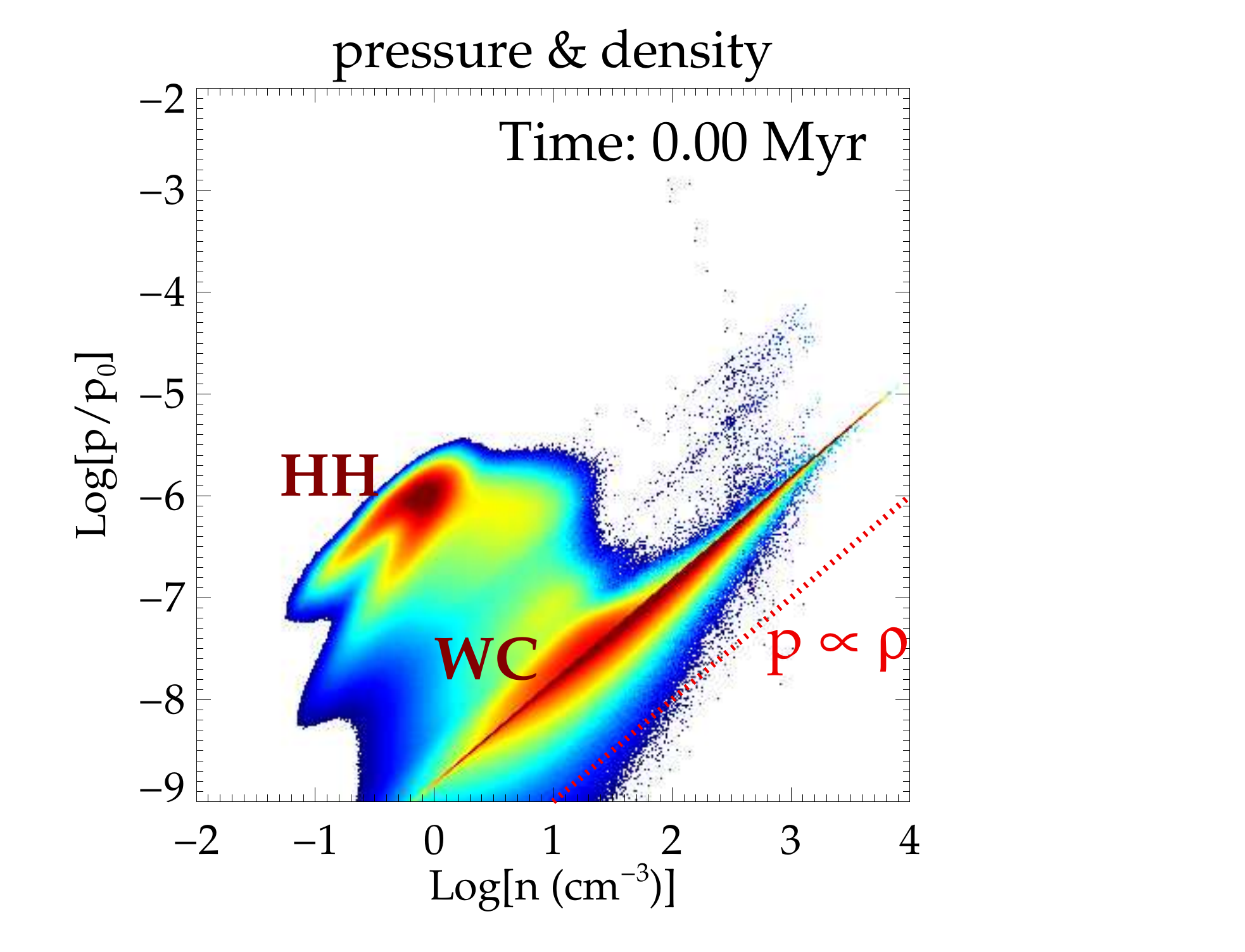}\vspace{-0.1cm}\hspace{-3cm}
	\includegraphics[width = 7.5cm, height = 5.6cm,keepaspectratio]
	{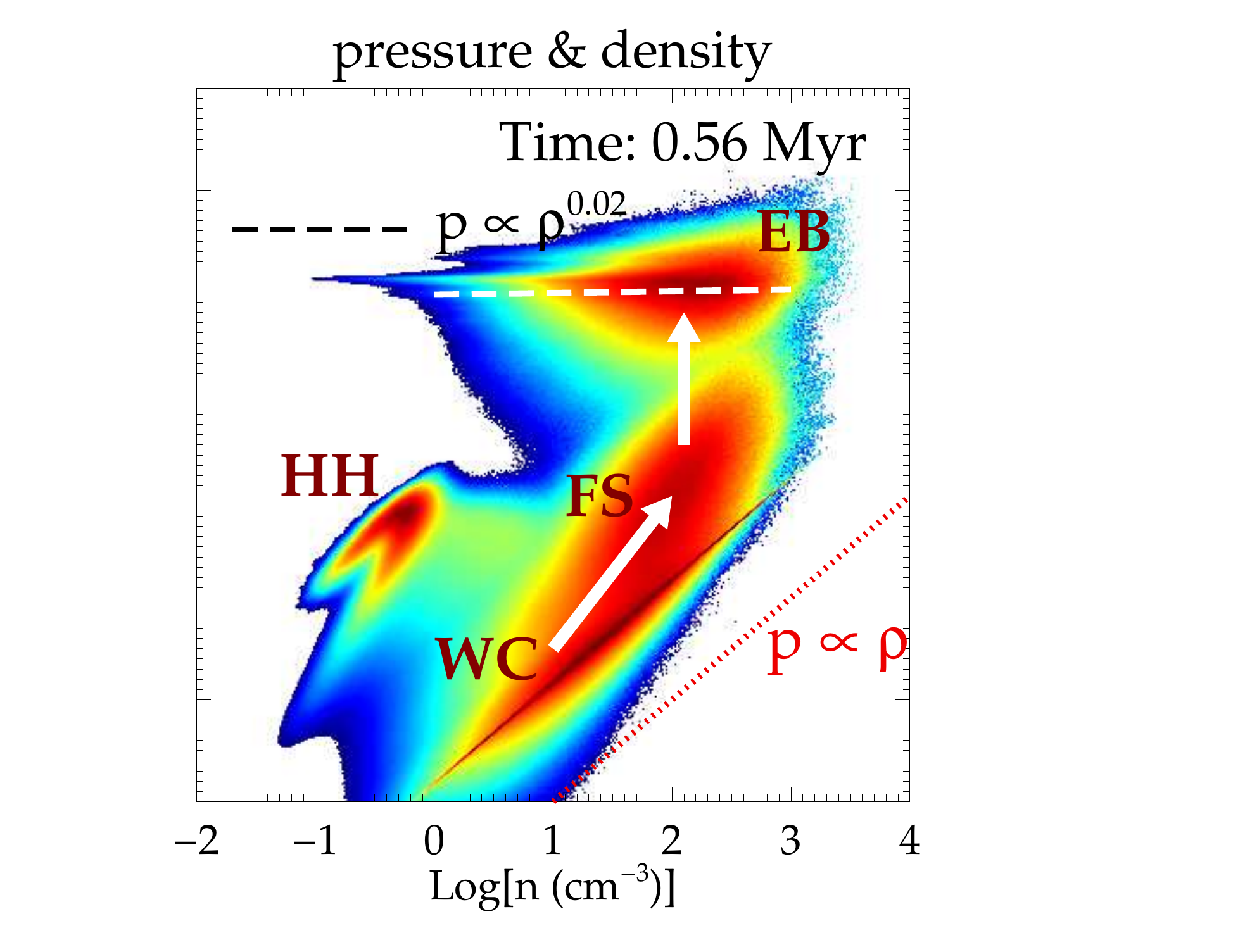}\vspace{-0.1cm}\hspace{-3cm}
	\includegraphics[width = 7.5cm, height = 5.6cm,keepaspectratio] 
	{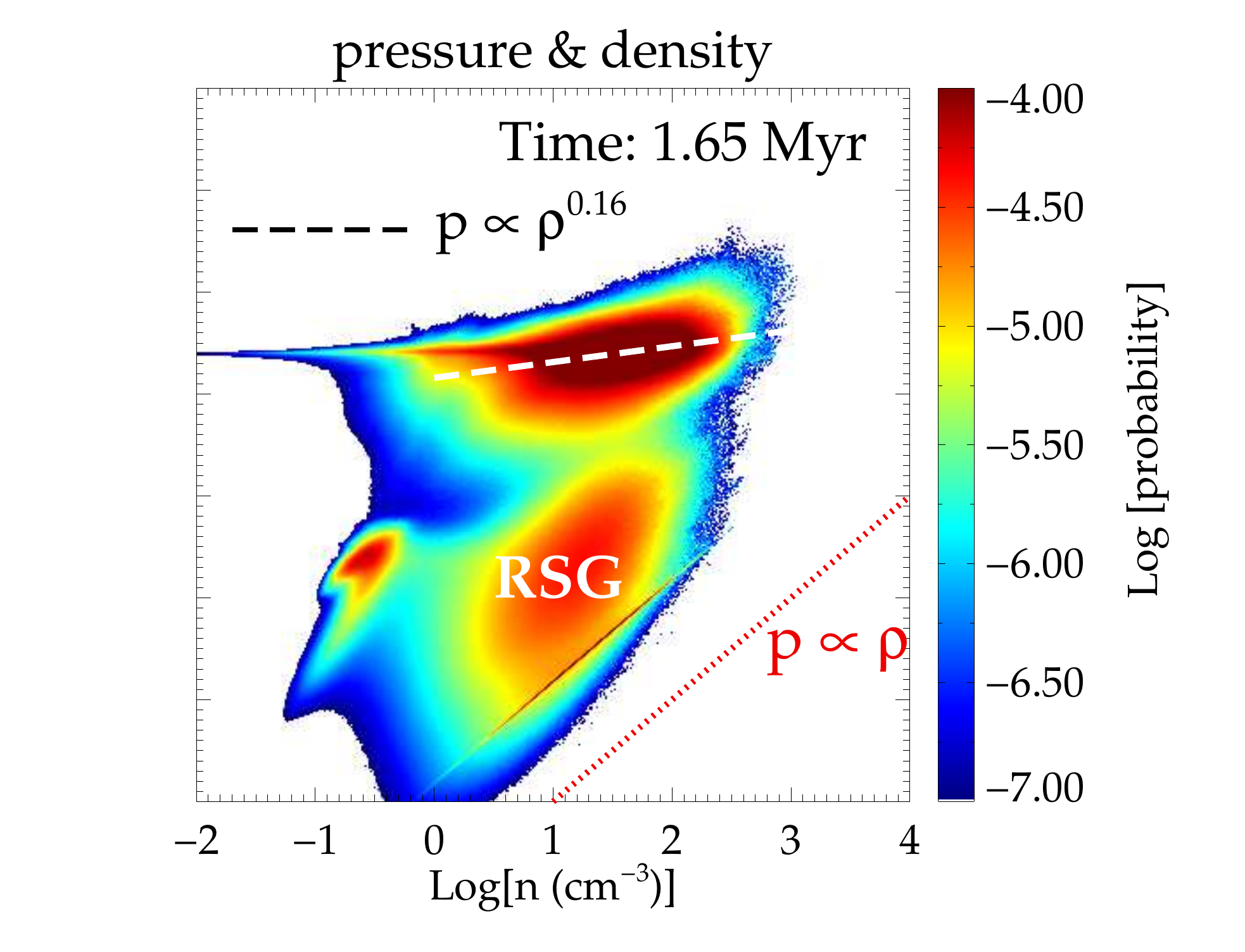}\vspace{-0.1cm}\linebreak
	\includegraphics[width = 7.5cm, height = 5.6cm,keepaspectratio]
	{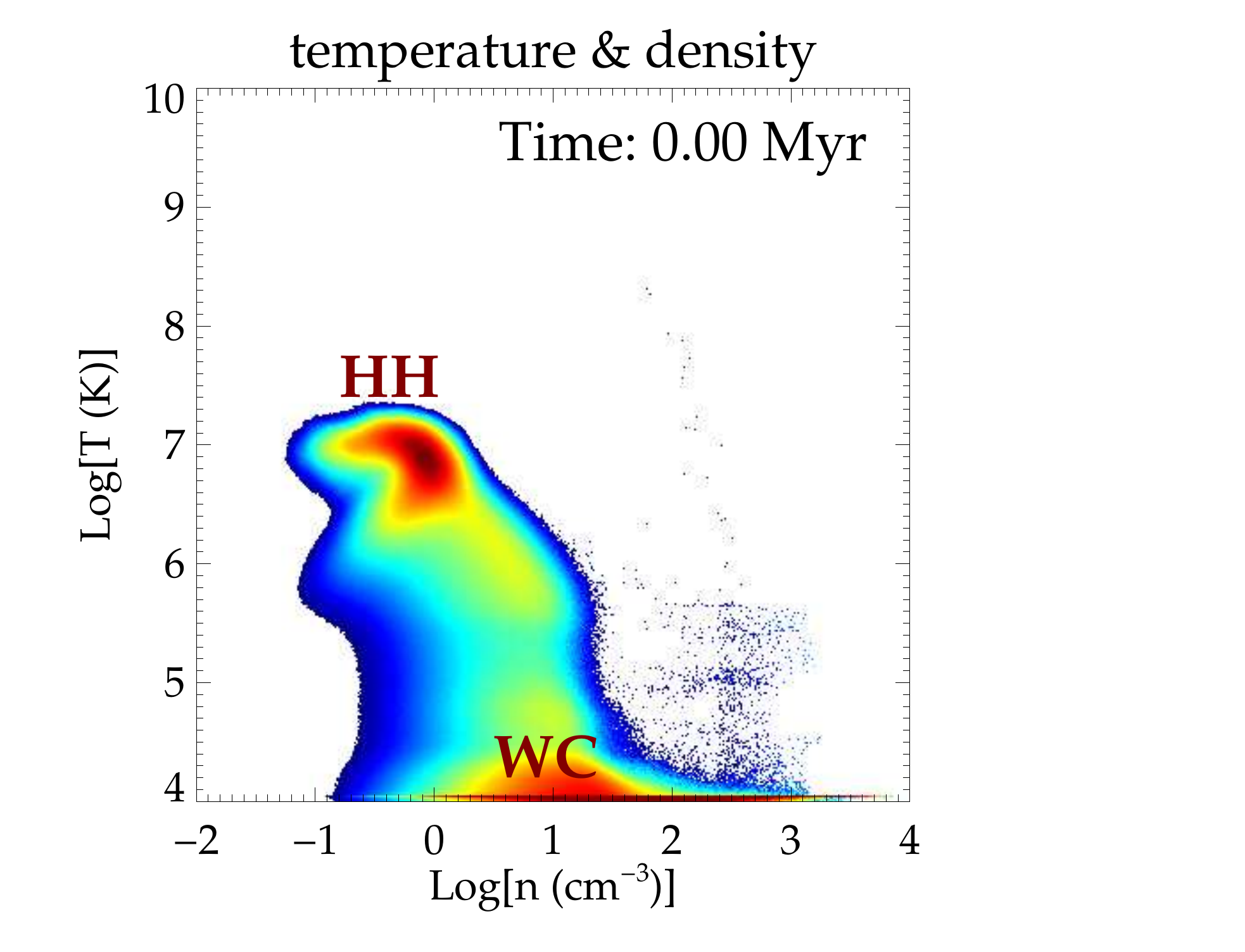}\vspace{-0.2cm}\hspace{-3cm} 
	\includegraphics[width = 7.5cm, height = 5.6cm,keepaspectratio]
	{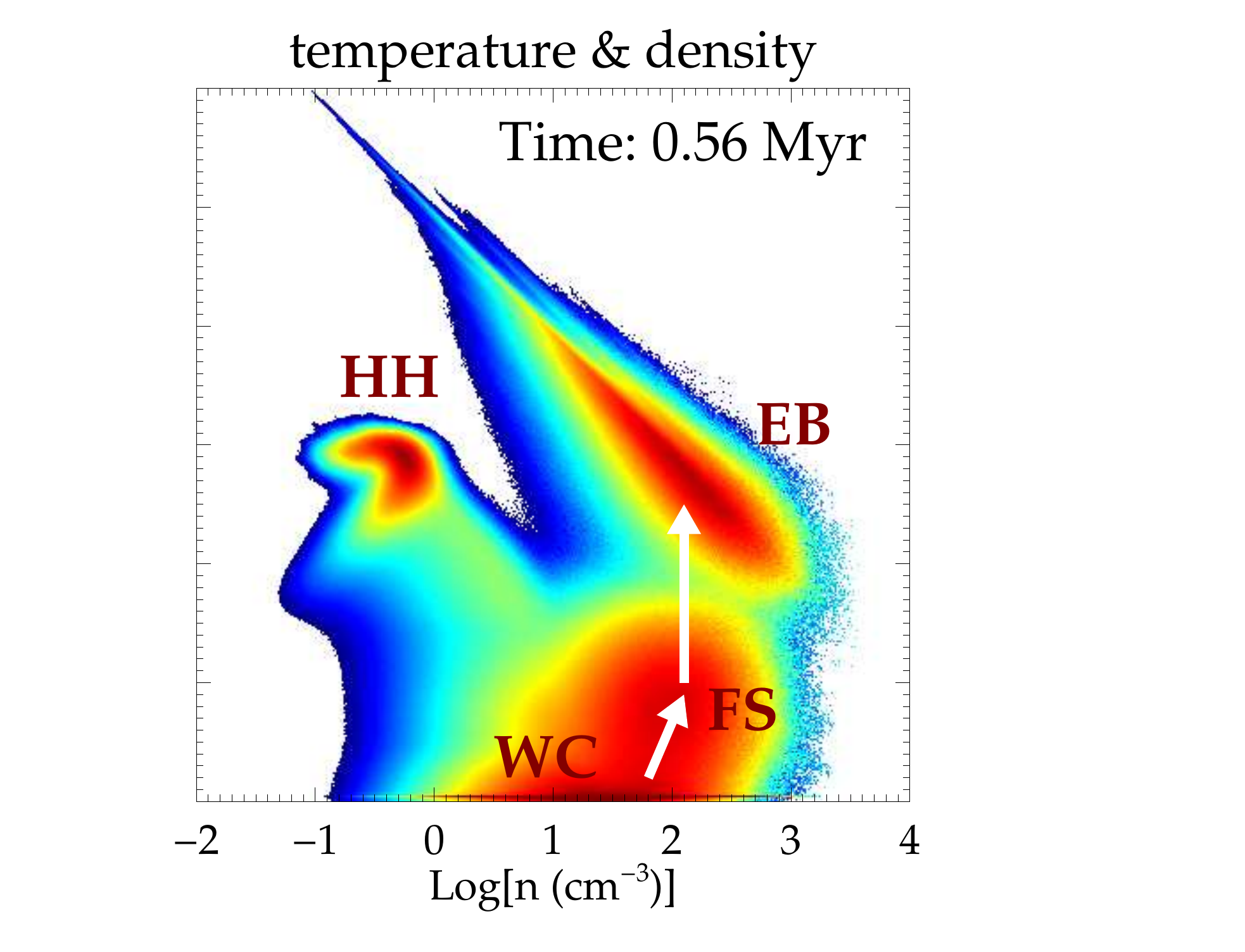}\vspace{-0.2cm}\hspace{-3cm}
	\includegraphics[width = 7.5cm, height = 5.6cm,keepaspectratio]
	{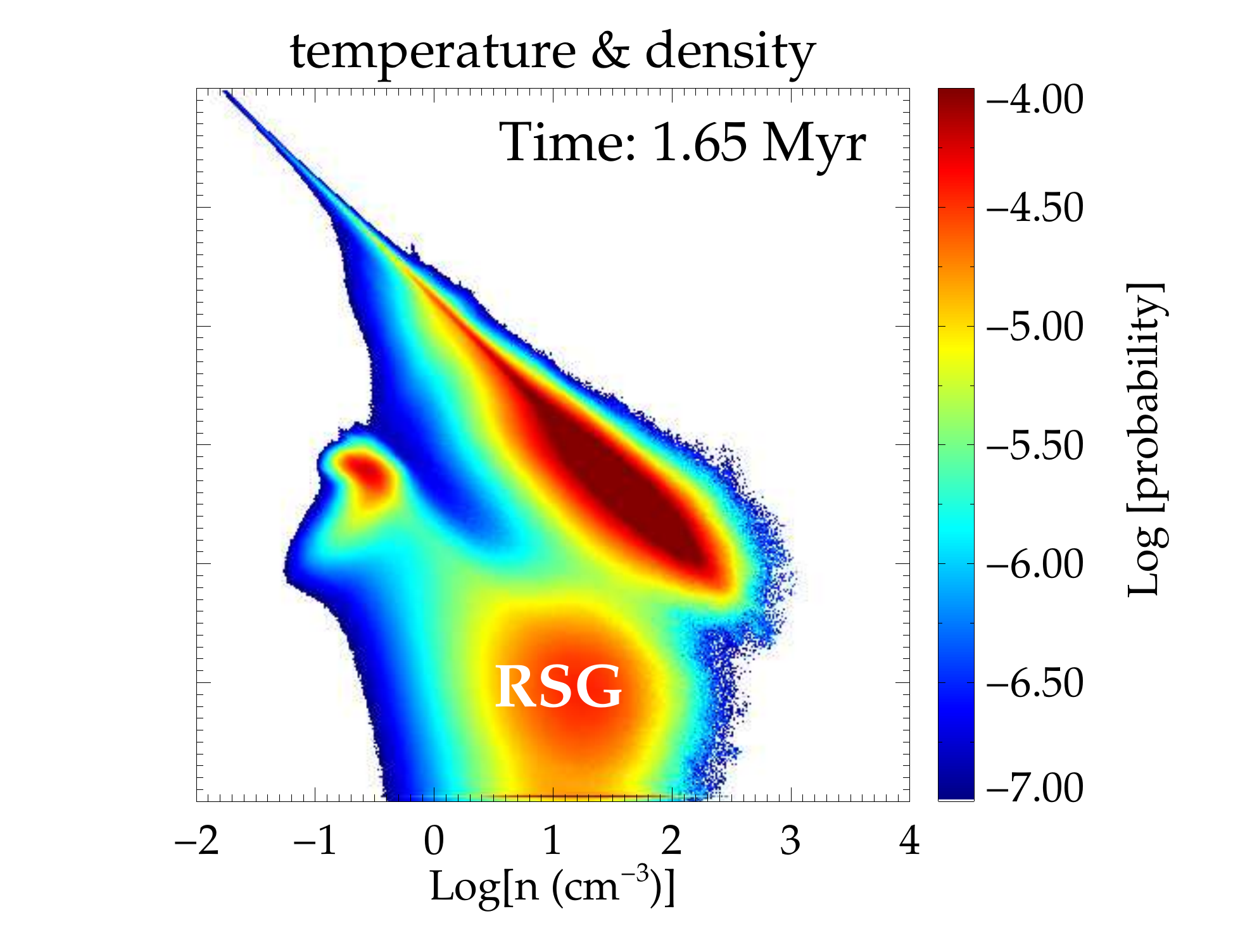}\vspace{-0.2cm}\linebreak
	\caption{\small Top: Mass weighted 2D probability distribution function (PDF) of pressure vs density at times corresponding to plots in Fig.~\ref{fig.jetsim.rho}. $p_0=9.2\times 10^{-4} \mbox{dynes cm}^{-2}$ is the scale pressure of the simulation. The color bar denotes the probability. The red-dotted line for an isothermal gas is presented to highlight that dense gas closely approximates an isothermal distribution at t=0. The white dashed line represents fits to the mean pressure for the phase corresponding to the energy bubble. The pressure is only weakly related to the density for this phase. Bottom:  Mass weighted 2D PDF of temperature vs density. The different ISM phases are labeled as described in Table~\ref{tab.ism.phases}}
	\label{fig.2DPDF.prs-rho.nw300}
\end{figure*}

Phase space diagrams are a useful way to understand the nature of the gas where different phases of the ISM driven by different physical mechanisms co-exist. In this section we discuss the evolution of the phase space of the ISM under the influence of the jet is depicted in Fig.~\ref{fig.2DPDF.prs-rho.nw300}. This figure depicts the two-dimensional mass weighted probability distribution function\footnote{The mass weighted PDFs are constructed by evaluating a two dimensional histogram of a quantity and counting the mass in each simulation cell normalised to the total mass as the weights for a histogram bin.} of $p$ vs $\rho$ and $T$ vs $\rho$. A mass weighted PDF gives preferential weights to the dense gas and hence is a more effective probe of the denser ISM than a volume weighted analysis. Table~\ref{tab.ism.phases} summarises the different phases and their labels. At $t=0$ the simulation domain has a two phase ISM consisting of a low density hot halo (``HH") and a warm dense turbulent filamentary  medium (``WC") at $T\sim 10^4$ K near the central $\sim 2$ kpc (see Fig.~\ref{fig.cldfilaments}).  Efficient atomic cooling renders the turbulent filaments nearly isothermal as the ISM is allowed to settle prior to the injection of the jet.

With the onset of the jet two other phases are clearly discernible. At first, a fast moving forward shock  sweeps through the ISM (as shown in Fig.~\ref{fig.jetsim.temp}) shocking the gas to $\sim 10^5$ K. The ISM shocked by the forward shock is labeled ``FS" in Fig.~\ref{fig.2DPDF.prs-rho.nw300}. The forward shock is followed by a slower, nearly adiabatically expanding high pressure energy bubble with $T \sim 10^6-10^7$ K. (labeled ``EB" in Fig.~\ref{fig.2DPDF.prs-rho.nw300}). The energy bubble is nearly homogeneous in pressure and depends only weakly on density as shown by the fits (white dashed lines in Fig.~\ref{fig.2DPDF.prs-rho.nw300}) to the mean pressure--density relationship of this phase. At later times, the effective polytropic index of the ISM in the bubble increases slightly as the pressure inside the bubble weakens and the denser filaments start to cool.  After the jet has broken out and decoupled from the turbulent ISM ($t \gtrsim 1.33$ Myr, corresponding to the third row of Fig.~\ref{fig.2DPDF.prs-rho.nw300}) there still remains a fraction of the gas shocked at $\sim 10^5$ K (``RSG"). This corresponds to the dense cores of the filaments which having survived shredding by the energy bubble have started to cool.

\item\emph{Polytropic turbulence:}
\begin{figure*}
	\centering
	\includegraphics[width = 7.5cm, keepaspectratio]
	{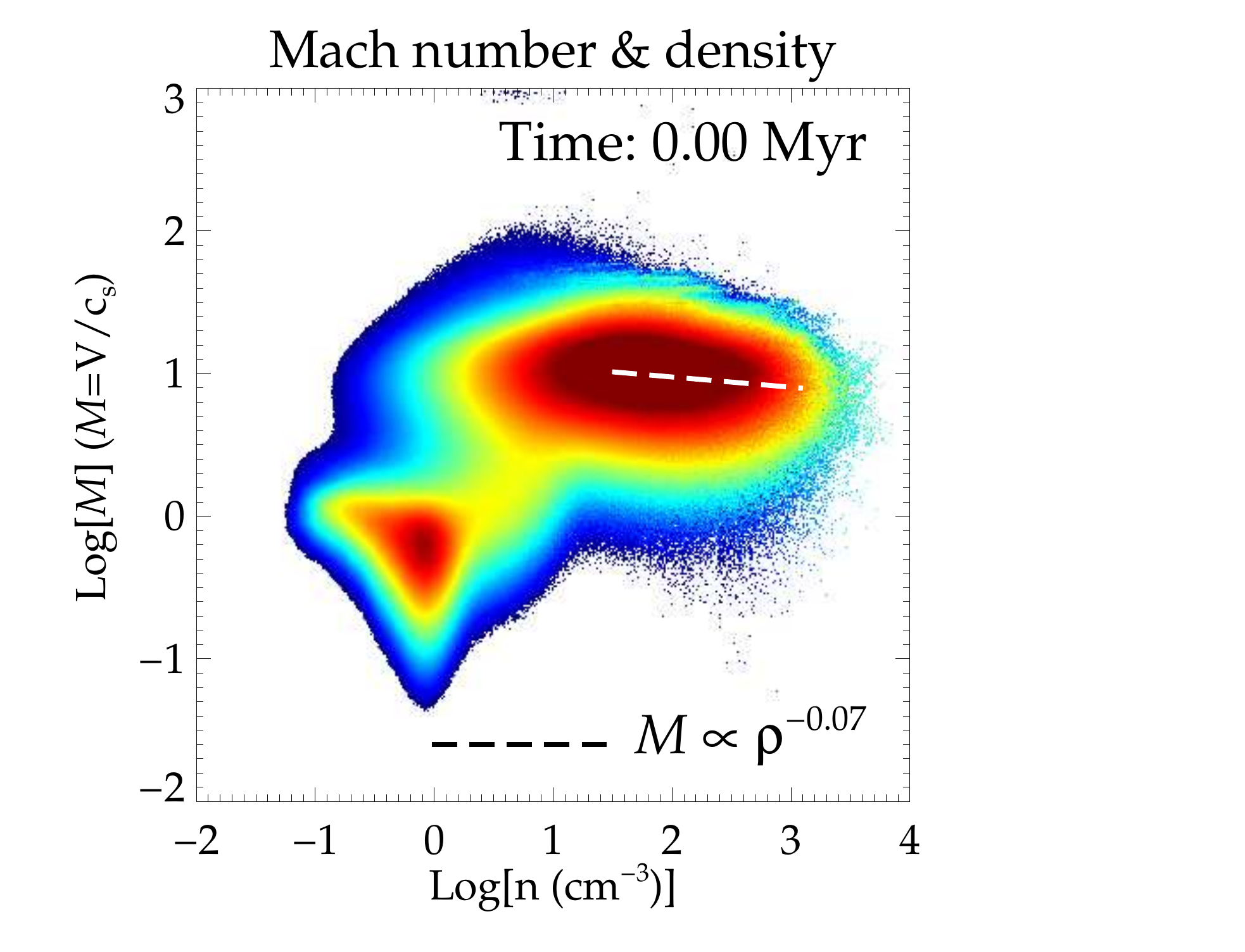}\hspace{-3cm}\vspace{-0.1cm}
	\includegraphics[width = 7.5cm, keepaspectratio]
	{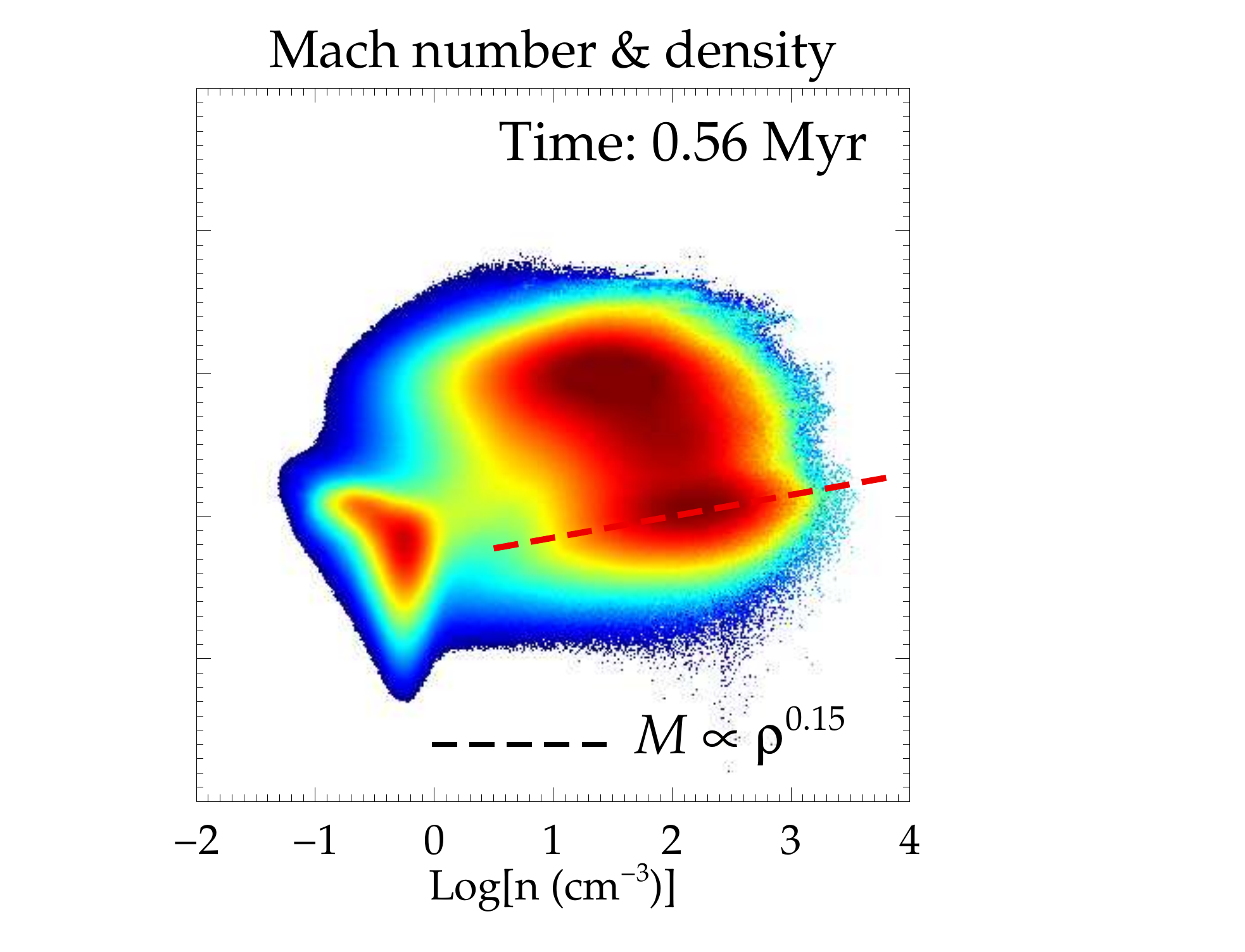}\hspace{-3cm}\vspace{-0.1cm}
	\includegraphics[width = 7.5cm, keepaspectratio] 
	{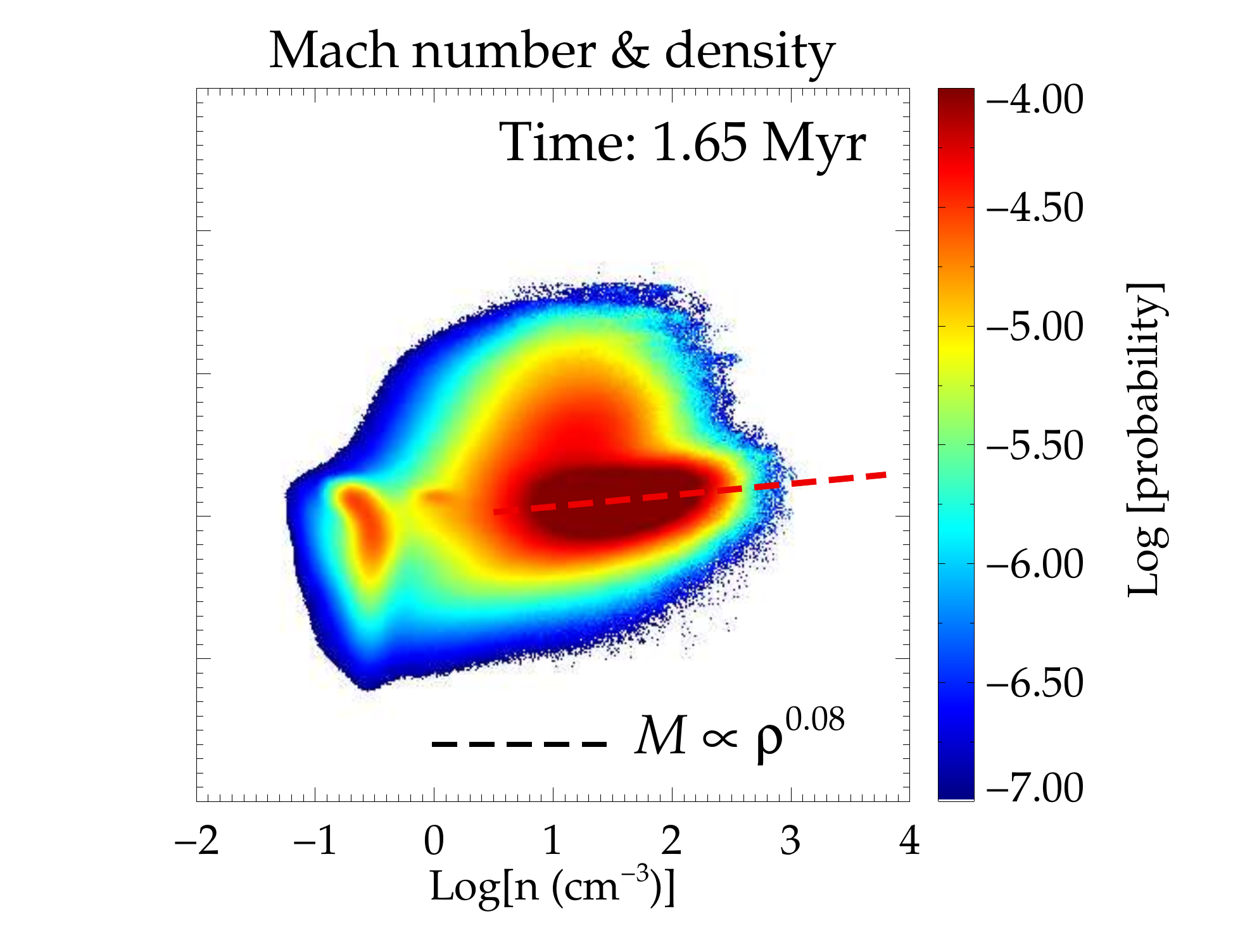}\vspace{-0.1cm}
	\caption{\small Mass weighted 2D PDF of $\log(\mach)$ vs $\log (\rho)$ at times corresponding to plots in Fig.~\ref{fig.jetsim.rho}. The dashed line represents fits to the mean pressure.}
	\label{fig.machhist}
\end{figure*}

In Fig.~\ref{fig.machhist} we show the 2D 
PDF of $\mach$ vs $\rho$, where $\mach$ is the mach number defined by $\mach=v/c_s$, $c_s$ being the sound speed. For the settling ISM, the dense filaments form a region of isothermal supersonic turbulence (left panel of Fig.~\ref{fig.machhist}). The Mach number is only weakly correlated with the density, similar to previously reported results from numerical simulations of supersonic isothermal turbulence \citep{kritsuk07,federrath10,federrath15}. Ideally for isothermal supersonic turbulence the mach number is independent of the density if the density fluctuations are completely random at all scales \citep{passot98a}. The weak correlations appearing in the numerical simulations result from strong shocks or rarefactions where the fundamental assumptions of the independence of the density fluctuations break down  \citep{vazquez94a,passot98a,federrath10}. With the onset of the jet, $\mach$ and $\rho$ are seen to be positively correlated with density. This is indicative of polytropic turbulence in a medium with polytropic index less than unity \citep{passot98a,federrath15}, where the PDF deviates from a lognormal distribution. 

\item \emph{Outflow velocity of different phases:}
\begin{figure*}
	\centering
	\includegraphics[width = 7.5cm, keepaspectratio] 
	{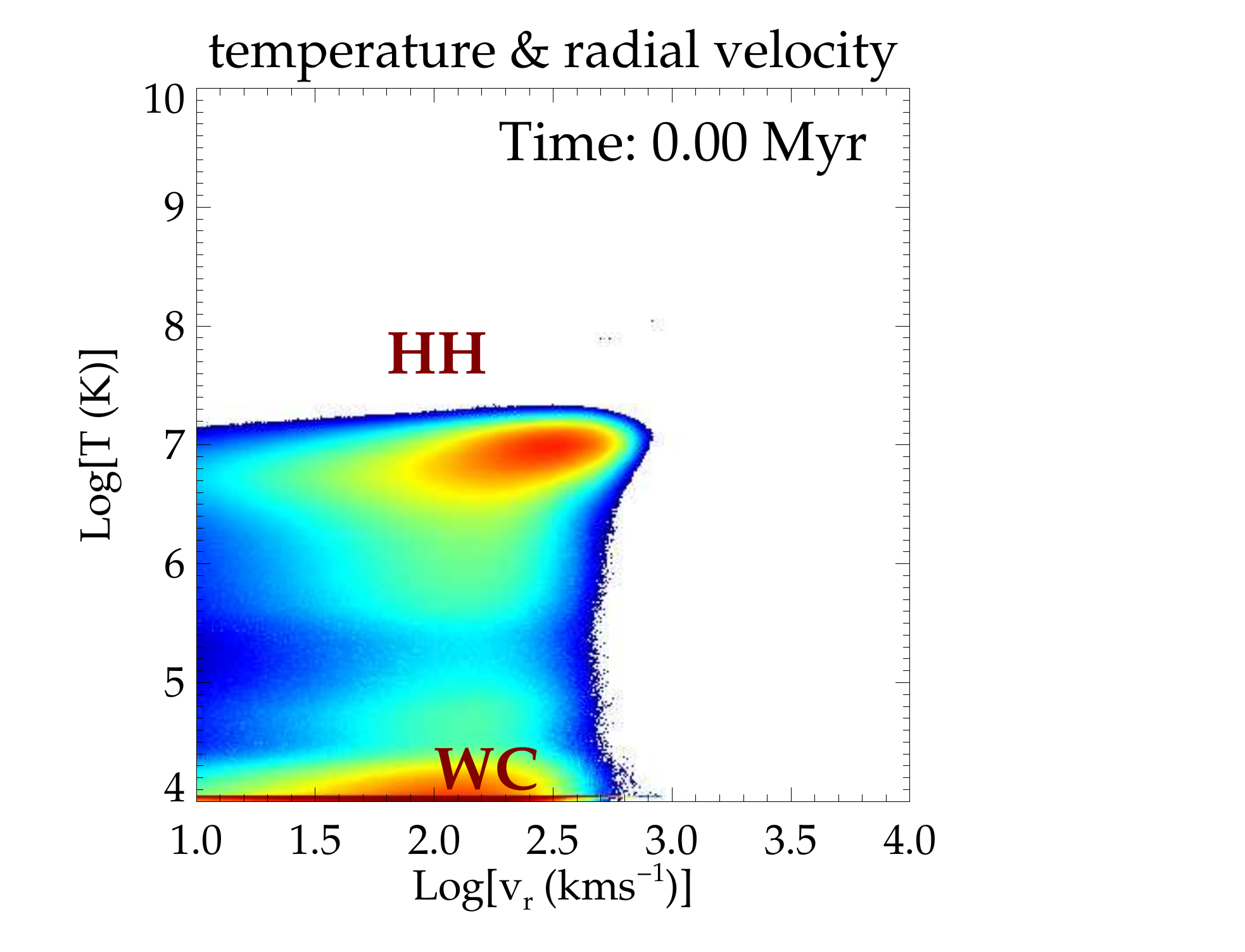}\vspace{-0.1cm}\hspace{-3cm}
	\includegraphics[width = 7.5cm, keepaspectratio] 
	{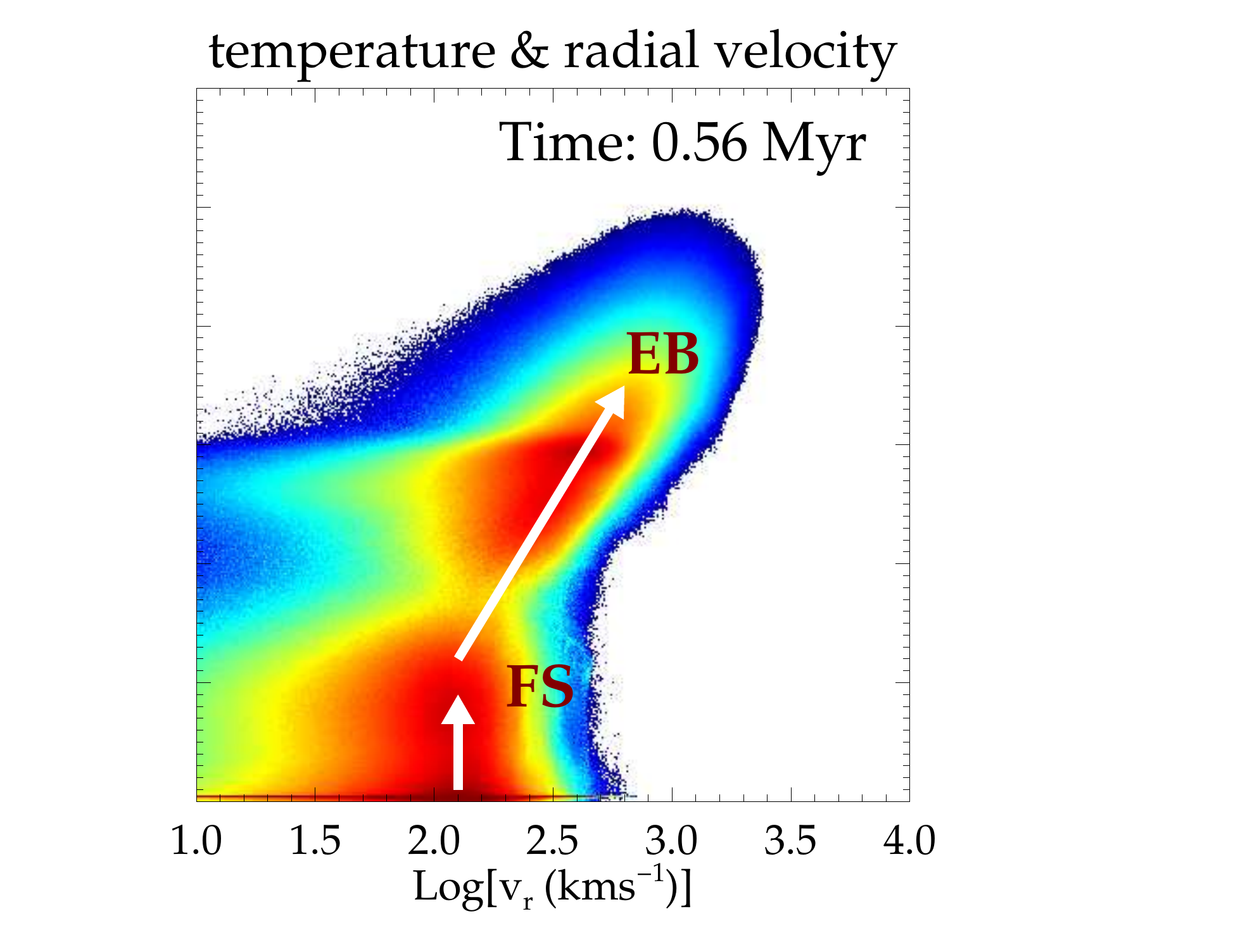}\vspace{-0.1cm}\hspace{-3cm}
	\includegraphics[width = 7.5cm, keepaspectratio] 
	{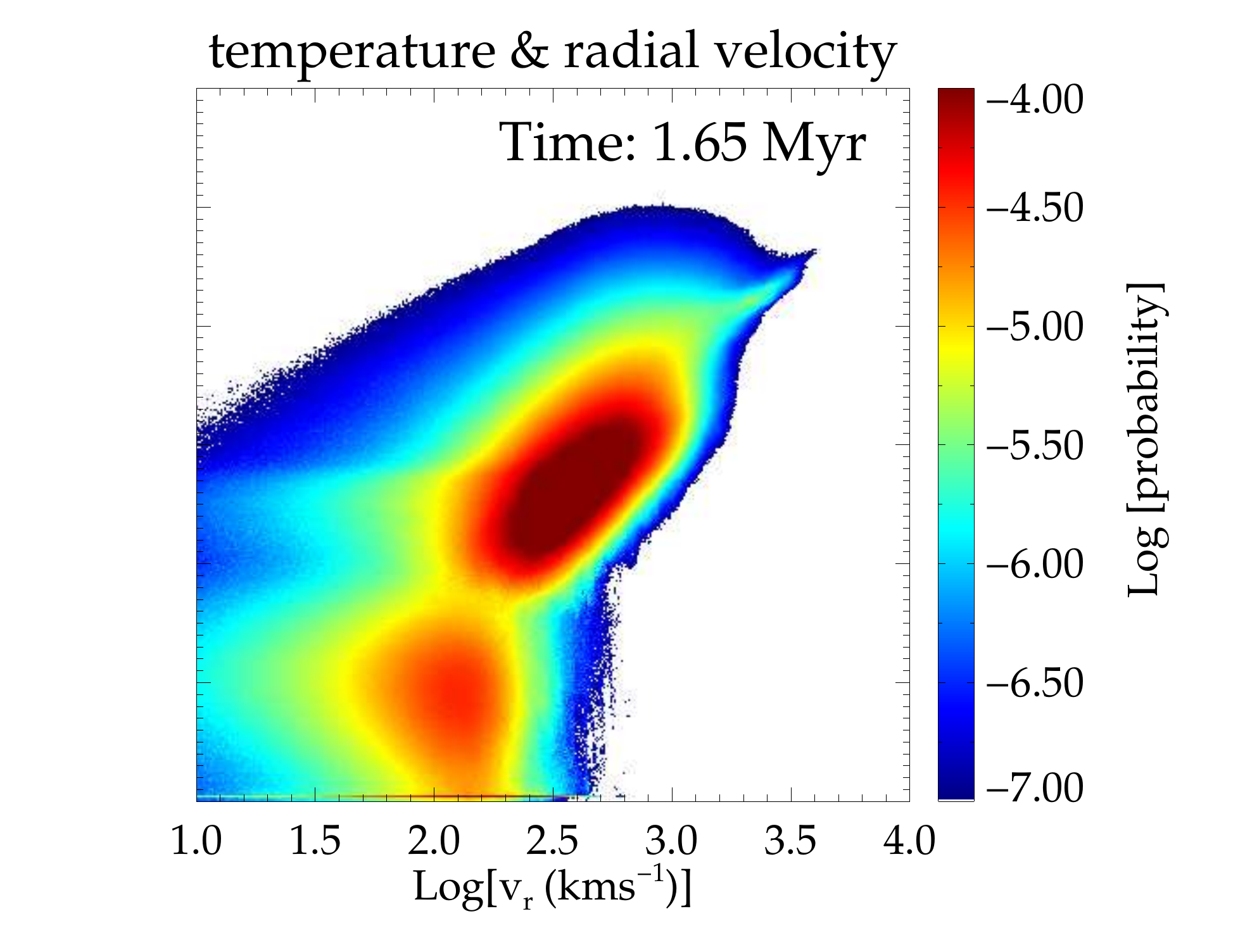}\vspace{-0.1cm}\linebreak
	\includegraphics[width = 7.5cm, keepaspectratio] 
	{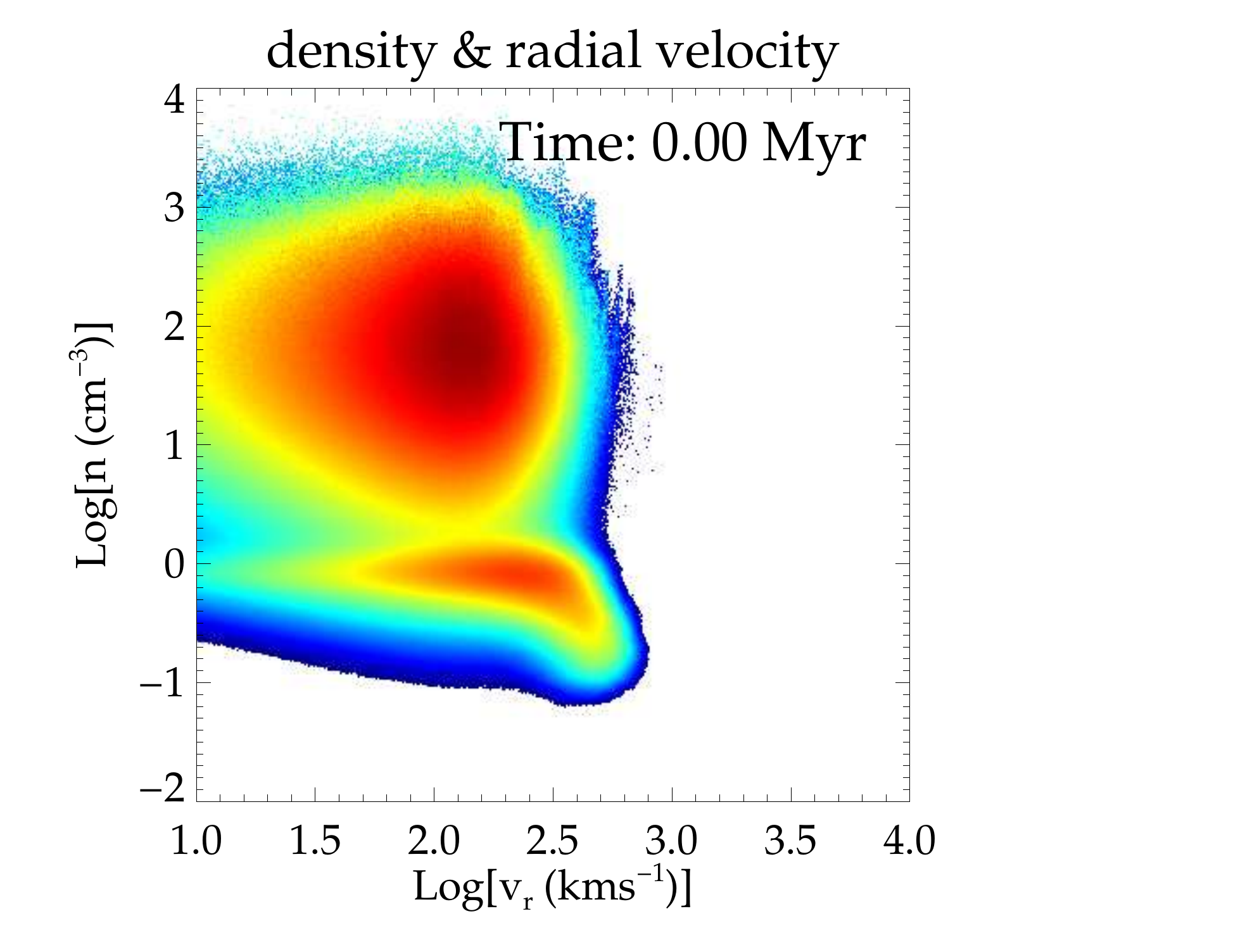}\vspace{-0.2cm}\hspace{-3cm}
	\includegraphics[width = 7.5cm, keepaspectratio] 
	{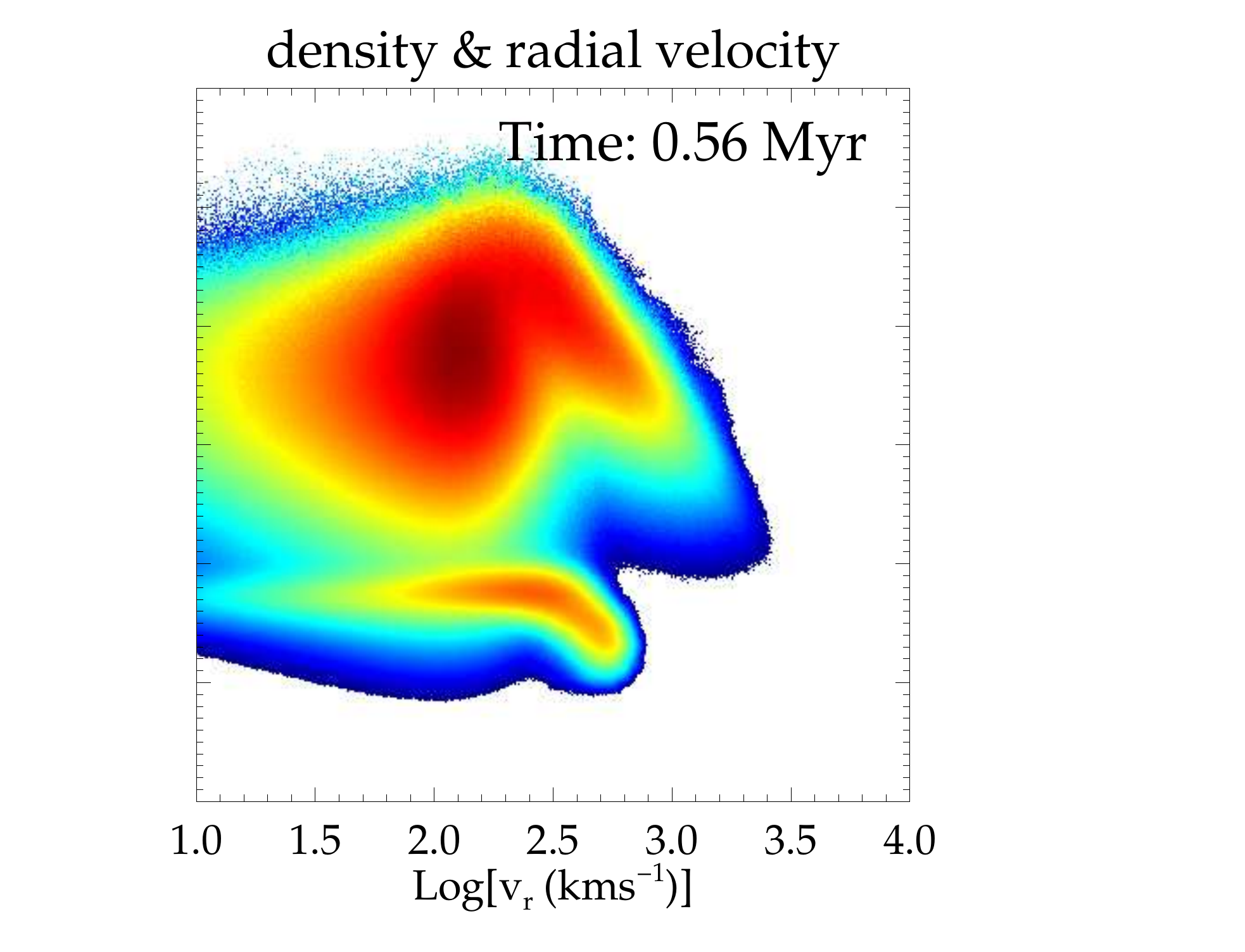}\vspace{-0.2cm}\hspace{-3cm}
	\includegraphics[width = 7.5cm, keepaspectratio] 
	{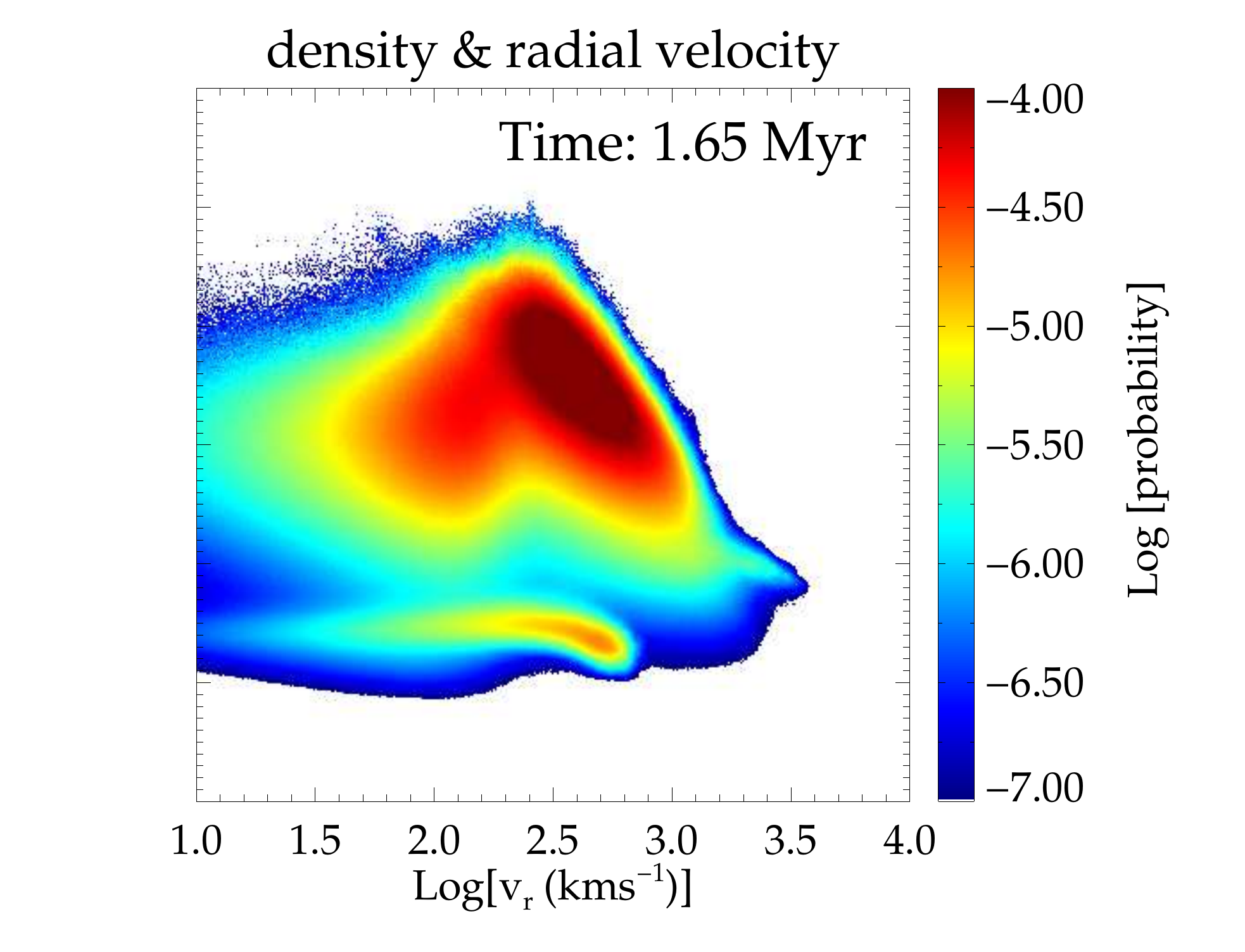}\vspace{-0.2cm}\linebreak
	\caption{\small  Mass weighted 2D PDF of temperature vs radial velocity (top) and density vs radial velocity (bottom) at times corresponding to plots in Fig.~\ref{fig.jetsim.rho}. After the jet injection, the warm dense medium is shocked to $\sim 10^5$ K by the forward shock without an appreciable increase in velocity. Dense gas is accelerated to radial velocities $\gtrsim 500 \kms$ by the hot pressure bubble.}	\label{fig.2DPDF.temp-vr.nw300}
\end{figure*}

Fig.~\ref{fig.2DPDF.temp-vr.nw300} shows the  temperature vs radial velocity ($|v_r|$) and density vs radial velocity ($|v_r|$) 2D probability density. We see that the passage of the forward shock results in heating up of the ISM to $\sim 10^5$ K as mentioned above without any appreciable increase in radial velocity. The clouds are primarily accelerated by the high pressure energy bubble to speeds of $\gtrsim 500 \kms$. In the right hand panels we see a tail of very high velocity $\gtrsim 1000 \kms$ with less mass weight. This corresponds to the low density cloud ablated material swept up by the energy bubble to high radial speeds, as shown in Fig.~\ref{fig.jetsim.rho}. 
\end{itemize}

\section{Dependence of morphology on jet power}\label{sec.diffpower}
\begin{figure*}
	\centering
	\includegraphics[width = 6.2cm,keepaspectratio] 
	{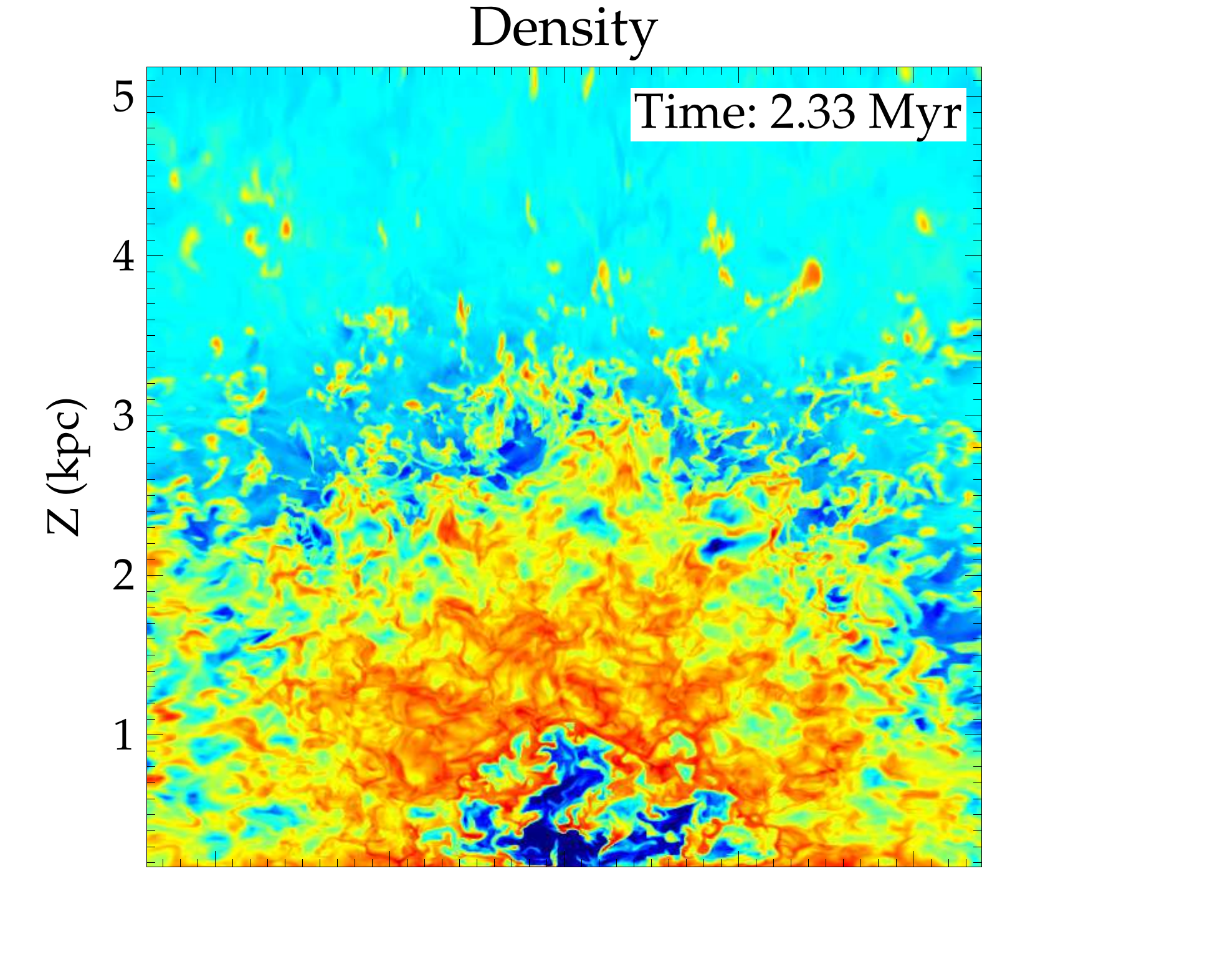} \vspace{-0.1cm}\hspace{-2cm}
	\includegraphics[width = 6.2cm,keepaspectratio] 
	{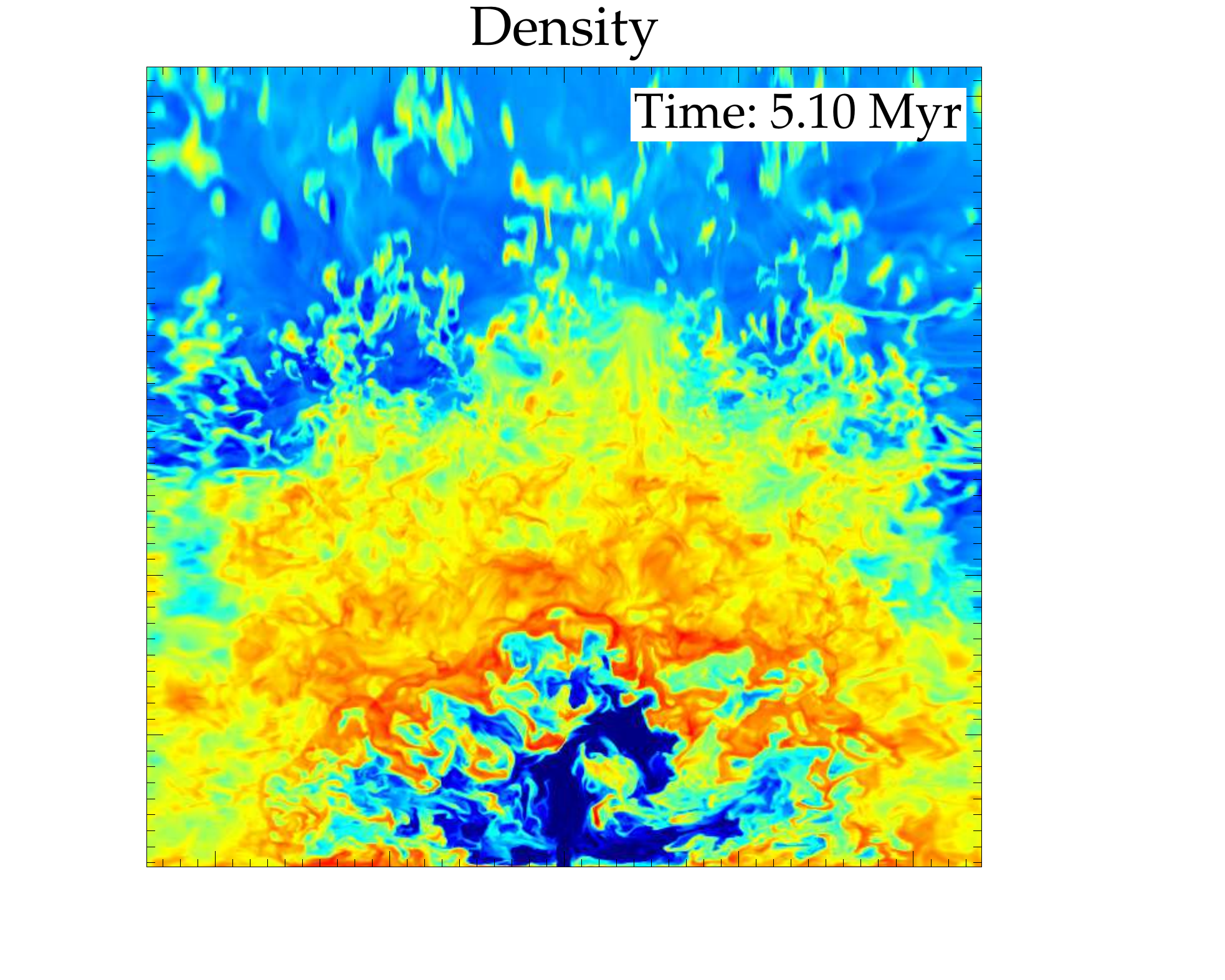} \vspace{-0.1cm}\hspace{-2cm}
	\includegraphics[width = 6.2cm,keepaspectratio] 
	{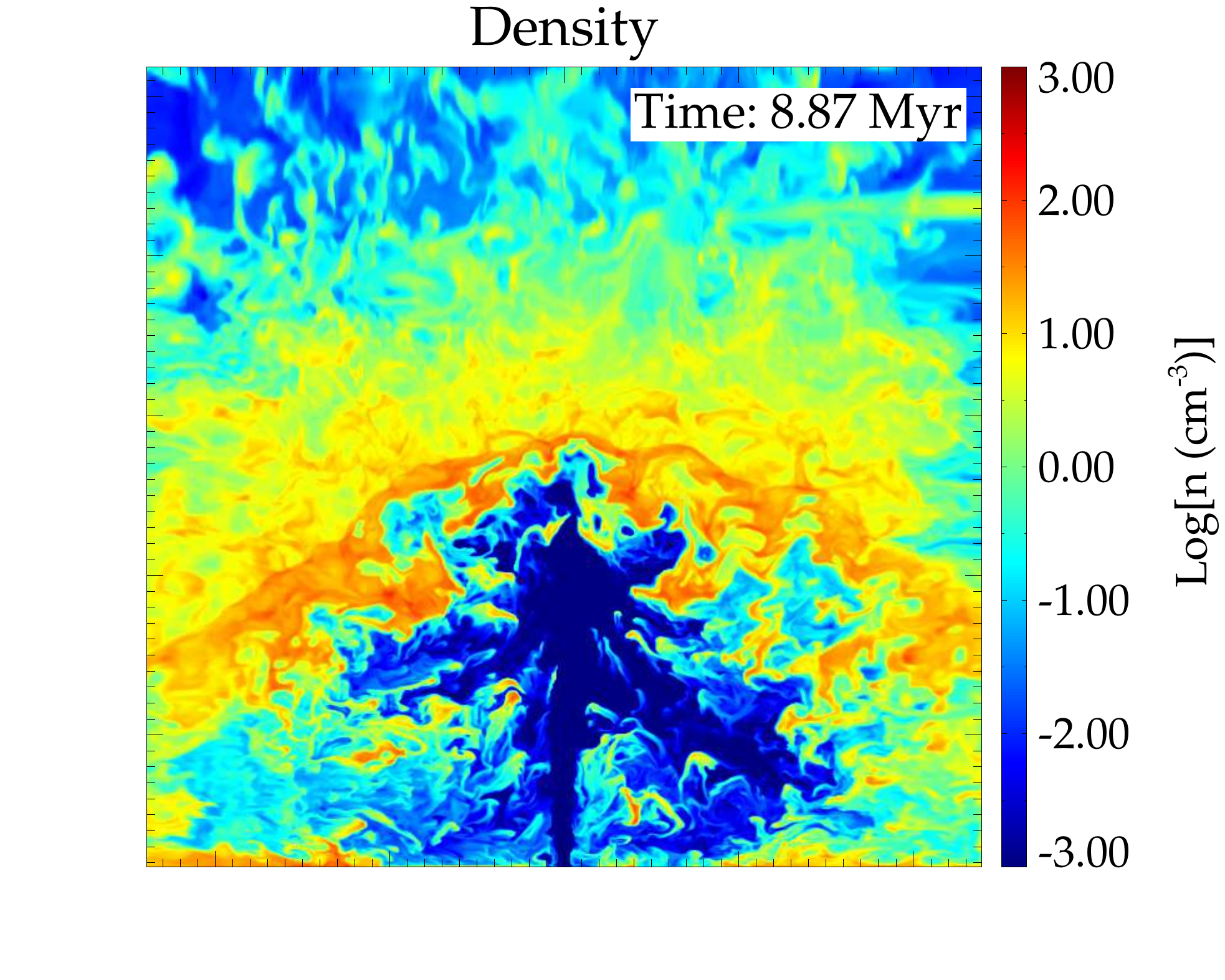} \vspace{-0.1cm}\linebreak
	\includegraphics[width = 6.2cm,keepaspectratio] 
	{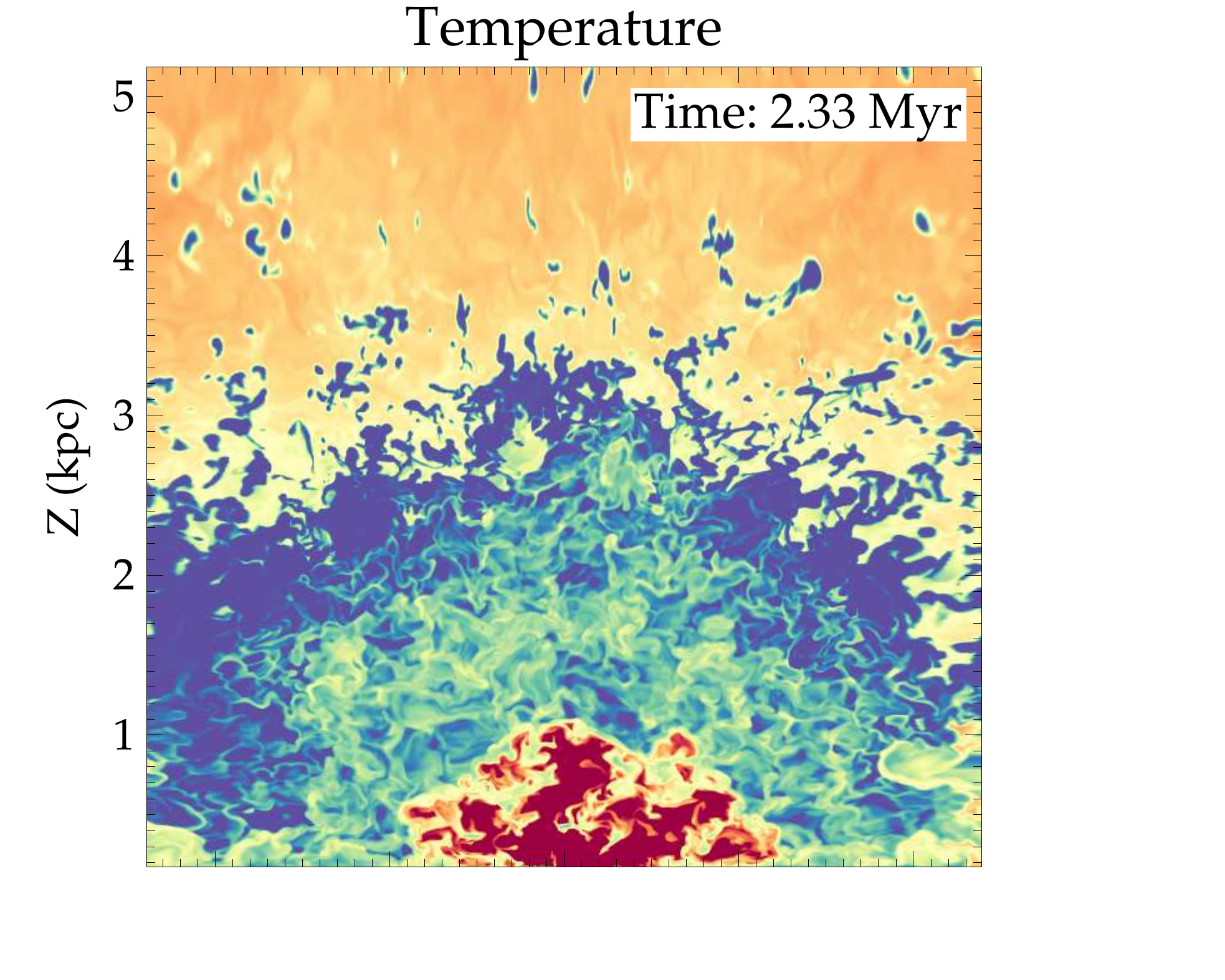} \vspace{-0.1cm}\hspace{-2cm}
	\includegraphics[width = 6.2cm,keepaspectratio] 
	{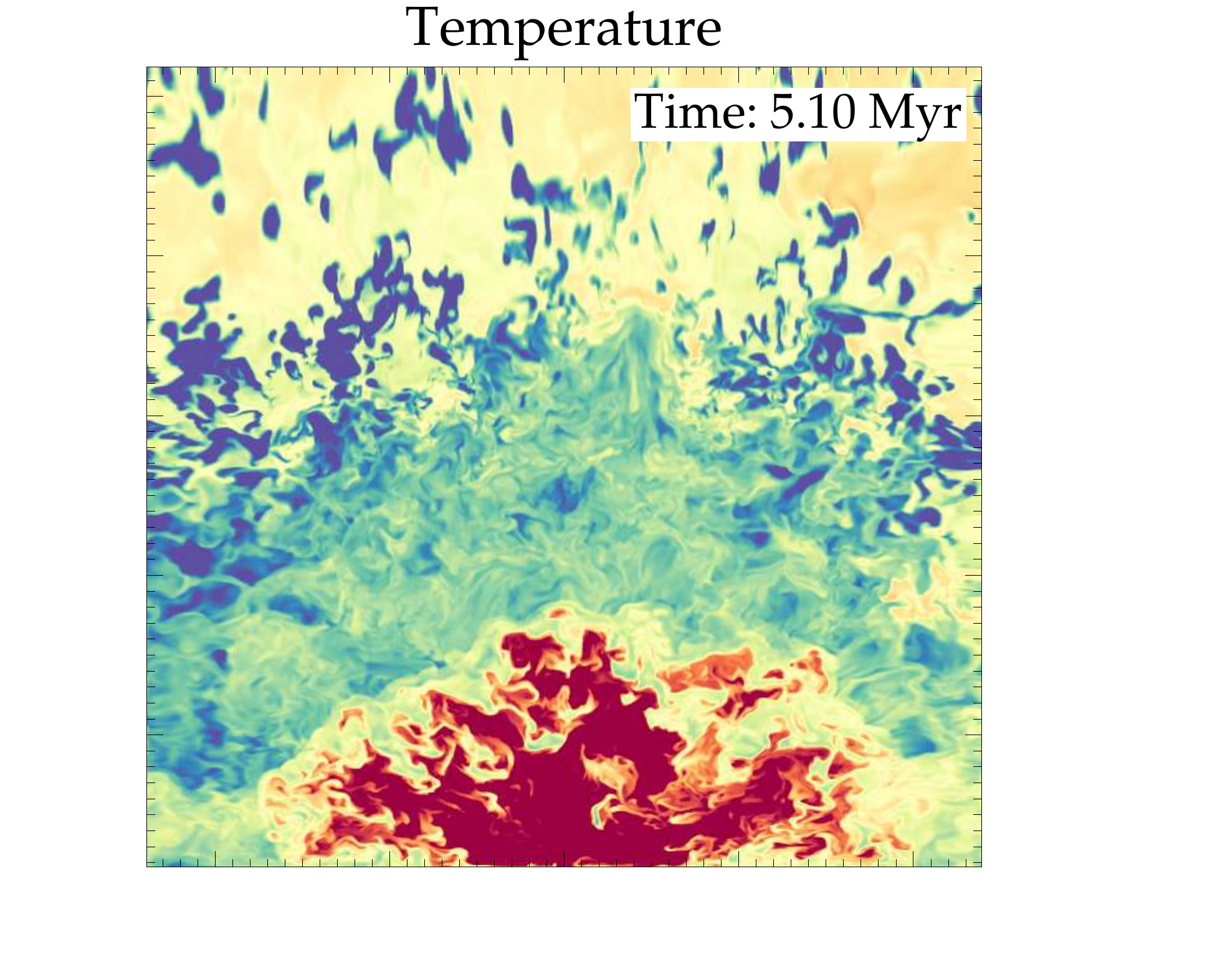} \vspace{-0.1cm}\hspace{-2cm}
	\includegraphics[width = 6.2cm,keepaspectratio] 
	{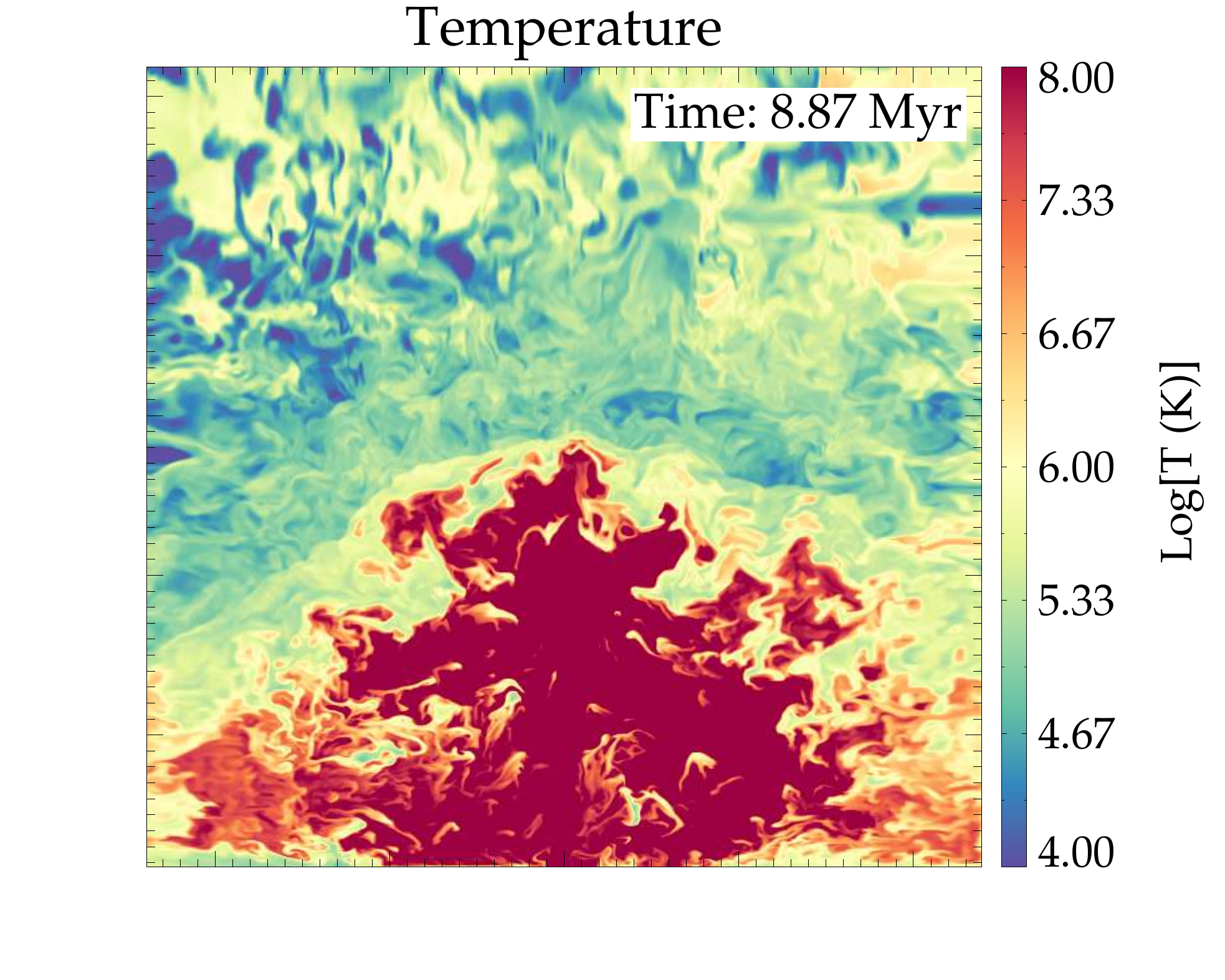} \vspace{-0.1cm}\linebreak
	\includegraphics[width = 6.2cm,keepaspectratio] 
	{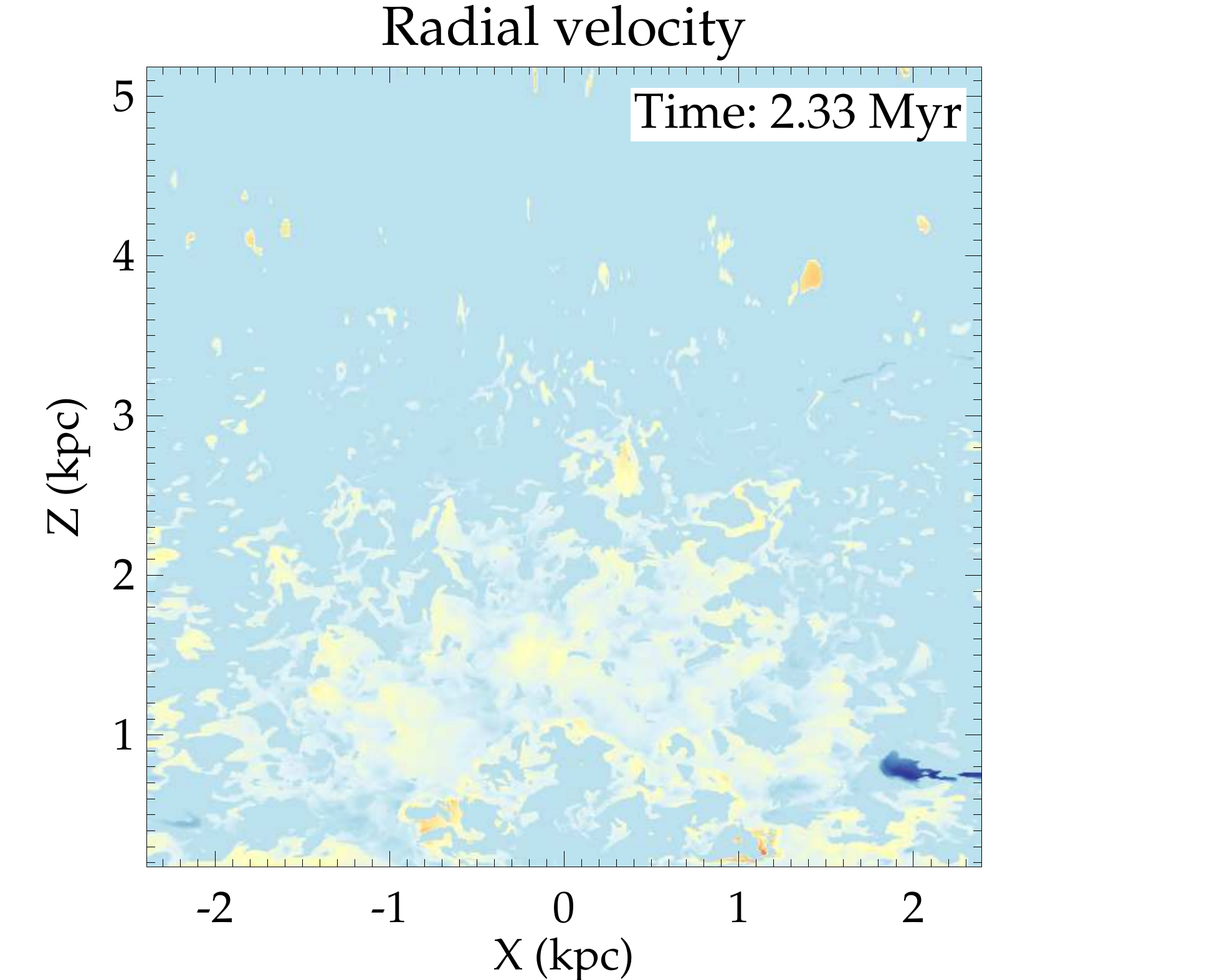}\vspace{-0.1cm} \hspace{-2cm}
	\includegraphics[width = 6.2cm,keepaspectratio] 
	{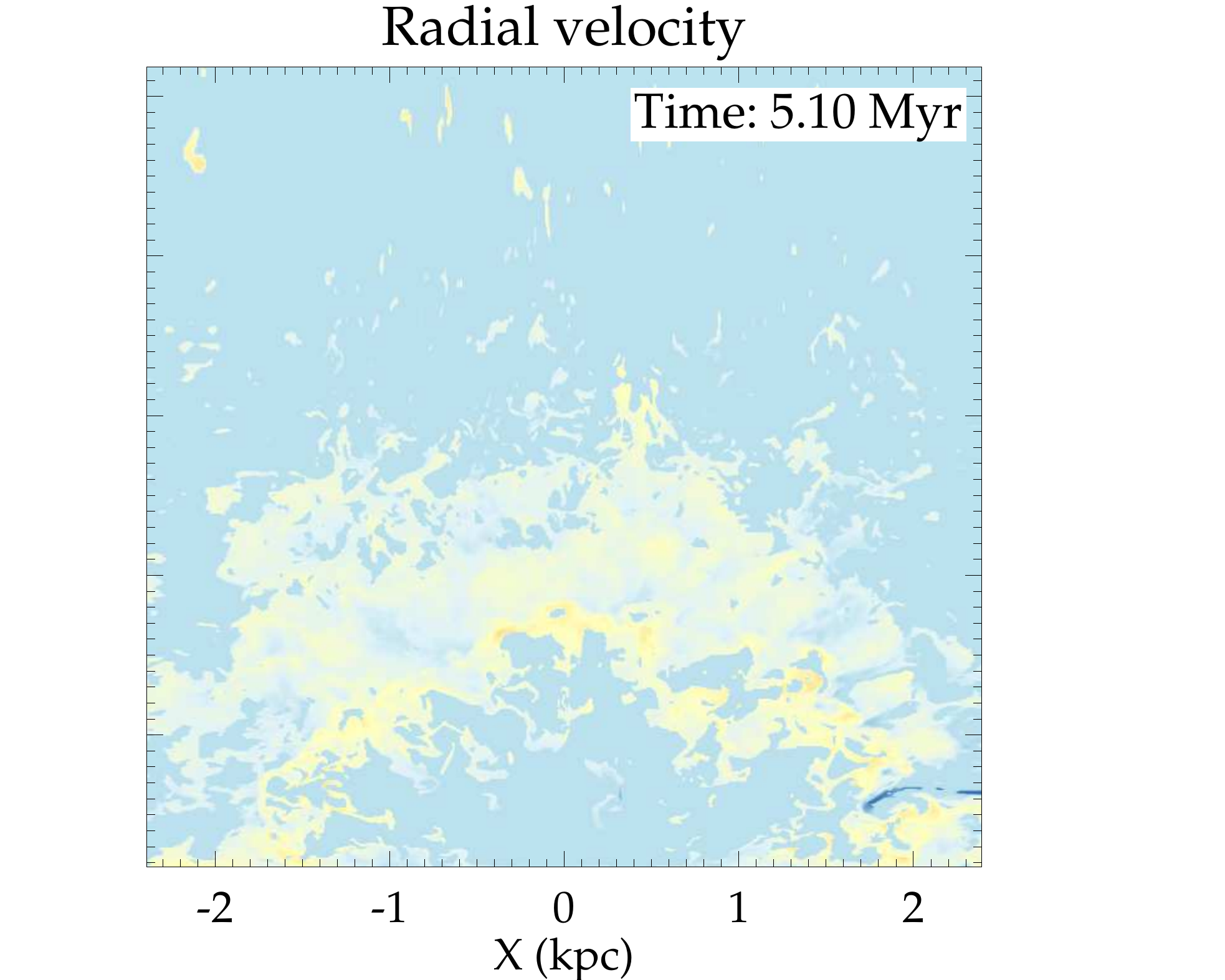}\vspace{-0.1cm}  \hspace{-2cm}
	\includegraphics[width = 6.2cm,keepaspectratio] 
	{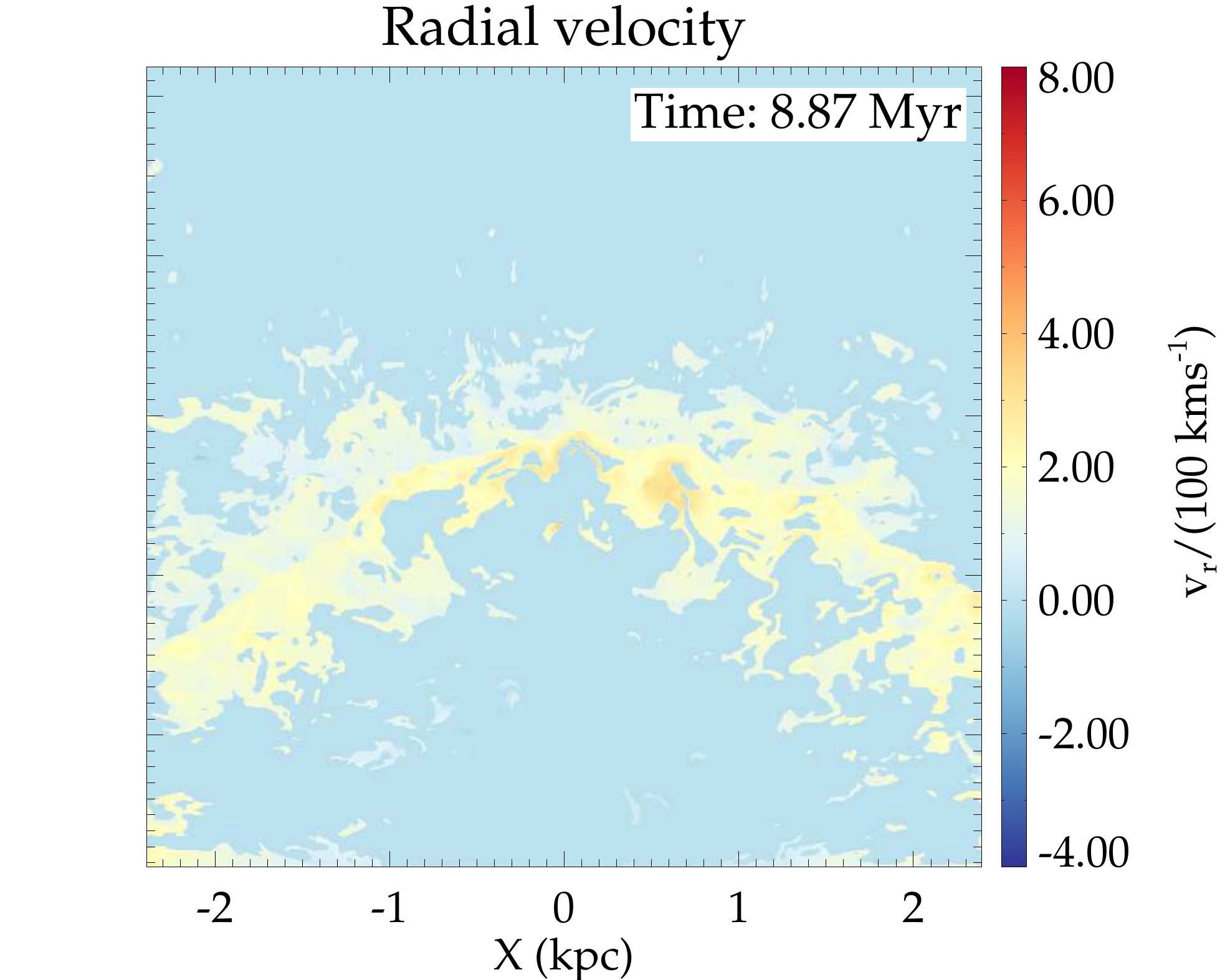}\vspace{-0.1cm}
	\caption{\small Evolution of density (in $\cc$), temperature (in Kelvin) and radial velocity (normalised to $100 \kms$) for sim.~D with $P_{\rm jet}=10^{43} \ergs$. The jet remains trapped in the ISM for a much longer time as compared to jets of higher power. The trapped energy bubble affects a larger volume of the ISM.}
	\label{fig.jetsim.f43}
\end{figure*}
\begin{figure*}
	\centering
	\hspace{-0.1cm}\includegraphics[width = 7.5cm,keepaspectratio] 
	{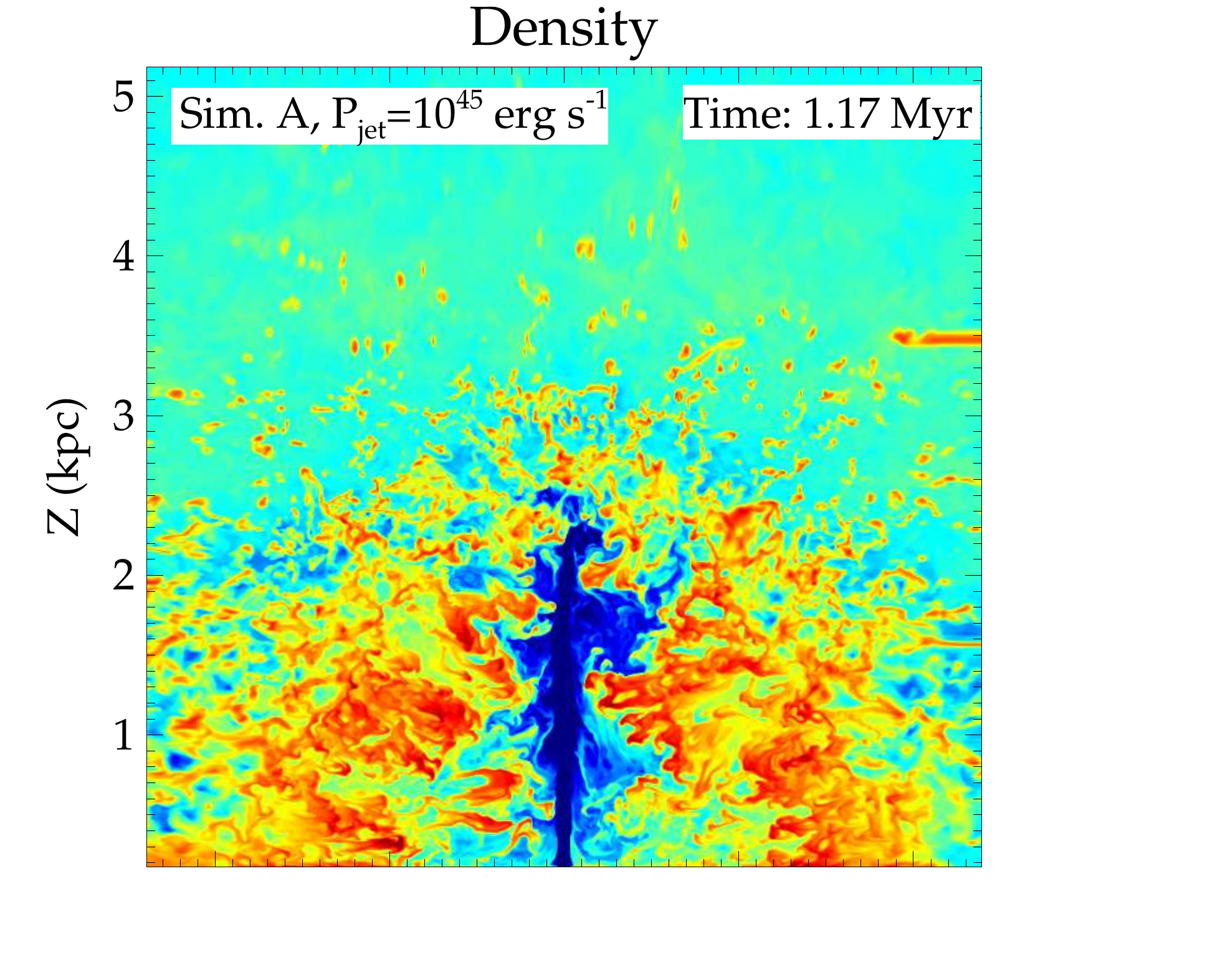}{\hspace{-2.3cm}}{\vspace{-0.5cm}}
	\includegraphics[width = 7.5cm,keepaspectratio] 
	{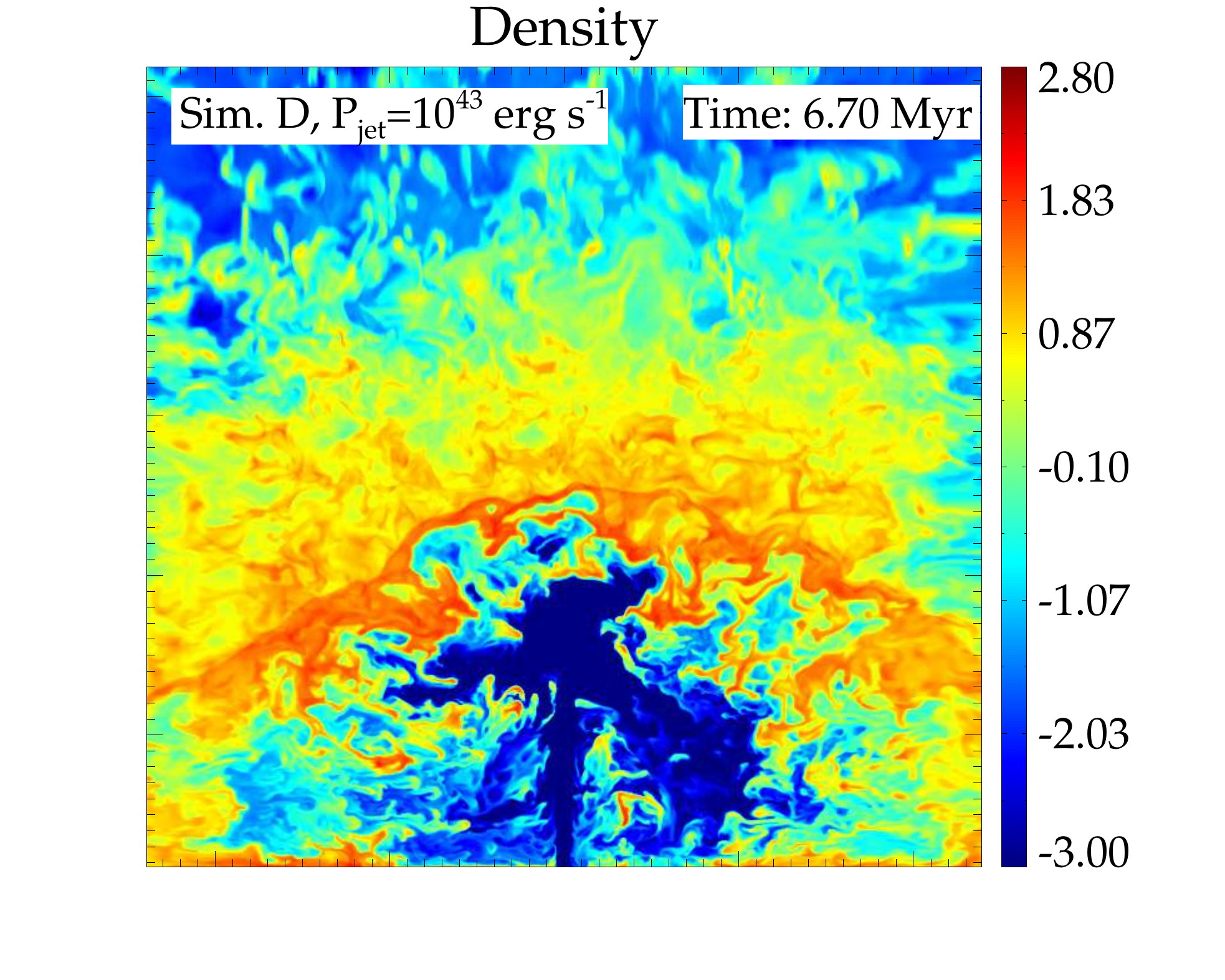}{\vspace{-0.5cm}}
	\includegraphics[width = 7.5cm, keepaspectratio] 
	{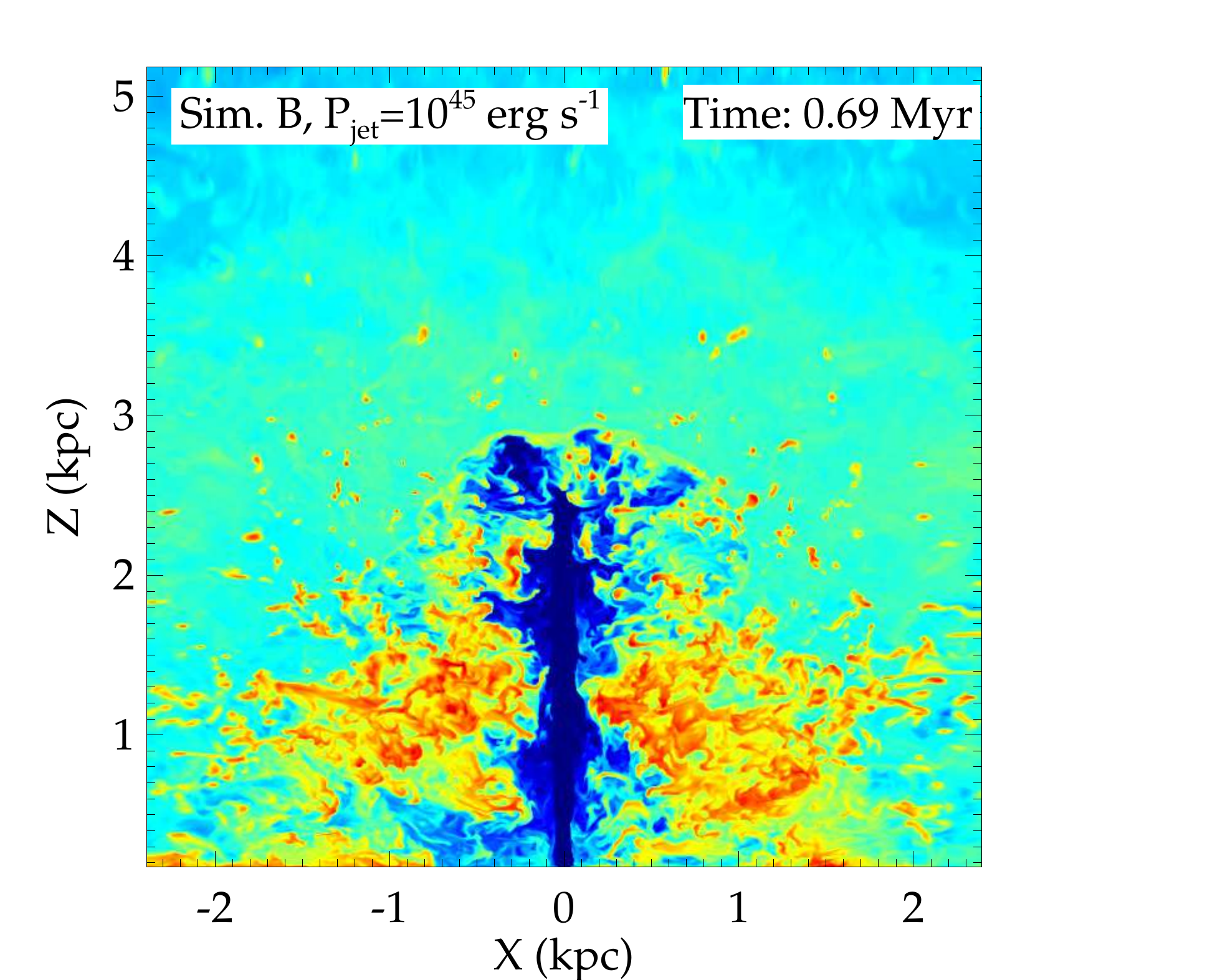}{\hspace{-2.3cm}}
	\includegraphics[width = 7.5cm, keepaspectratio] 
	{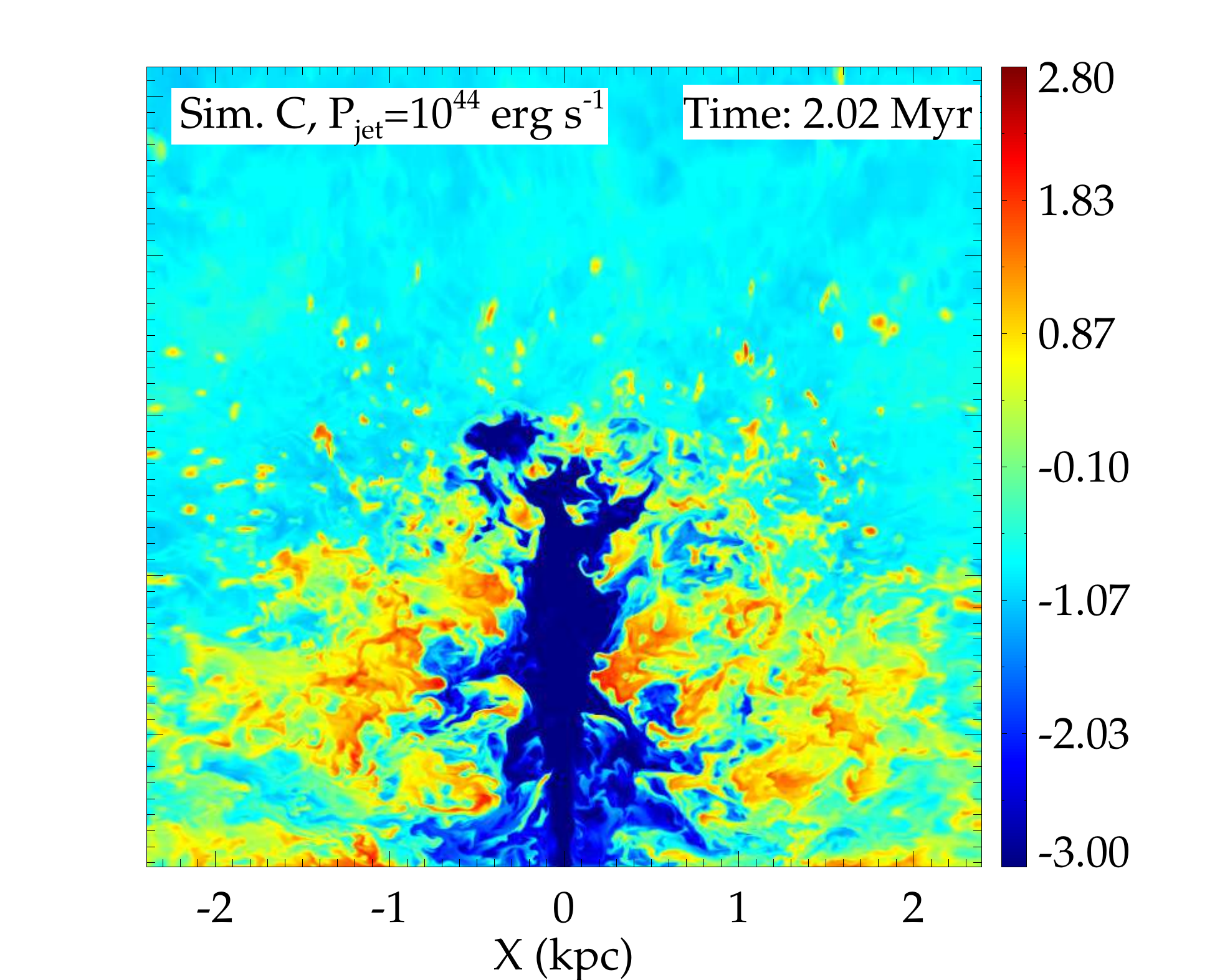}
	\caption{\small Density (in cm$^{-3}$) at the x-z plane for different simulations. Upper panel: Jets of power $10^{45}$ erg s$^{-1}$ (left, Sim.~A) and  $10^{43}$ erg s$^{-1}$ (right, Sim.~D)
passing through the same initial ISM derived from a relaxed turbulent fractal with $n_{w0}=300 \cc$ (see Sec.~\ref{sec.initialisation} and Sec.~\ref{sec.settling} for details). Lower panels: Jets of power $10^{45}$ erg s$^{-1}$ (left, Sim.~B) and  $10^{44}$ erg s$^{-1}$ (right, Sim.~C), for an ISM initialised with $n_{w0} \sim 150 \cc$.}
	\label{fig.jetrhocompare}
\end{figure*}
\begin{figure}
	\centering
	\includegraphics[width = 7cm, height = 8cm,keepaspectratio] {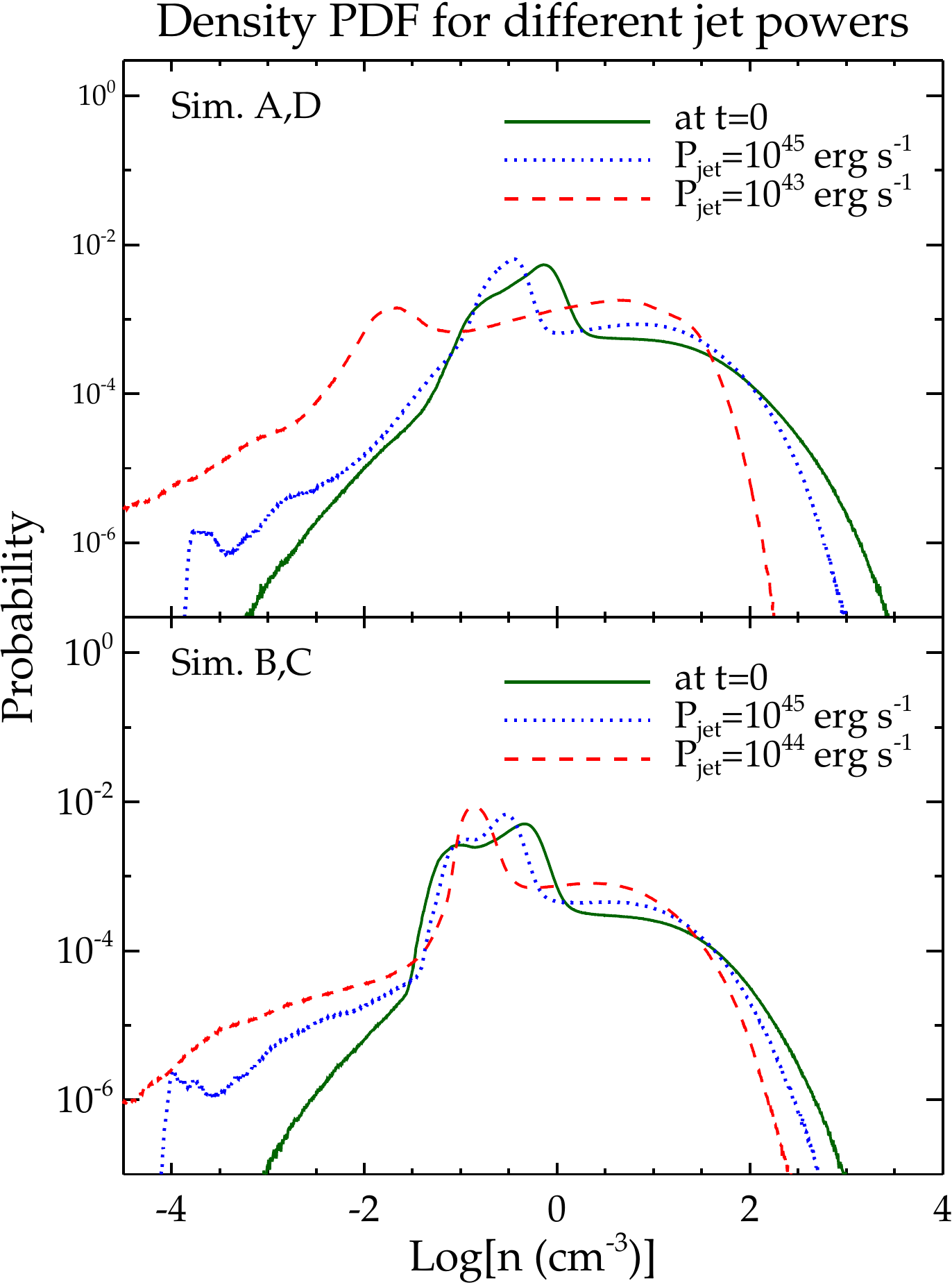}
	\caption{\small The density PDF of the ISM of simulations corresponding to Fig.~\ref{fig.jetrhocompare}. The PDF of the ISM before the injection of the jet is represented in black. The jets of lower power are more effective in destroying the high density cloud cores.}
	\label{fig.jetpowercomparepdf}
\end{figure}
\begin{table}
\centering
\caption{Coefficients of the fit to the density PDF in Fig.~\ref{fig.jetrhocompare}}
\label{tab.comparepdffit}
\begin{tabular}{| l | l | l | l | l |}
\hline
   &      						& \multicolumn{3}{c|}{Sim. A, D} \\		
No & Parameter						&  t=0      &  $P_{45}$   &  $P_{45}/100 $     \\
\hline
1  & $\bar{\rho}$ in $\cc$				& 21.38     &  16.51			   &  7		\\
2  & $\sigma _\rho $	in $\cc$			& 52.68     &  31.27			   &  9.32     \\
3  & $\eta$ 						& 0.14	    &  0.11                        &  0.37	\\
4  & $\tilde{\rho} _{>}$ in $\cc$			& 50	    &  35.54			   &  20.19     \\
\hline
   &      						& \multicolumn{3}{c|}{Sim. B, C} \\		
No & Parameter						&  t=0      &  $P_{45}$   &  $P_{45}/10$     \\
\hline
1  & $\bar{\rho}$ in $\cc$				& 12.32      &  9.17			   &  3.98	\\
2  & $\sigma _\rho $	in $\cc$			& 24.74      &  15.88			   &  7.52     \\
3  & $\eta$ 						& 0.19	     &  0.15                       &  0.28	\\
4  & $\tilde{\rho} _{>}$ in $\cc$			& 34.2	     &  26.63			   &  21     \\
\hline
\end{tabular} \\
\begin{tablenotes}
\small
\item (a) $P_{45} = P_{\rm jet}=10^{45} \ergs$
\end{tablenotes}
\end{table}
\begin{figure*}
	\centering
	\includegraphics[width = 7cm,keepaspectratio] {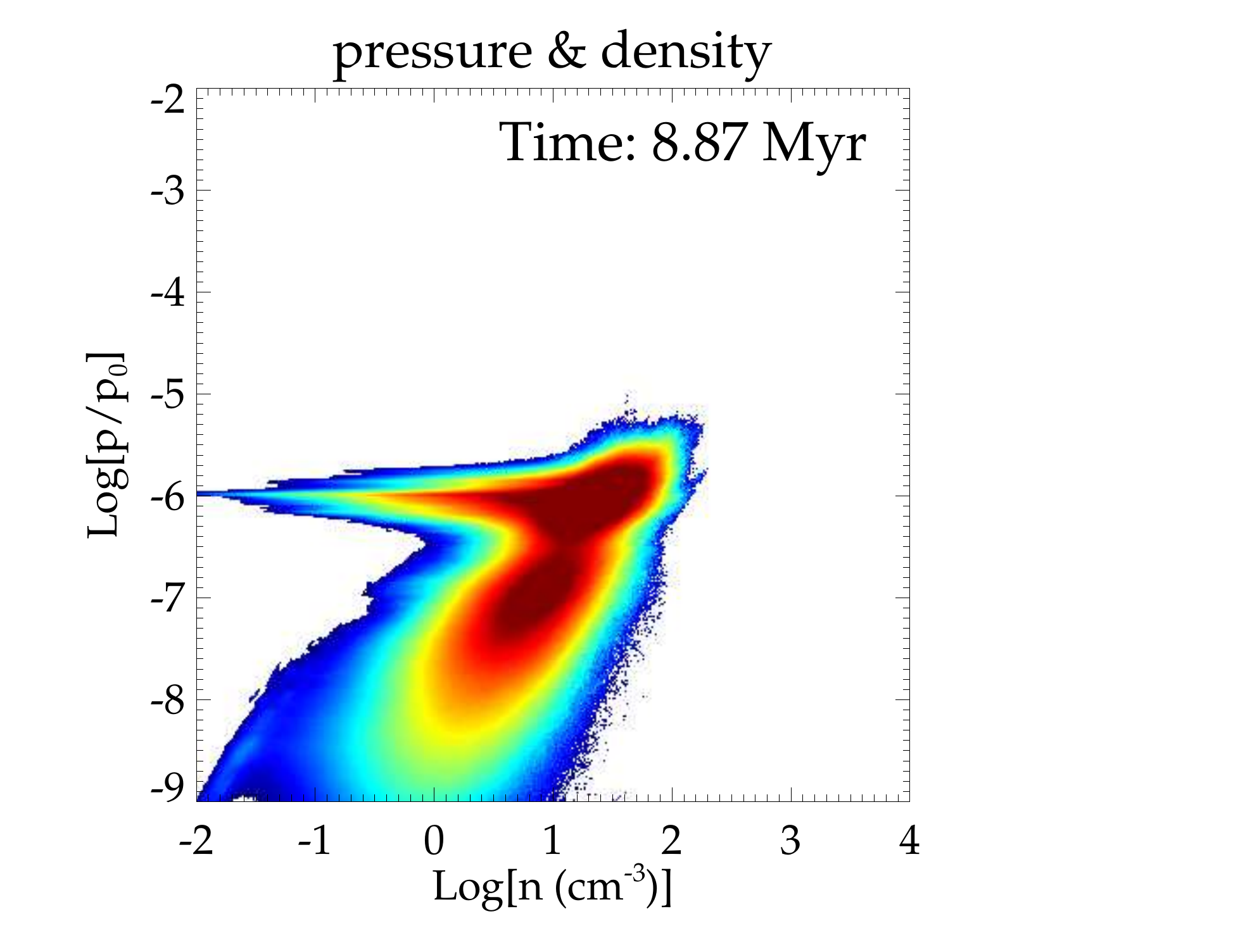}\hspace{-2cm}
	\includegraphics[width = 7cm,keepaspectratio] {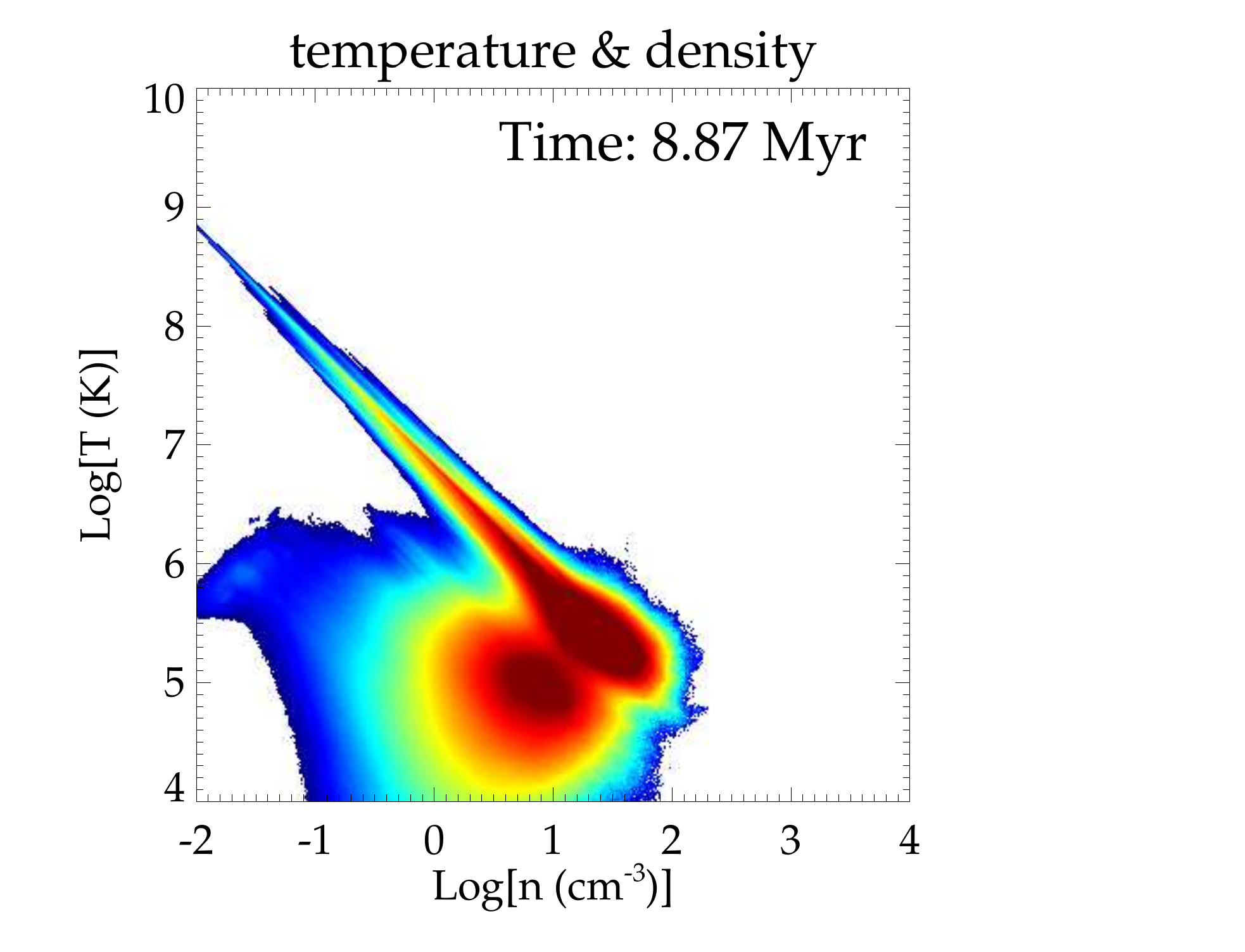}\hspace{-2cm}
	\includegraphics[width = 7cm,keepaspectratio] {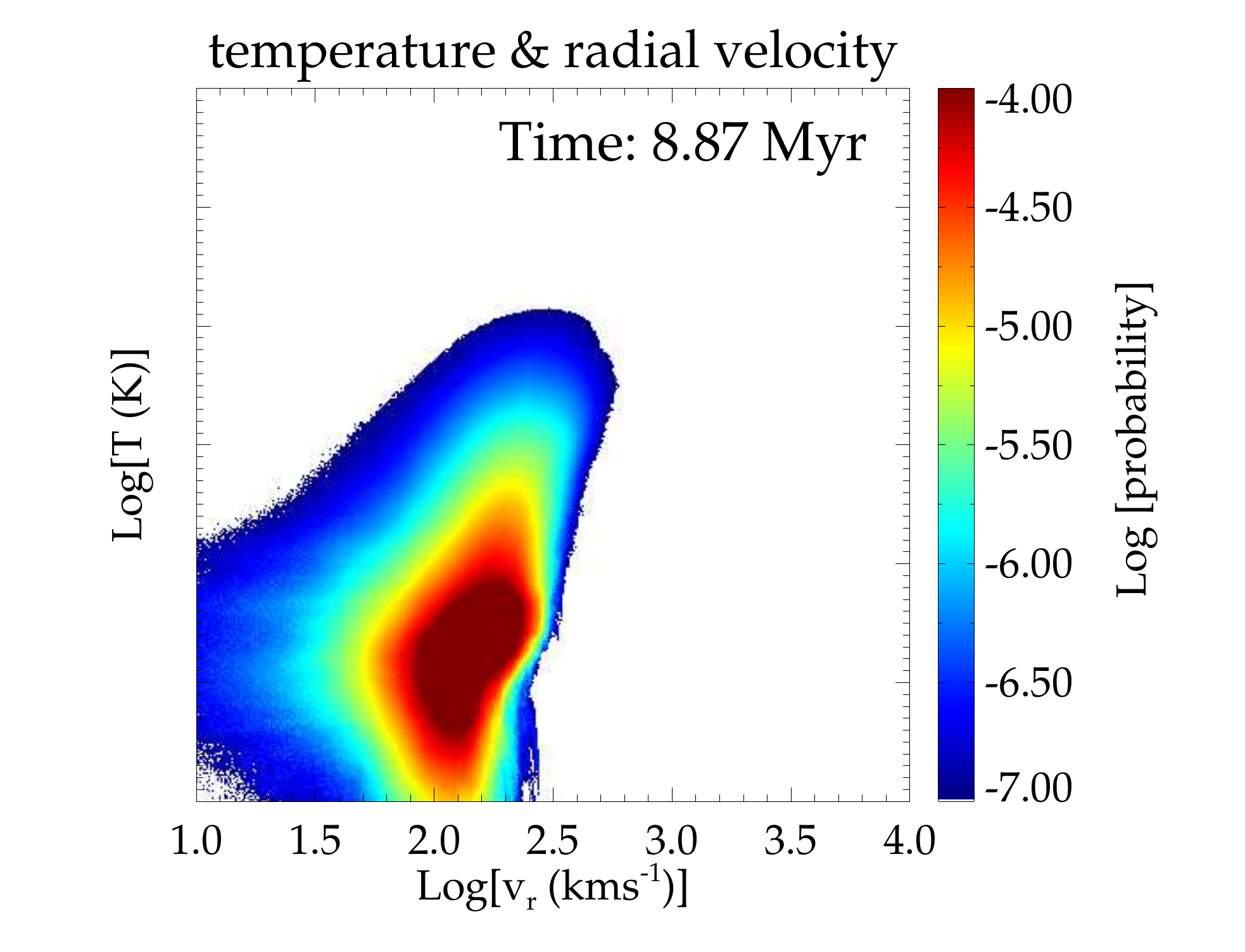}
	\caption{\small 2D PDF for Sim.~D with $P_{\rm jet}=10^{43}\ergs$ at $\sim 9$ Myr. The evolution of the ISM is significantly different from that of the high power jets, as inferred by comparing similar plots presented in Fig.~\ref{fig.2DPDF.prs-rho.nw300} - Fig.~\ref{fig.2DPDF.temp-vr.nw300}. A significant amount of the gas remains in the mildly hot phase ($\sim 10^5$ k) from the forward shock. The temperature-velocity plot shows very little outward radial
acceleration of the gas.}
	\label{fig.2DPDF.F43}
\end{figure*}
The coupling of the jet with the ISM depends significantly on the jet power. Jets with higher power, although more efficient in driving  powerful outflows, drill through the ISM more rapidly causing less shredding of the cloud cores situated a kiloparsec away from the jet axis. Low-power jets, on the other hand, remain trapped in the ISM for a longer time (see Fig~\ref{fig.jetsim.f43}), interacting with the ISM over a much larger volume. In Fig.~\ref{fig.jetrhocompare} where we compare the density for jets of different power at a time when the jet head is approximately at the same height for left and right panels. 
Jets with a power $\sim 10^{45} \ergs$ evacuate a central cavity of radius $\sim 0.5$ kpc, whereas simulation C with  $P_{\rm jet}=10^{44} \ergs$  
a larger cavity ($r \sim 1$ kpc) is evacuated. The situation is more marked for simulation D with $P_{\rm jet}=10^{43} \ergs$ where a larger central cavity is evacuated leaving only a few of the densest cores. The density PDFs of the simulation snap shots in Fig.~\ref{fig.jetrhocompare} are presented in Fig.~\ref{fig.jetpowercomparepdf}. The coefficients of the fits to the density portion of the density PDF following eq.~\ref{eq.hopkinsfit} is presented in Table~\ref{tab.comparepdffit}. For simulations with low power jets the mean density and dispersion of the density PDF is less than half of their counterparts with $P_{\rm jet} = 10^{45} \ergs$. This indicates more shredding of the dense cores by the trapped energy bubble. 

The evolution of the phase space is also significantly different for simulation~D, as shown in Fig.~\ref{fig.2DPDF.F43}. Unlike the high power jets, the phase space of the ISM cannot be distinctly divided into a high pressure energy bubble and gas shocked by the forward shock. The energy bubble is depicted by a horizontal branch in the left panel of Fig.~\ref{fig.2DPDF.F43}.  Comparing with the left panel of Fig.~\ref{fig.2DPDF.prs-rho.nw300}  depicting the initial condition of the ISM before the jet injection, we find the bubble for sim.~D to be very weakly over pressured. Slow evolution of the bubble facilitates cooling of the shocked gas. Most of the gas swept up by the bubble is shocked to $T\sim 10^5-10^6$ K, lower than that of the high power jets. The weak bubble is unable to accelerate the clouds to high velocities (as shown in the right panel of Fig.~\ref{fig.2DPDF.F43}). Thus, although the bubble is weak and does not drive significant outflows apparently indicative of
weak feedback, its effect on the density distribution is significant, since it is trapped for a longer time inside the ISM.

\section{Energetics of the jet disturbed ISM}\label{sec.energetics}
\subsection{Kinetic energy imparted to the ISM}
\begin{figure}
	\centering
	\includegraphics[width = 8cm, height = 9cm,keepaspectratio] {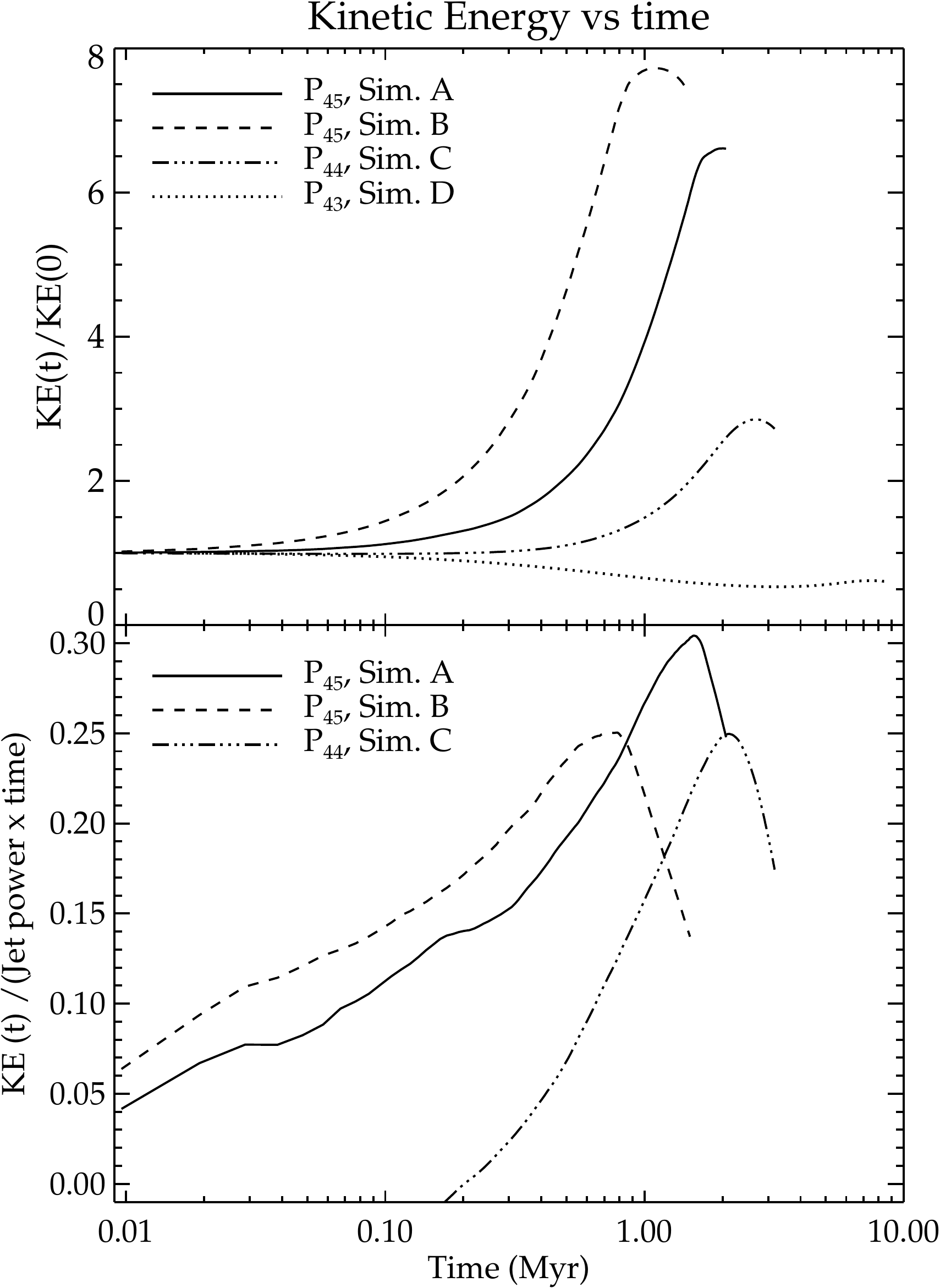}
	\caption{\small The evolution of the kinetic energy of the dense ISM ($\rho > 1 \cc$)} with time. The top panel shows the fractional increase of the kinetic energy of the ISM from its initial value. The lower panel plots the kinetic energy at a given time normalised to the total energy injected by jet till that time, indicating the efficiency of coupling of the jet.
	\label{fig.KEevolve}
\end{figure}
For any jet power, the jet couples strongly with the ISM initially as it drills through the dense medium. Fig.~\ref{fig.KEevolve} shows the evolution of the kinetic energy of the dense ISM  ($\rho > 1 \cc$) for different simulations. The kinetic energy rises with time and then decreases after the jet breaks out and decouples from the ISM. For a jet of power $\sim 10^{45}\ergs$ the kinetic energy of the ISM increases by a factor of 6 or more, whereas for $P_{\rm jet} \sim 10^{44}\ergs$ the kinetic energy increases by a factor of 3. 
The efficiency of jet feedback is better illustrated by plotting the ratio of the kinetic energy of the medium to the total integrated energy input of the jet as a function of time (lower panel in Fig.~\ref{fig.KEevolve}). We see that $\sim25-30\%$ of the jet energy is injected into the ISM, for jets of power $\gtrsim 10^{44} \ergs$. This is similar to the previous results of \citep{oneill05a,gaibler09a,wagner12a,hardcastle13a}, where intially the jet is shown to impart $\sim 20-30 \%$ of its energy as kinetic energy to the ISM, while the rest is deposited as internal energy, a fraction of which will be radiated away. More detailed analysis of the energy budget will be presented in a future work.

The evolution of the kinetic energy is however very different for a low power jet as in Sim.~D with $P_{\rm jet}=10^{43} \ergs$, due to the weakly over-pressured nature of the bubble. Although the jet strongly interacts with the ISM in its immediate surroundings, the total kinetic energy of the ISM decreases with time. This is because the jet evolves very slowly and at the initial stages the outer layers of the turbulent ISM are not affected. Even at later stages, since the weakly over pressured bubble evolves very slowly, the bulk kinetic energy of the gas decreases as a result of atomic cooling. Thus, simply computing the ratio of the total kinetic energy of the ISM to the energy injected by the jet as an indicator of the effect of feedback of the jet on the ISM is misleading in this case. The effect of the jet on the density PDF is a better probe of jet-ISM coupling for such cases.

\subsection{Galactic fountains and the effect on star formation}
\begin{figure}
	\centering
	\includegraphics[width = 6.8cm, height = 6.8cm,keepaspectratio] {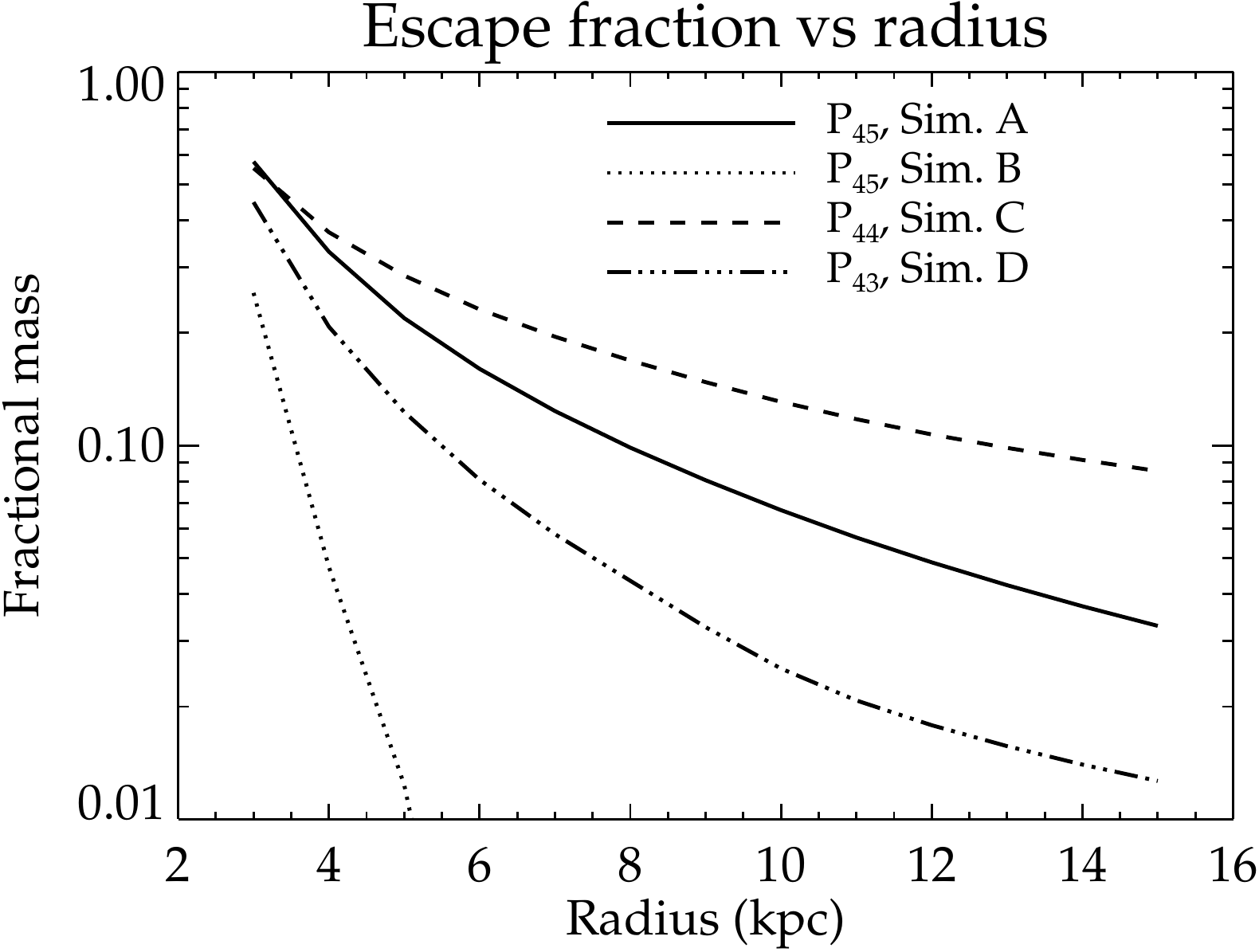}
	\caption{\small The figure shows the maximum radial distance a gas with a given velocity may reach under the influence of gravity, expressed as a fraction of the total ISM mass. }
	\label{fig.escapefrac}
\end{figure}
Although the high power jets ($P_{\rm jet} \gtrsim 10^{45} \ergs$) launch very fast outflows with speeds $\gtrsim 500 \kms$, the fraction of total mass ejected from the influence of the galaxy's potential is small. In Fig.~\ref{fig.escapefrac} we show the maximum radial distance the accelerated gas may reach by assuming a ballistic trajectory for the gas and solving
$\phi(r_{\rm max}) - \phi(r_0) = v_0^2/2$, for the maximum radius $r_{\rm max}$, where $r_0$ and $v_0$ are the initial radius and radial velocity, respectively.
 In this way we estimate the fractional mass of the ISM that will reach a given distance. The escape fraction is computed after the jet break out when the jet has decoupled from the ISM and the kinetic energy of the ISM saturates as shown in Fig.~\ref{fig.KEevolve}. We see that only a small fraction of the gas ($\lesssim 5\%$) goes beyond 10 kpc. Simulation~B with lower initial mean density has a larger fraction of mass ejected, as the less dense gas is easier to accelerate. For $P_{\rm jet} = 10^{43}\ergs$, all of the gas is retained within $\sim 5$ kpc. This indicates that although powerful jets can launch fast outflows, the fractional mass loss is very small and is slightly higher for an ISM with lower mean density.

The implication of this result is that the gas is not completely dispersed even though it is accelerated to a radial speed exceeding the velocity dispersion. Instead it forms a  galactic fountain in which the gas eventually falls back into the central regions of the galaxy. There may be a temporary inhibition of star formation as a result of the driving of gas out to a few kpc and the production of turbulence.

\begin{figure*}
	\centering
	\includegraphics[width = 20cm, height = 5.8cm,keepaspectratio] 
	{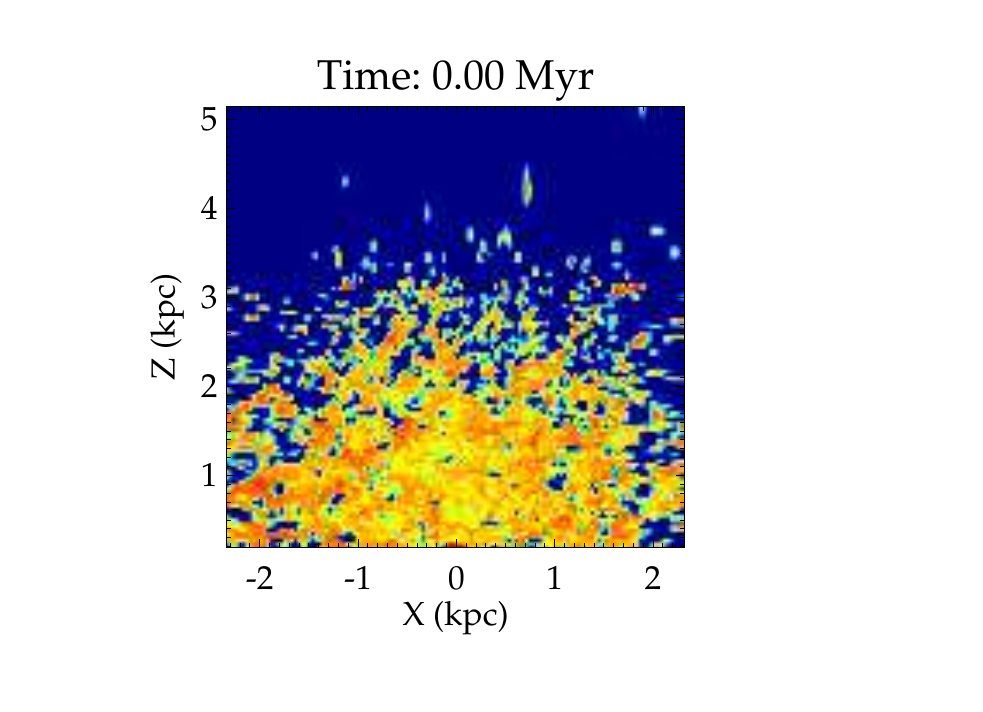}\hspace{-4cm}\vspace{-0.2cm}
	\includegraphics[width = 20cm, height = 5.8cm,keepaspectratio] 
	{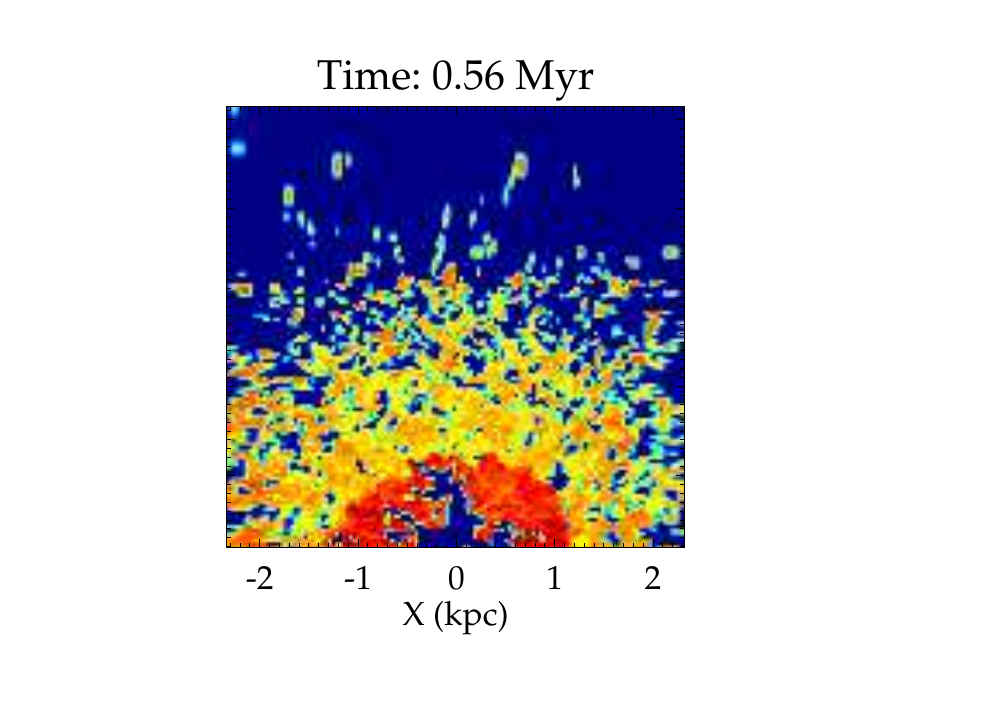}\hspace{-4cm}\vspace{-0.2cm}
	\includegraphics[width = 20cm, height = 5.8cm,keepaspectratio] 
	{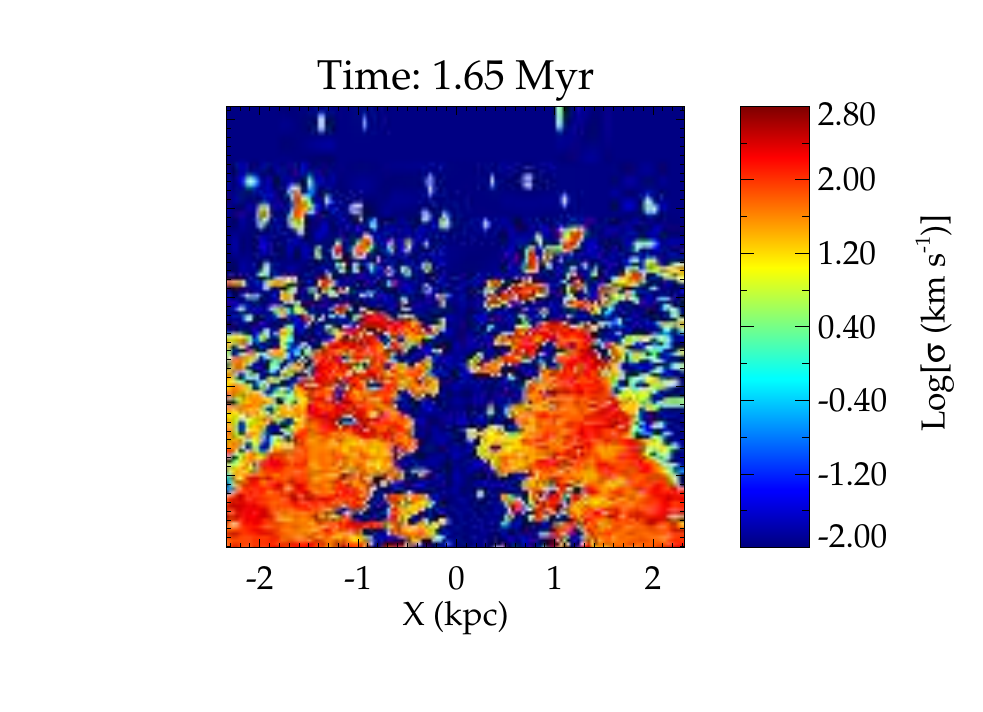}\vspace{-0.2cm}
	\caption{\small The velocity dispersion map at the $x-z$ plane for Simulation A.  The color bar represents $\log(\sigma)$, $\sigma$ being the mean of the dispersion of the three velocity components $\left(\sigma^2=\displaystyle\sum^{3}_{i=1} \sigma ^2_i/3 \right)$.}
	\label{fig.dispersion}
\end{figure*}
Thus, turbulent kinetic energy injected by the jet is probably a more important regulator of star formation activity than mass loss by outflows, as also reported recently in the observational papers by \citet{lanz15a} and \citet{alatalo15a}. We have calculated the mean velocity dispersion in our simulations by computing the variance of the three velocity components from adjacent cubes of dimensions $5\times5\times5$ cells in the  simulation domain. The jet-driven energy bubble significantly increases the turbulent velocity dispersion of the swept up gas, as shown in Fig.~\ref{fig.dispersion}.  The mass in the swept up shell is seen to have significantly high velocity dispersion indicating the existence of strong turbulent motions, which can significantly affect star formation \citep{krumholz05a,federrath12}. Detailed quantitative analysis of the effect of jets on the star formation rate in the host galaxy will be addressed in a future work.

\section{Summary and discussion}\label{sec.summary}
Let us now summarize the main results from this work:
\begin{enumerate}
\item \emph{Filamentary nature of settling ISM:}

The simulations of the settling, turbulent ISM result in elongated filaments of dense gas, rather than spheroidal clouds. The filaments are caused by shearing of dense clouds and turbulent mixing; this feature  is also seen in simulations of driven turbulence \citep[see][and references therein]{federrath15,federrath12}. Recent observations with high spatial resolution of high redshift  galaxies also report a filamentary nature of the extended halo of gas \citep{swinbank15a,swinbank15b,wisotzki15}. After $\sim 2$ Myr, corresponding approximately to half the dynamical time $\sim r_B/\sigma _B$, the filaments settle to form a turbulent central region of warm gas ($T\sim10^4$K) extending to about $\sim 2$ kpc and a hot halo ($T\sim10^7$K). Fig.~\ref{fig.fillfactor} shows the volume filling factor of the ISM, with the dense gas ($n>10 \cc$) having a filling factor $\lesssim 0.1$.

\item \emph{Evolution of the energy bubble and multi-phase ISM:}

The jet launches a high pressure energy bubble, which sweeps through the ISM. The bubble is preceded by a forward shock, which  heats the filaments to temperatures $\sim 10^5$ K but does not accelerate them (Fig.~\ref{fig.2DPDF.temp-vr.nw300}). The high pressure energy bubble progresses more slowly, shearing  and accelerating the filamentary fragments, creating fast radial outflows ($\sim 500 \kms$ for $P_{\rm jet}=10^{45} \ergs$). The shocks driven by the energy bubble create a multi-phase ISM of mildly hot ($T\sim 10^5$ K) gas in the outer layer of the forward shock. The forward shock is followed by an adiabatically expanding hot energy bubble ($T\sim 10^6-10^7$ K). Observational evidence of such shocked multi-phase ISM have been obtained from extended X-ray emission from radio loud galaxies \citep{kraft03a,croston07a,mingo11a,wang12a}. Clouds trapped in the energy bubble are shredded with their outer layers flowing out in fast, hot, low density outflows (with $T>10^6$ K, $n<10 \cc$, $v_r \sim 1000 \kms$). The dense central cores are radially accelerated to velocities of $\sim 500 \kms$. Such velocities are in agreement with observations of jet driven outflows \citep[see for example][and references therein]{wagner12a,collet16a}.

\item \emph{Feedback from low power jets:}

A significant result from this work is the effect of low power jets on the ISM of the host galaxy. High power jets, although more effective in launching faster outflows, are less destructive of the ISM since they efficiently drill through the ISM. Low power jets lack sufficient momentum to readily pierce the ISM and remain trapped for a longer time. This results in a more lateral spreading of the trapped energy bubble which causes enhanced shearing of the ISM filaments (as shown in Fig.~\ref{fig.jetrhocompare} and Fig.~\ref{fig.jetpowercomparepdf}). 
Such persistent coupling of a trapped  jet with the ambient ISM  will result in constant stirring of the turbulent ISM, inhibiting star formation in the process. This agrees with recent suggestions of suppressed star formation in some systems with a weak radio jet, such as NGC 1266 \citep{nyland13a,alatalo15a} and some  molecular hydrogen emission galaxies with weak radio jets \citep[MOHEGS,][]{ogle07a,ogle10a,lanz15a}.

As noted in Sec.~\ref{sec:intro}, the radio luminosity function implies that the distribution of 1.4~GHz radio power, $P_{1.4}$, peaks at around the FRI/FRII break at $10^{24.6} \> \rm W \> Hz^{-1}$ \citep{mauch07}. Approximately this corresponds to $P_{\rm jet} \sim 10^{42-43} \> \rm ergs \> s^{-1}$. Thus, given our results from the simulation with $P_{\rm jet} = 10^{43} \> \rm ergs \> s^{-1}$, we expect that low-powered jets with $ P_{\rm jet} \la 10^{43}\> \rm ergs \> s^{-1}$ should play a significant role in affecting the evolution of the ISM and star formation in the host galaxy.


\item \emph{Efficiency of feedback:}

The jet significantly couples to the ISM within the central few kpc before it breaks out into the ambient halo. From Fig.~\ref{fig.KEevolve} we see that nearly $\sim 30\%$ of the jet energy is transferred as kinetic energy to the ISM for high power jets. This measure of coupling efficiency\footnote{$E_{\rm kin}/P_{\rm jet} t$, $E_{\rm kin}$ being the kinetic energy of the dense gas ($n > 1 \cc$.)} is independent of jet power and density, as long as the jet creates a sufficiently over-pressured bubble. This agrees with previous results of \citet{wagner12a}.

\item \emph{Small net mass loss:}

Only a few percent of the dense gas mass is ejected from the galaxy to large distances (see Fig.~\ref{fig.escapefrac}). Most of the mass affected by the energy bubble is expected to rain back down into the galaxy's potential on free-fall time scales -- typically of the order of a few tens of Myr. This supports the galactic fountain scenario of jet driven feedback \citep[similar to][]{oppenheimer10a,dave12a}. The jets may cause temporary quenching of star formation by launching local outflows and making the ISM  turbulent, but the ejected mass will fall back and may be available for star formation after a few tens of Myr. The effect of such repeated cyclic explosive episodes and its connection to the AGN duty cycle needs to be explored in future work.

\end{enumerate}

\section{Acknowledgements}
This research was supported by the Australian Research Council through the
Discovery Project, \emph{The Key Role of Black Holes in Galaxy Evolution},
DP140103341. We thank Christoph Federrath, Matt Lehnert and Nicole Nesvadba for
useful discussions. We thank the HPC and IT teams at the National Computational
Infrastructure, the ANU, the Pawsey Supercomputing Centre and RSAA for their
help and support in carrying out the simulations and subsequent analysis.
We acknowledge  constructive comments by the referee, which assisted us in improving 
the original manuscript.

\appendix
\section{Probability density functions}\label{sec.hopkinsfit}

The lognormal distribution has proven to provide excellent description of the density probability distribution function (PDF) for simulations of isothermal turbulence \citep{kritsuk07,federrath10,li03a}. The distribution is defined as a Gaussian in $s=\ln \rho$ with a mean $-\sigma _s^2/2$ and variance $\sigma _s^2$:
\begin{equation}\label{eq.lognormal}
P_V(s) = \frac{1}{\sigma _s\sqrt{2 \pi}} \exp \left[-\frac{(s+\sigma _s^2/2)^2}{2 \sigma _s^2} \right]
\end{equation}
The subscript V refers to volume weighted PDF, which is considered primarily in this work while describing the 1D density PDFs. The above definition of the PDF satisfies the following the following two conditions of normalisations:
\begin{align}\label{eq.PDFnorm}
&\int_{-\infty}^\infty P_V(\ln \rho)d\ln \rho = 1 \nonumber \\
&\int_{-\infty}^\infty \rho P_V(\ln \rho) = \tilde{\rho} \, \, \mbox{(mean density)} \\
\end{align}
The mean of the distribution is
\begin{align}\label{eq.mean.LN}
\langle \ln \rho \rangle &= -\frac{\sigma _s^2}{2} \nonumber \\
\langle \rho \rangle &= \bar{\rho}  \\
\end{align}
and the variance in $\rho$ is
\begin{equation}\label{eq.var.LN1}
\sigma ^2_{\rho} = \bar{\rho}^2\left[\exp \left(\sigma _s^2 \right) -1 \right]
\end{equation}

In the presence of strong shocks, the density PDF shows significant departure from a true lognormal, especially in the high and low density tails. An improved function proposed by \citet{hopkins13a} gives a better description of the density PDF in the presence of intermittency:
\begin{align}\label{eq.hopkinsfit}
P_V(s) &= I_1\left(2\sqrt{\lambda u(s)}\right) \exp\left[-\lbrace\lambda + u(s)\rbrace \right]\sqrt{\frac{\lambda}{u(s) \eta ^2}} \nonumber \\
u(s) &= \frac{\lambda}{1+\eta} - \frac{s}{\eta} \, \, (u \geq 0) \, ; \, s=\ln \left(\rho/\bar{\rho}\right) \nonumber \\
\lambda &= \frac{\sigma _s^2}{2 \eta ^2} 
\end{align} 
Here $I_1$ is the modified Bessel function of the first kind. The PDF in eq.~\ref{eq.hopkinsfit} is defined by three parameters: the mean density $\bar{\rho}$ which is defined by \ref{eq.PDFnorm}, the dispersion ($\sigma _s$) and a parameter $\eta$ defining the degree of departure from a lognormality. For $\eta=0$, eq.~\ref{eq.hopkinsfit} reduces to the standard expression of a lognormal distribution (eq.~\ref{eq.lognormal}). The mean density of the improved function is same as in eq.~\ref{eq.mean.LN} above. The variance in $\rho$ is given by:
\begin{equation}\label{eq.var.LN2}
\sigma ^2_{\rho} = \bar{\rho}^2\left[\exp \left(\frac{\sigma _s^2}{1 + 3 \eta + 2\eta ^2} \right) -1 \right]
\end{equation}

\section{Adiabatic expansion of an energy bubble}\label{append.bubble}
For a spherical bubble with radius $R_B$, expanding with uniform pressure ($p_B$), driven by a constant input energy flux from a jet or wind ($P_j$), the energy equation can be written as
\begin{align}
&\frac{d}{dt} \left[\frac{4 \pi}{3} \frac{p_B}{\gamma -1} R_B^3 \right] 
 + 4 \pi R_B^2 p_B \frac{d R_B}{dt}  = P_j - L_{\rm cool} \label{eq.energy1} \\
&\frac{d}{dt} \left[p_B R^3 \gamma \right] = \frac{3(\gamma -1)}{4 \pi} P_j R_B^{3(\gamma -1)} \label{eq.energy2} 
\end{align}
where $\gamma$ is the adiabatic index and $L_{\rm cool}$ is energy loss from atomic cooling. The first term in the LHS of eq.~\ref{eq.energy1} is the change of internal energy inside the volume, while the second term is the work done by the expanding bubble. Since we are considering an adiabatically expanding bubble, we do not consider the cooling losses in eq.~\ref{eq.energy2}. If the radius expands as $R_B \propto t^\alpha$, following eq.~\ref{eq.energy2}, we find that the pressure to evolve as 
\begin{equation}\label{eq.presevolve}
p_B \propto t^{1-3\alpha}.
\end{equation}
For a self similarly expanding  bubble in a ISM of constant density ($\rho_0$), the radius and pressure evolve as \citep{castor75a,weaver77a}:
\begin{equation}\label{eq.rad.adiabatic}
R_B \propto \left(\frac{P_j }{\rho _0} \right)^{1/5} t^{3/5} \, ; \quad p_B \propto t^{-4/5}
\end{equation}
However in a multiphase ISM the bubble expansion is slower than the adiabatic case due to cooling losses and turbulent mixing \citep[e.g.][]{rosen14a}. For Fig.~\ref{fig.volbubble} we find the radius to evolve as $R_B \propto t^{0.55}$. For an adiabatically expanding bubble, this implies (following eq.~\ref{eq.presevolve}) that the pressure should vary as $p_B \propto t^{-0.65}$. This approximately agrees with the initial  evolution of the mean pressure of the bubble (shown in Fig.~\ref{fig.meanpres}), for the simulations with high power jets ($P_{\rm jet} \gtrsim 10^{44} \ergs$). This indicates that for jets with higher power, an over-pressured bubble is formed which initially evolves as an adiabatically expanding spherical bubble till jet break out.

\def\apj{ApJ}%
\def\mnras{MNRAS}%
\def\aap{A\&A}%
\def\apjl{ApJ}
\def\physrep{PhR}
\def\apjs{ApJS}
\def\pasa{PASA}
\def\pasj{PASJ}
\def\nat{Nature}
\def\memsai{MmSAI}
\def\aj{AJ}%
\def\aaps{A\&AS}%
\def\iaucirc{IAU~Circ.}%
\def\sovast{Soviet~Ast.}%
\def\apss{Ap\&SS}

\bibliographystyle{mnras}
\bibliography{dmrefs}

\end{document}